\documentclass[authoryear]{elsbook}
\usepackage{primer,url,multirow}
\usepackage{elsbook-main}

\usepackage{multirow}
\usepackage[dvips]{epsfig}
\usepackage{slashbox}
\usepackage{rotating}
\usepackage[dvips]{graphicx}
\usepackage{graphicx}
\usepackage{boxedminipage}
\usepackage{color}
\usepackage{amsfonts}
\usepackage{amssymb}
\usepackage{graphics,amsfonts,amssymb,epsfig,latexsym,amsfonts,graphicx,amssymb}

\usepackage[dvips,colorlinks=true, backref]{hyperref}

\newcommand{\bfdelta}{{\mbox{\boldmath $\delta$}}}

\newcommand{\bfnu}{{\mbox{\boldmath $\nu$}}}

\newcommand{\bflambda}{{\mbox{\boldmath $\lambda$}}}
\newcommand{\bfeta}{{\mbox{\boldmath $\eta$}}}
\newcommand{\bfkappa}{{\mbox{\boldmath $\kappa$}}}

\newcommand{\bU}{\mathbf{U}}
\newcommand{\bV}{\mathbf{V}}
\newcommand{\bfUpsilon}{{\mbox{\boldmath $\Upsilon$}}}

\def\BibTeX{{\rm B\kern-.05em{\sc i\kern-.025em b}\kern-.08em
    T\kern-.1667em\lower.7ex\hbox{E}\kern-.125emX}}

\title{Signal Processing and Optimal Resource Allocation for the Interference
Channel}
\author{Mingyi Hong and Zhi-Quan Luo}

\begin{document}
\maketitle

\section{Introduction}\label{secIntroduction}
\subsection{Resource Allocation in Communication Networks}\label{subSecResourceAllocation}

Resource allocation is a fundamental task in the design and management of
communication networks. For example, in a wireless network, we must
judiciously allocate transmission power, frequency bands,
time slots, and transmission waveforms/codes across multiple
interfering links in order to achieve high system performance while
ensuring user fairness and quality of service (QoS). The same is
true in wired networks such as the Digital Subscriber Lines (DSL).

The importance of resource allocation can be attributed to its key role
in mitigating multiuser interference. The latter is the main performance limiting factor for
heterogeneous wireless networks where the number of interfering macro/pico/femto base stations
(BS) can be very large.
In addition, resource allocation provides an efficient utilization of
limited resources such as transmission power and communication
spectrum. These resources are not only scarce but also expensive.
In fact, wireless system operators typically spend billions of
dollars to acquire licences to operate certain frequency bands. Moreover, the rising cost of
electricity for them to operate the infrastructure has already surpassed the salary cost to employees in some countries.
Thus, from the system operator's perspective,
efficient spectrum/power utilization directly leads to high investment return and low operating cost
(see e.g., \cite{goldsmith05} and \cite{cave07}). The transmission power of a
mobile terminal is another scarce resource.
In this case, careful and efficient  power allocation is the key to effectively prolong the
battery life of mobile terminals.


Current cellular networks allocate orthogonal resources to users.
For example, in a time-division multiplex access (TDMA) or a frequency-division
multiplex access (FDMA) network, users in the same cell transmit
in different time slots/frequency bands, and users in the
neighboring cells transmit using orthogonal frequency channels.
Although the interference from neighboring cells is suppressed, the
overall spectrum efficiency is reduced,
as each BS only utilizes a fraction of the available spectrum.
According to a number of recent studies \cite{fcc02,sahai09}
current spectrum allocation strategies are not
efficient, as at any given time and location, much of the allocated
spectrum appears idle and unused. Moreover, users in cell edges still suffer
from significant interference from non-neighboring cells, or pico/femto cells. In addition,
for cell edge users the signal power from their own
cells are typically quite weak. All of these factors can adversely affect their service quality.

%
%
To improve the overall system performance as well as user
fairness, future wireless standards \cite{3gpp09} advocate the notion of a heterogeneous network, in
which low-power BSs and relay nodes are densely deployed to provide
coverage for cell edge and indoor users.
This new paradigm of network design brings
the transmitters and receivers closer to each other, thus is able to
provide high link quality with low transmission power (\cite{damnjanovic11} and
\cite{chandrasekhar08}).
Unfortunately, close proximity of many transmitters and receivers also introduces
substantial in-network interference, which, if not properly managed, may significantly affect
the system performance. Physical layer techniques such as multiple input multiple output (MIMO) antenna
arrays and multiple cell coordination will be crucial for effective resource allocation and
interference management in heterogeneous networks.

An effective resource allocation scheme should allow not only flexible coordination among BS
nodes but also sufficiently distributed
implementation. Coordination is very effective for interference mitigation
among interfering nodes (e.g., Coordinated Multi-Point (CoMP)),
but is also costly in terms of signalling overhead. For example, CoMP requires
full BS coordination as well as the sharing of transmit data among all cooperating BSs.
In contrast, a distributed resource allocation requires far less signaling overhead and no data sharing, albeit at the cost of
a possible performance loss. For in-depth discussions of various
design issues in heterogenous networks, we recommend the recent articles and books
including \cite{Foschini06}, \cite{gesbert07}, \cite{gesbert10}, \cite{Sawahashi10}, \cite{damnjanovic11},
\cite{Martin-Sacristan2009} and \cite{khan09}.

In this article, we examine several design and complexity aspects of the
optimal physical layer resource allocation problem
for a generic interference channel (IC). The latter is a natural model for
multi-user communication networks. In particular, we characterize the computational
complexity, the convexity as well
as the duality of the optimal resource allocation problem. Moreover, we summarize various
existing algorithms for resource allocation and discuss their complexity and
performance tradeoff. We also mention various open research problems throughout the article.

\subsection{Notations}

Throughout, we use bold upper case letters to denote matrices,
bold lower case letters to denote vectors, and regular lower case letters to denote
scalars. For a symmetric (or Hermitian) matrix $\mathbf{X}$, the notation $\mathbf{X}\succeq \mathbf{0}$ (or
 $\mathbf{X}\succ \mathbf{0}$) signifies $\mathbf{X}$ is positive semi-definite (or definite). Users are denoted by subscripts, while frequency tone indices are denoted by superscripts. Zero mean normalized complex Gaussian distributions are denoted by ${\cal CN}(\mathbf{0},\mathbf{I})$.

\subsection{Interference Channels}\label{subSecInterferenceChannel}

\begin{figure*}[htb]
    \begin{minipage}[t]{0.48\linewidth}
    \centering
    {\includegraphics[width=
0.6\linewidth]{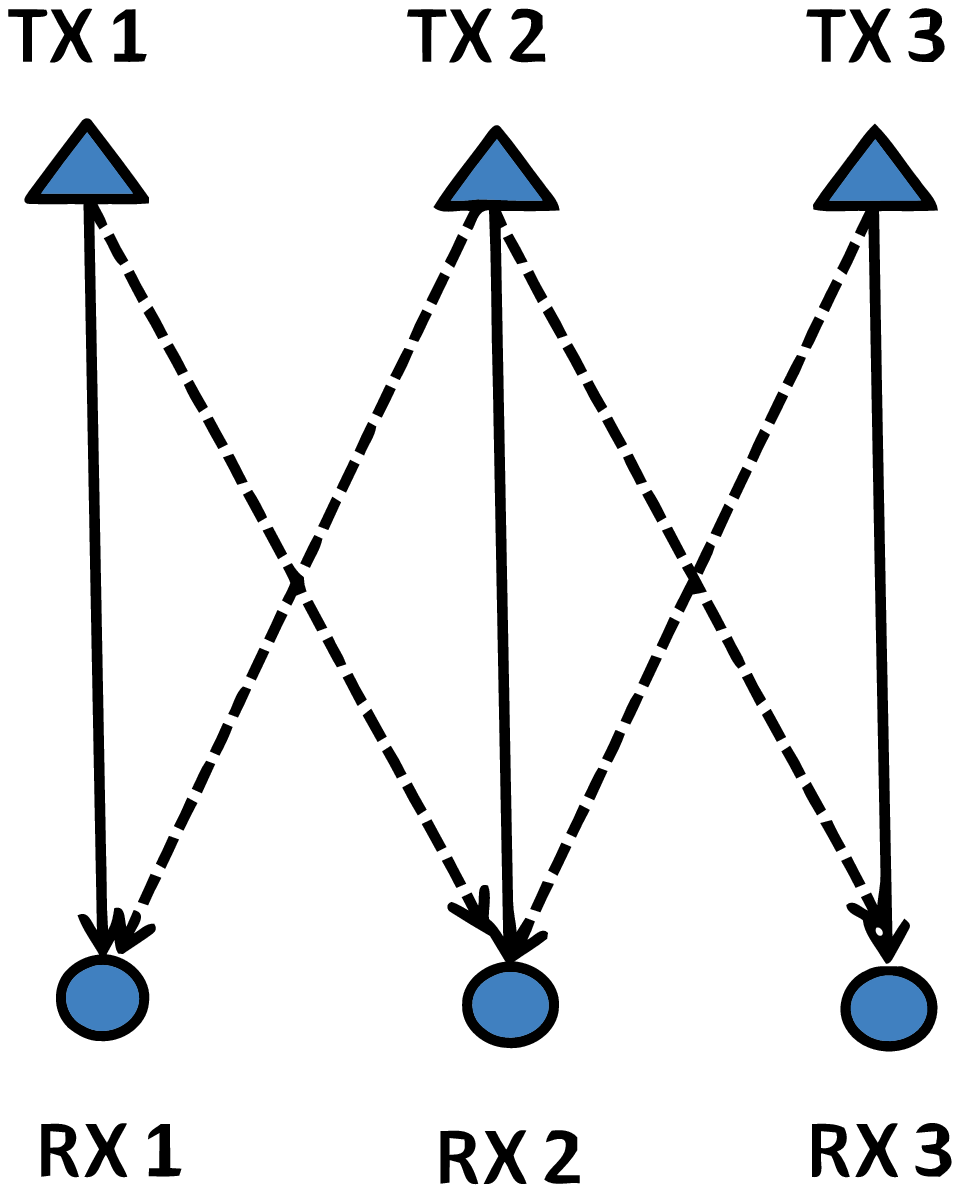} \caption{\small The Interference
Channel model. The solid lines represent the direct channels, while
the dotted lines represent the interfering channels.}\label{figIC} }
    \end{minipage}
    \hfill
        \begin{minipage}[t]{0.48\linewidth}
    {\includegraphics[width=
0.8\linewidth]{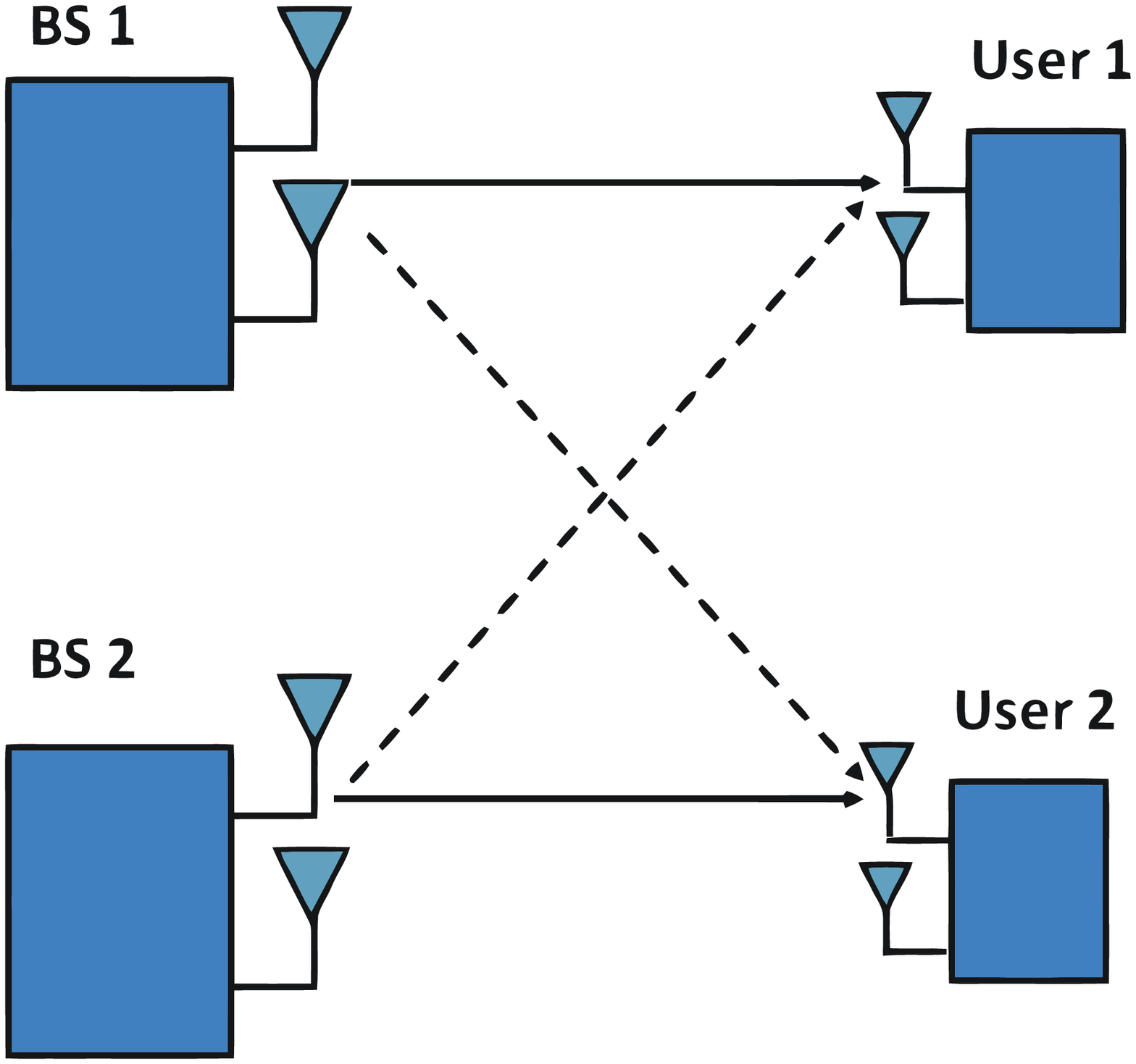} \caption{\small Illustration of a
2-user multiple input multiple output (MIMO) interference channel.
}\label{figMIMO} }
\end{minipage}
\end{figure*}

An interference channel (IC) represents a communication network in
which multiple transmitters simultaneously transmit to their
intended receivers  in a common channel. See Fig. \ref{figIC} for a
graphical illustration of the IC. Due to the shared communication
medium, each transmitter generates interference to all the other
receivers. The IC model can be used to study many practical
communication systems. The simplest example is a wireless ad hoc
network in which transmitters and their intended receivers are
randomly placed. When all these nodes are equipped with multiple
antenna arrays, the channel becomes a MIMO IC. See Fig.\
\ref{figMIMO} for a graphical illustration of a 2-user MIMO IC. If
each transmitter and receiver pair communicates over multiple
parallel subchannels, the resulting overall channel model becomes a
{\it parallel IC}. This parallel IC model can be used to describe
communication networks employing Orthogonal Frequency Division
Multiple Access (OFDMA) where the available spectrum is divided into
multiple independent tones/channels. Networks of this kind include
the DSL network or the IEEE 802.11x networks.

\begin{figure*}[ht]
    \begin{minipage}[t]{0.48\linewidth}
    \centering
     {\includegraphics[width=
0.9\linewidth]{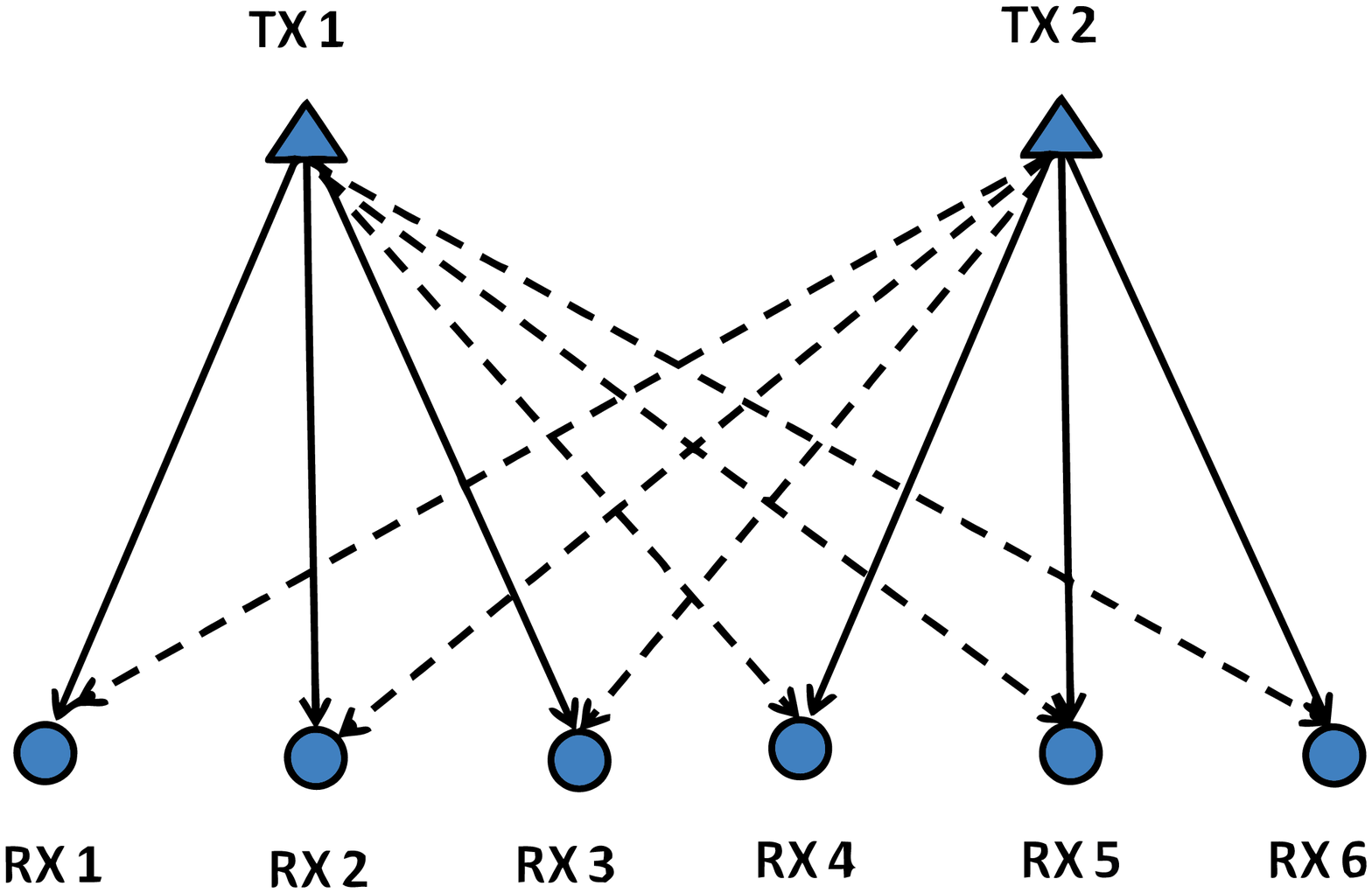} \caption{\small The Interfering
Broadcast Channel model. The solid lines represent the direct
channels, while the dotted lines represent the interfering
channels.}\label{figIBC} }
\end{minipage}\hfill
    \begin{minipage}[t]{0.48\linewidth}
    \centering
    {\includegraphics[width=
0.9\linewidth]{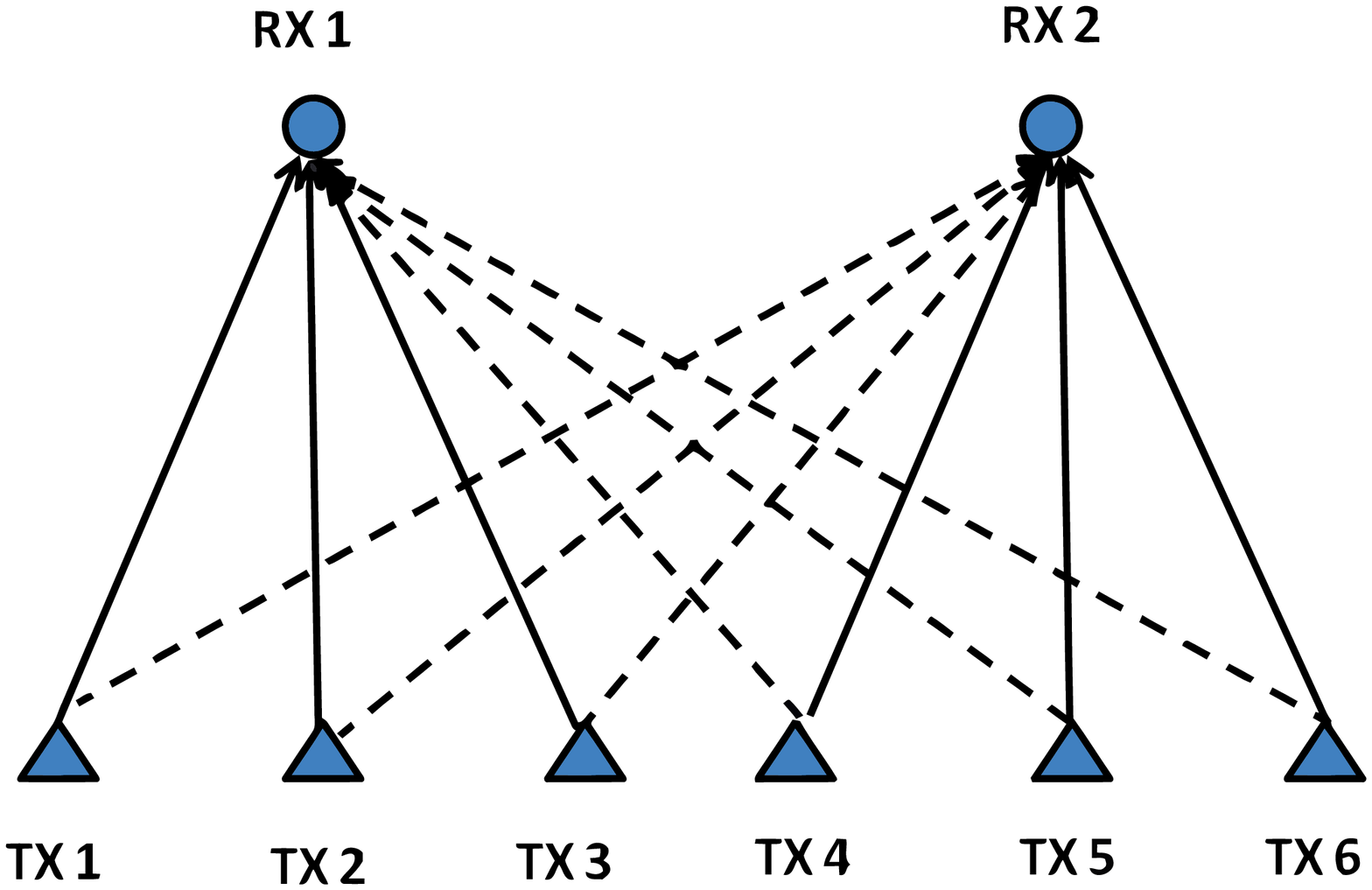} \caption{\small The Interfering
Multiple Access Channel model. The solid lines represent the direct
channels, while the dotted lines represent the interfering
channels.}\label{figIMAC} }
\end{minipage}
    \end{figure*}

Another practical network is the
multi-cell heterogenous wireless network. In the downlink of such network a
set of interfering transmitters (BSs) simultaneously transmit to
their respective groups of receivers. This channel is hitherto referred as
an interfering broadcast channel (IBC). The
uplink of this network can be modeled as an interfering multiple
access channel (IMAC). See Fig.\ \ref{figIBC} and Fig.\
\ref{figIMAC} for graphical illustrations of these two channel
models. Note that both IMAC and IBC reduce to an IC when there is
only a single user in each cell.

Our ensuing discussions will be focussed on these channel models. We will
illustrate key computational challenges associated with optimal resource allocation and
suggest various practical resource allocation approaches to overcome them.

\subsection{System Model}\label{subSystemModel}
We now give
mathematical description for three types of IC model -- the scalar,
parallel and MIMO IC models. Let us
assume that there are $K$ transmitter and receive pairs in the system,
and we refer to each transceiver pair as a user. Let
$\mathcal{K}=\{1,\cdots, K\}$ denote the set of all the users.

\subsubsection{Scalar IC Model}

In a scalar IC model each user transmits and receives a scalar
signal. Let $x_k\in\mathcal{C}$ denote user $k$'s transmitted signal,
and let $p_k=|x_k|^2$  denote its power. Let
 $\bar{p}_k$ denote user $k$'s power constraint:
 $p_k\le \bar{p}_k$. Let $z_k\sim\mathcal{CN}(0,1)$ denote user
$k$'s normalized complex Gaussian noise with unit variance. Note that we have normalized the power
of the noise to unity. Let $H_{lk}\in\mathcal{C}$ denote the channel
between transmitter $l$ and receiver $k$. Then user $k$'s received
signal $y_k\in\mathcal{C}$ can be expressed as
\begin{eqnarray}
y_k=\underbrace{H_{kk}x_k}_{\textrm{user $k$'s intended
signal}}+\underbrace{\sum_{l\ne k}H_{lk}x_l}_{\mbox{multiuser
interference}}+z_k.
\end{eqnarray}
The signal to interference plus noise ratio (SINR) for user $k$ is defined as
\begin{eqnarray}
\mbox{SINR}_k=\frac{|H_{kk}|^2p_k}{1+\sum_{l\ne k}|H_{lk}|^2 p_l}\label{eqSINRScalar}.
\end{eqnarray}
We denote the collection of all the users' transmit powers as $\mathbf{p}=\left[p_1,\cdots,p_K\right]^{T}$.

\subsubsection{Parallel IC Model}

In a parallel IC model,  the spectrum is divided into $N$
independent non-overlapping bands, each giving rise to a parallel subchannel.
Let $\mathcal{N}$ denote the set
of all  subchannels. Let $x^n_k\in\mathcal{C}$ denote the transmitted
signal of user $k$ on channel $n$, and let $p^n_k=|x^n_k|^2$ denote
its power. We use $\bar{p}_k$ to denote user $k$'s power budget so that
$\sum_{n=1}^{N}p^n_k\le \bar{p}_k$. Let $H^n_{lk}\in\mathcal{C}$
denote the channel coefficient between the transmitter of user $l$
and the receiver of user $k$ on channel $n$. Let $z^n_k\sim
\mathcal{CN}(0,1)$ denote the Gaussian channel noise. The received signal of user $k$ on subchannel $n$,
denoted as $y^n_k\in\mathcal{C}$, can be expressed as
\begin{eqnarray}
y^n_k=\underbrace{H^n_{kk}x^n_k}_{\textrm{user $k$'s intended
signal}}+\underbrace{\sum_{l\ne
k}H^n_{lk}x^n_l}_{\mbox{multiuser
interference}}+z^n_k.\label{eqParallelChannelIO}
\end{eqnarray}
We define the collection of user $k$'s transmit power as
$\mathbf{p}_k=\left[p^1_k,\cdots,p^N_k\right]^{T}$, and define all
the users' transmit powers as $\mathbf{p}=
\left[\mathbf{p}^{T}_1,\cdots,\mathbf{p}^{T}_K\right]^{T}$.

\subsubsection{MIMO IC Model}

In a  MIMO IC model the receivers and transmitters are
equipped with $N_r$ and $N_t$ antennas, respectively. Let
$\mathbf{x}_k\in\mathcal{C}^{N_t}$ and
$\mathbf{y}_k\in\mathcal{C}^{N_r}$ denote the transmitted and
received signal of user $k$. Let $\mathbf{H}_{lk}\in\mathcal{C}^{N_r\times
N_t}$ represent the channel gain coefficient matrix between
transmitter $l$ and receiver $k$.

Suppose each user $k$ transmits/receives $d_k$ data streams, and let
$\mathbf{s}_k\in\mathcal{C}^{d_k\times 1}$ and
$\widehat{\mathbf{s}}_k\in\mathcal{C}^{d_k\times 1}$ denote the
transmitted symbols and the received {\it estimated} symbols,
respectively. Assume that the data vector $\mathbf{s}_{k}$ is
normalized so that $\mathbb{E}[\mathbf{s}_{k} \mathbf{s}_{k}^H] =
\mathbf{I}$, and that the data signals for different users are
independent from each other. Throughout this article, we will focus
on {\it linear strategies} in which users use beamformers to
transmit and receive data symbols. Let
$\mathbf{V}_k\in\mathcal{C}^{N_t\times d_k}$ and
$\mathbf{U}_k\in\mathcal{C}^{N_r\times d_k}$ denote the transmit and
receive beamformers, respectively. Let
$\mathbf{z}_k\sim\mathcal{CN}(0,\mathbf{I}_{N_r})$ denote the
normalized complex Gaussian noise vector at receiver $k$, where
$\mathbf{I}_{N_r}$ is the $N_r\times N_r$ identity matrix.  Then the
transmitted and received signal for user $k$ can be expressed as
\begin{eqnarray}
&&\mathbf{x}_k=\mathbf{V}_k\mathbf{s}_k. \label{eqTransmitBeamFormer}\\
&&\mathbf{y}_k=\underbrace{\mathbf{H}_{kk}\mathbf{x}_k}_{\textrm{user
$k$'s intended signal}}+\underbrace{\sum_{l\ne
k}\mathbf{H}_{lk}\mathbf{x}_l}_{\textrm{multiuser
interference}}+{\mathbf{z}_k}
\label{eqMIMO}\\
&&\widehat{\mathbf{s}}_k=\mathbf{U}^H_k\mathbf{y}_k \label{eqReceivBeamformer}.
\end{eqnarray}

Let $\mathbf{Q}_{k}=E[\mathbf{x}_k\mathbf{x}_k^{H}]$ denote the
covariance matrix of the transmitted signal of user $k$. We assume that
each transmitter has an averaged total power budget of the form
\begin{eqnarray}
\mbox{Tr}(\mathbf{Q}_k)\le \bar{p}_k,~k=1,\cdots,{K}\label{eqMIMOPowerBudget}.
\end{eqnarray}

When we have a single stream per user, $\mathbf{V}_k$ and $\mathbf{U}_k$
reduce to vectors $\mathbf{v}_k \in\mathcal{C}^{N_t\times 1}$ and $\mathbf{u}_k\in\mathcal{C}^{N_r\times 1}$.
In this case the SINR for user $k$'s stream can be defined as
\begin{eqnarray}
\mbox{SINR}_k=\frac{|\mathbf{u}^{H}_k\mathbf{H}_{kk}\mathbf{v}_k|^2}
{\|\mathbf{u}_k\|^2+\sum_{l\ne k}|\mathbf{u}^{H}_k\mathbf{H}_{lk}\mathbf{v}_l|^2}\label{eqMIMOSINR}.
\end{eqnarray}

Multiple Input Single Output (MISO) IC is a special case of  MIMO IC
in which the receivers only have a single antenna. In this case each
user can only transmit a single stream ($d_k=1$,
$s_k\in\mathcal{C}$), and the beamforming matrix $\mathbf{V}_k$
reduces to a beamforming vector
$\mathbf{v}_k\in\mathcal{C}^{N_t\times 1}$. The channel coefficient
matrix $\mathbf{H}_{kl}$ becomes a {\it row vector}
$\mathbf{h}_{kl}$, the received signal $\mathbf{y}_k$ reduces to a
scalar, which can be expressed as
\begin{eqnarray}
{y}_k=\underbrace{\mathbf{h}_{kk}\mathbf{x}_k}_{\textrm{user $k$'s
intended signal}}+\underbrace{\sum_{l\ne
k}\mathbf{h}_{lk}\mathbf{x}_l}_{\textrm{multiuser
interference}}+{{z}_k} \label{eqMISO}.
\end{eqnarray}
The SINR for each user $k$ can be expressed as
\begin{eqnarray}
\mbox{SINR}_k=\frac{|\mathbf{h}_{kk}\mathbf{v}_{k}|^2} {1+
\sum_{l\ne k} |\mathbf{h}_{lk}\mathbf{v}_{l}|^2}\label{eqMISOSINR}.
\end{eqnarray}
The power budget constraint becomes
\begin{eqnarray}
\|\mathbf{v}_k\|^2\le \bar{p}_k,~k=1,\cdots,K.\label{eqMISOPowerBudget}
\end{eqnarray}


\section{Information-Theoretic Results}\label{secInformationTheory}
\subsection{Capacity Results for IC Model}\label{subSecCapacity}
In this subsection we briefly review some information theoretical
results related to the capacity of the interference channel.

Consider a single user point to point additive white Gaussian noise
(AWGN) scalar channel in the following form
\begin{eqnarray}
y=H x+z
\end{eqnarray}
where $x$, $y$, $H$, $z$ are the transmitted signal, the received
signal, the channel coefficient and the Gaussian noise,
respectively. Assume that the noise is independently distributed as
$z\sim \mathcal{CN}(0,1)$, and that the signal has a power
constraint $ |x|^2\le \bar{p}$. An achievable transmission rate $R$
for this channel is defined as the rate that can be
transmitted and decoded with diminishing error probability. The capacity of a channel $C$ is the supremum of all
achievable rates. Let us define the signal to noise ratio (SNR) of
the channel as $\mbox{SNR}={\bar{p}|H|^2}$, then the capacity of the
Gaussian channel  is given by
\begin{eqnarray}
C(\mbox{SNR})= \log_2\left(1+\mbox{SNR}\right)~\textrm{bit per
transmission}.
\end{eqnarray}
We refer the readers' to the classic books such as \cite{cover05} and
\href{http://classx.stanford.edu/View/Subject.php?SubjectID=2011_Q1_EE376_Lec}{the
online course} for an introductory treatment of information theory.

Now consider a 2-user interference channel
\begin{eqnarray}
y_1&=&H_{11}x_1+H_{21}x_2+z_1\nonumber\\
y_2&=&H_{22}x_2+H_{12}x_1+z_2\label{eqTwoUser}.
\end{eqnarray}
The capacity region of this channel is the set of all achievable
rate pairs of user $1$ and user $2$. Unlike the previous point to
point channel, the complete characterization of the capacity region in this simplest
2-user IC case is an open problem in information theory.
The largest achievable rate region for the interference channel is the
Han-Kobayashi region \cite{han81}, and it is achieved using
superposition coding and interference subtraction. Recently,
\cite{etkin07} showed that this inner region is within one bit of the capacity region for
scalar ICs. The capacity of the scalar interference channel under
strong or very strong interference has been found in
\cite{carleial75}, \cite{sato81} and \cite{han81}. In particular, in the very strong
interference case, i.e., $\frac{|H_{21}|^2}{{|H_{11}|^2}}\ge
1+\bar{p}_1$ and $\frac{|H_{12}|^2}{{|H_{22}|^2}}\ge 1+\bar{p}_2$,
the capacity region is given as
\begin{eqnarray}
R_k\le\log_2\left(1+{|H_{kk}|^2 \bar{p}_k}\right),~k=1,2
\end{eqnarray}
where $R_k$ is the transmission rate for user $k$. This result
indicates that in very strong interference case the capacity is not
reduced. The references \cite{shang10} and \cite{Chung07} include recent results
that establish the capacity region for more general MIMO and parallel ICs
in the strong interference case. However, for the general case where the
interference is moderate, the capacity region remains
unknown.

The capacity of a communication
channel can be approximated by the notion of {\it degrees of freedom}. Recall that in the high SNR regime
the capacity of a point to point link can be expressed as
\begin{eqnarray}
C(\mbox{SNR})=d\log_2(\mbox{SNR})+o(\log_2(\mbox{SNR})).
\end{eqnarray}
In this case we say the channel has $d$ degrees of freedom. In a 2-user
interference channel, the {\it degrees of freedom region} can be
characterized as follows. Let the sum transmit power across all the transmitters
be $\rho$, and let $R_k(\rho)$ denote the transmission rate
achievable for user $k$. Then the capacity region $C(\rho)$ of this
2-user channel is the set of all achievable rate tuples $(R_1(\rho),
R_2(\rho))$. The degree of freedom region $\mathcal{D}$ for this
channel approximates the capacity region, and is defined as (see \cite{Cadambe08})
\begin{eqnarray}
\mathcal{D}&=&\Bigg\{(d_1, d_2)\in \mathcal{R}^2_{+}: \forall (w_1,
w_2)\in \mathcal{R}^2_{+}, w_1 d_1+w_2 d_2 \nonumber\\
&\le& \lim_{\rho\to\infty}\sup\left[ \sup_{(R_1(\rho), R_2(\rho))\in
C(\rho)}[w_1 R_1(\rho)+w_2 R_2(\rho)]\frac{1}{\log_2(\rho)}\right]
\Bigg\}.
\end{eqnarray}

The goal of resource allocation is to achieve the optimal performance established by information theory, subject to resource budget constraints. Unfortunately, optimal strategies for achieving the information theoretic limits are often unknown, too difficult to compute or too complicated to implement in practice. For practical considerations, we usually rely on simple transmit/receive strategies (such as linear beamformers) for resource allocation, with the goal of attaining an approximate information theoretic performance bounds. The latter can be in terms of the degrees of freedom or some approximate capacity bounds which we describe next.

\subsection{Achievable Rate Regions When Treating Interference as Noise}\label{subSecOptimalTheoretical}

Due to the difficulties in characterizing the capacity region
and the optimal transmit/receive strategy for
a general interference channel, many works in the literature study
simplified transmit/receive strategies and the corresponding achievable
rate regions. One such simplification,
which is well motivated from practical considerations, is to assume that
low-complexity single user receivers are used and that the multiuser
interference is treated as additive noise. The authors of \cite{shang09, shang11}
show that treating interference as noise in a Gaussian IC
actually achieves the sum-rate channel capacity if the channel
coefficients and power constraints satisfy certain conditions.
These results serve as a theoretical justification for this simplification. In the rest of this article
we will treat interference as noise at the receivers. Let us first review some achievable rate
region results for different IC models with this simplified assumption.

\subsubsection{Definition of Rate Region}\label{subsubSecRateRegion}
Consider the  2-user scalar IC
(\ref{eqTwoUser}). The users' transmission powers are constrained by $0\le
p_1\le \bar{p}_1$ and $0\le p_2\le \bar{p}_2$, respectively. The
following rates are achievable when the users treat their respective
interference as noise
\begin{eqnarray}
R_1(p_1, p_2)&=\log_2\left(1+\mbox{SINR}_1\right)\nonumber\\
R_2(p_1, p_2)&=\log_2\left(1+\mbox{SINR}_2\right)\nonumber
\end{eqnarray}
where the term $\mbox{SINR}_k$ has been defined in (\ref{eqSINRScalar}).
The {\it directly achievable} rate region $\bar{\mathcal{R}}$ is defined as the union of the
achievable rate tuples $(R_1(p_1,p_2), R_2(p_1,p_2))$
\begin{eqnarray}
\bar{\mathcal{R}}=\left\{(R_1(p_1,p_2), R_2(p_1,p_2)): 0\le p_1\le \bar{p}_1, 0\le p_2\le
\bar{p}_2\right\}.\label{eqRateTwoUser}
\end{eqnarray}
The { directly achievable} rate region represents the set of achievable rates when
the transmitters are not able to synchronize with each other (\cite{hui85}). If transmitter
synchronization is possible, time-sharing among the extreme points of the directly
achievable rate region can be performed. In this case, the achievable rate region becomes
the convex hull of the directly achievable rate region (\ref{eqRateTwoUser}). Sometimes for convenience,
we will refer the directly achievable rate regions simply as {\it rate regions}. The exact meaning of the rate region should be clear from the corresponding context.

For a parallel IC model, user $k$'s achievable rate on channel $n$,
$R^n_k$, can be expressed as
\begin{eqnarray}
R^n_k(p^n_{1},\cdots,p^n_K)=\log_2\left(1+\frac{|H^n_{kk}|^2
p^n_k}{1+ \sum_{l\ne k}|H^n_{lk}|^2
p^n_l}\right).\label{eqParallelChannelLinkCapacity}
\end{eqnarray}
User $k$'s achievable sum rate is the sum of the rates achievable on
all the channels
\begin{eqnarray}
R_k=\sum_{n=1}^{N}R^n_k(p^n_{1},\cdots,p^n_K).\label{eqparallelChannelIndividualCapacity}
\end{eqnarray}
The directly achievable rate region $\bar{\mathcal{R}}$ in this case can be expressed as
\begin{eqnarray}
\bar{\mathcal{R}}=\left\{(R_1, \cdots, R_K): \sum_{n=1}^{N}{p^n_k}\le
\bar{p}_k, p_k\ge 0, \forall~k\in\mathcal{K}\right\}.
\end{eqnarray}

For a MIMO IC model, user $k$'s achievable rate when treating all other
users' interference as noise is
\begin{eqnarray}
R_k(\mathbf{Q}_1,\cdots,\mathbf{Q}_K)&=\log_2\det\left(\mathbf{I}_{N_{r}}+\mathbf{H}_{kk}\mathbf{Q}_{k}\mathbf{H}_{kk}^{H}
\left(\mathbf{I}_{N_r}+ \sum_{l\ne k}
\mathbf{H}_{lk}\mathbf{Q}_{l}\mathbf{H}_{lk}^{H}\right)^{-1}\right)\label{eqMIMORate}.
\end{eqnarray}
 The directly achievable rate region $\bar{\mathcal{R}}$ can be
expressed as
\begin{eqnarray}
\bar{\mathcal{R}}=\left\{(R_1,\cdots,R_K):\mbox{Tr}(\mathbf{Q}_k)\le
\bar{p}_k,~\mathbf{Q}_k\succeq 0,~\forall~k\in\mathcal{K}\right\}.
\end{eqnarray}

\subsubsection{Characterization of the Directly Achievable Rate Regions}

Resource allocation requires a good understanding of the achievable
rate regions. The (directly achievable) rate regions of the 2-user
and the more general $K$-user scalar IC have been recently
characterized in \cite{charafeddine07,charafeddine09}. We briefly
elaborate the 2-user rate region and its properties. Let $\Phi(p_1,
p_2)$ denote a point in the rate region with $x,y$ coordinates
representing $R_1(p_1, p_2)$ and $R_2(p_1, p_2)$, respectively. Let
$\bar{p}_1=\bar{p}_2=\bar{p}$. Define two functions $\Phi_1(p_2)=
\Phi(\bar{p}, p_2)$, and $\Phi_2(p_1) = \Phi(p_1, \bar{p})$. Then
the boundary of the 2-user rate region consists of the union of two
axis and the following two curves
\begin{eqnarray}
\Phi_1(
p_2)&=&\log_2\left(1+\frac{\frac{|H_{22}|^2}{|H_{21}|^2}(|H_{11}|^2\bar{p}-(2^{R_1}-1))}{(2^{R_1}-1)(1+|H_{12}|^2
\bar{p})}\right),~0\le p_2\le\bar{p} \label{eqFrontier1}\\
\Phi_2(p_1)&=&\log_2\left(1+\frac{|H_{22}|^2\bar{p}}{1+\frac{|H_{12}|^2}{|H_{11}|^2}
(1+|H_{21}|^2\bar{p})(2^{R_1}-1)}\right),~0\le
p_1\le\bar{p}\label{eqFrontier2}.
\end{eqnarray}
Each of the above two curves consists of the set of rates achievable by
one transmitter using its full power, while the other
transmitter sweeping over its range of transmit powers. The
convexity of this 2-user directly achievable rate region is studied in
\cite{charafeddine09}. The following two
conditions are sufficient to guarantee the convexity of the directly
achievable rate region
\begin{eqnarray}
\frac{\partial^2\Phi_1(p_2)}{\partial R^2_1}\bigg|_{p_2}<0,
~\forall~0\le p_2\le \bar{p}\label{eqRegionConvex1}\\
\frac{\partial^2\Phi_2({p}_1)}{\partial R^2_1}\bigg|_{p_1}<0,
~\forall~0\le p_1\le \bar{p}.\label{eqRegionConvex2}
\end{eqnarray}

\begin{figure*}[ht]
    {\includegraphics[width=
1\linewidth]{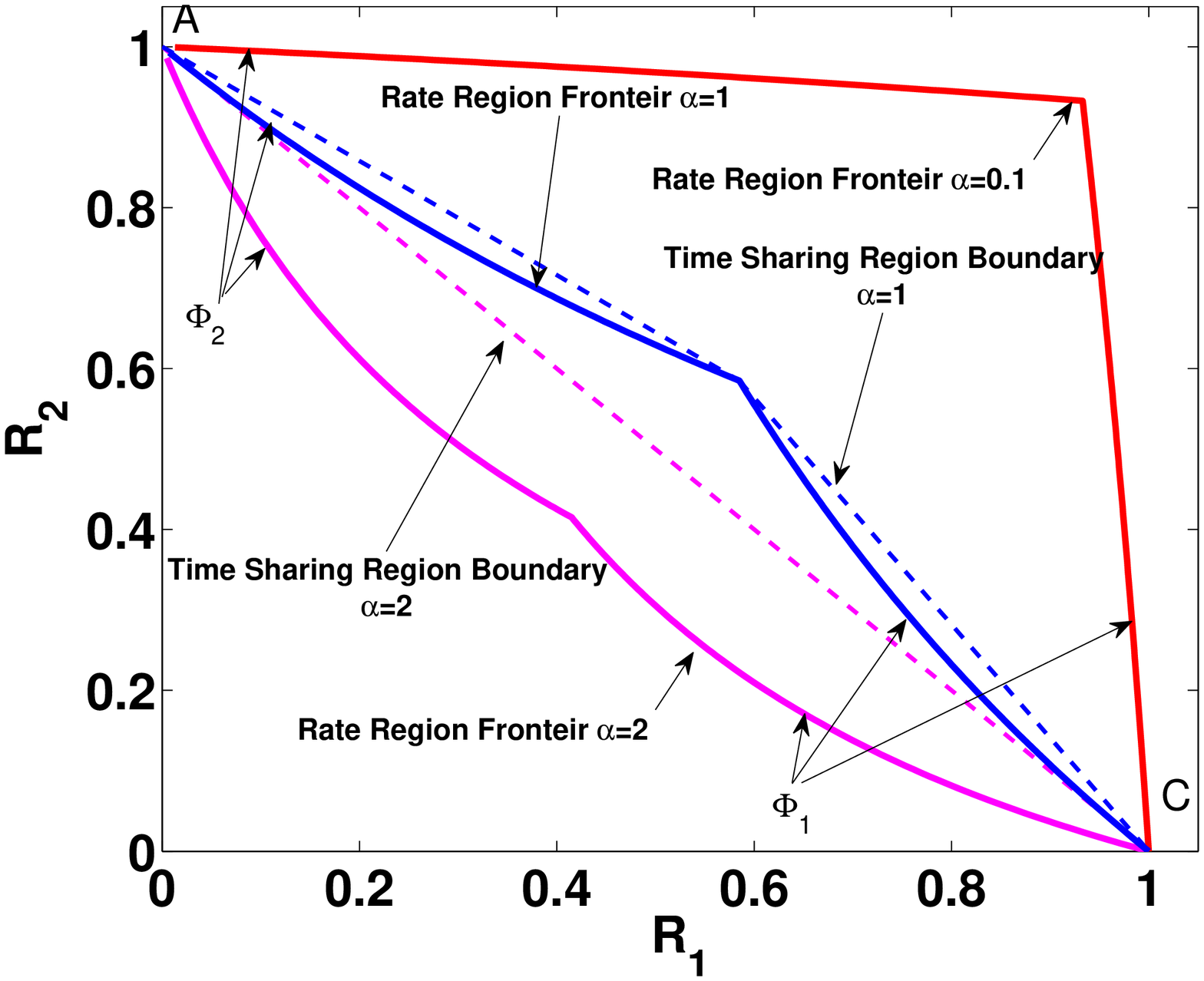} \caption{ {\small
The rate regions  for a 2-user IC with different interference
conditions. $\bar{p}_1=\bar{p}_2=1$, $|H_{11}|^2=|H_{22}|^2=1$,
$|H_{12}|^2=|H_{21}|^2=\alpha$. At point $A$ and $C$, a single user
transmits using full power. The solid lines are the directly
achievable rate frontier. The dotted lines represents the rate
boundary that can be achieved by time sharing. Note that the time
sharing boundary for $\alpha=0.1$ is the same as the rate region
frontier. }}\label{figRateRegionTransition} }
\end{figure*}

In particular, a necessary condition for (\ref{eqRegionConvex1})-(\ref{eqRegionConvex2}) is
\begin{eqnarray}
|H_{22}|^2|H_{12}|^2\bar{p}(1+|H_{21}|^2\bar{p})-|H_{11}|^2(1+|H_{22}|\bar{p})<
0
\end{eqnarray}
which requires that the maximum possible interference to be sufficiently small. As
the interference increases, the directly achievable rate regions
become non-convex. Fig. \ref{figRateRegionTransition} shows the
transition of the directly achievable rate regions as well as the
time-sharing regions when the interference levels change from strong
to weak. Clearly, when the interference is strong ($\alpha=2$ in
this figure), orthogonal transmission such as TDMA or FDMA is
optimal.

The same authors also characterize the achievable rate regions
for the general $K$-user case. However the conditions for the convexity of
the $K$-user regions are not available and deserve investigation. These conditions
can be useful in solving resource allocation problems for an interference network.

More generally, it remains an open problem to derive
a complete characterization of the (directly achievable) rate region for a
parallel IC. The exact conditions for its convexity (or the lack of)
are still unknown, although it is clear that the rate region will be convex if the
interference coefficients are sufficiently small.

Several efforts have been devoted to
characterizing certain interesting points (such as sum-rate optimal point) on the Pareto boundary of
the rate region. 
\cite{Hayashi2009} have shown that in a parallel IC
model with channel gains satisfying the following strong
interference conditions
\begin{eqnarray}
\frac{|H^n_{lk}|^2}{|H^n_{kk}|^2}>\frac{1}{2},~\mbox{and}~
\frac{|H^n_{lk}|^2}{|H^n_{kk}|^2}
\frac{|H^n_{kl}|^2}{|H^n_{ll}|^2}>\frac{1}{4}\left(1+\frac{1}{C-1}\right)^2,~\forall~n\in\mathcal{N},
(l,k)\in\mathcal{K}\times\mathcal{K}
\end{eqnarray}
where $C\ge 2$ is the minimum number of subchannels used by any
user, then the sum rate maximization point can only be achieved
using an FDMA strategy. In the special case of 2-user $N$ channel
model, the following condition is sufficient for the optimality of
FDMA strategy
\begin{eqnarray}
\frac{|H^n_{12}|^2}{|H^n_{22}|^2}
\frac{|H^n_{21}|^2}{|H^n_{11}|^2}>\frac{1}{4}\left(1+\frac{1}{C-1}\right)^2,~\forall~n\in\mathcal{N}.
\end{eqnarray}

The MIMO IC model is even more general than the
parallel IC, hence its achievable rate region is also difficult to
characterize. To see this, assuming that $N_t=N_r=N$; let all the
channel matrices be diagonal:
$\mathbf{H}_{lk}=\mbox{diag}\left([H^1_{lk},\cdots,H^{N}_{lk}]\right)$,
$(l,k)\in\mathcal{K}\times\mathcal{K}$; let all the transmission
covariances be diagonal as well:
$\mathbf{Q}_k=\mbox{diag}\left([p^1_k,\cdots,p^N_k]\right),~k\in\mathcal{K}$. In this
simplified model, user $k$'s transmission rate reduces to
\begin{eqnarray}
R_k(\mathbf{Q}_1,\cdots,\mathbf{Q}_K)&=&\log_2\det\left(\mathbf{H}_{kk}\mathbf{Q}_{k}\mathbf{H}_{kk}^{H}
\left(\mathbf{I}_{N_r}+ \sum_{l\ne k}
\mathbf{H}_{lk}\mathbf{Q}_{l}\mathbf{H}_{lk}^{H}\right)^{-1}+\mathbf{I}_{N_{r}}\right)\nonumber\\
&=&\sum_{n=1}^{N}\log_2\left(1+\frac{p^n_k
|H^n_{kk}|^2}{1+\sum_{l\ne k} |H^n_{lk}|^2p^n_l}\right)
\label{eqRateMIMODiagonal}
\end{eqnarray}
which is exactly the rate expression for the $N$ channel $K$ user
parallel IC as expressed in (\ref{eqParallelChannelLinkCapacity})
and (\ref{eqparallelChannelIndividualCapacity}).

\cite{Larsson08} and \cite{Jorswieck08misogame} have
characterized the achievable rate region of a 2-user MISO IC. In this case,
user $k$'s achievable transmission rate reduces to
\begin{eqnarray}
R_k(\mathbf{v}_1,\cdots,\mathbf{v}_k)=\log_2\left(1+\mbox{SINR}_k\right)\label{eqMISORateSINR}
\end{eqnarray}
where the SINR for user $k$ is defined in (\ref{eqMISOSINR}).

Define the maximum-ratio transmission (MRT) and the zero forcing (ZF) beamformers for both users as
\begin{equation}\label{eqMRTBeamFormer}
\begin{array}{lll}
&\displaystyle \mathbf{v}^{{MRT}}_1=\bar{p}_1\frac{\mathbf{h}^{H}_{11}}{\|\mathbf{h}^{H}_{11}\|},
~
&\displaystyle \mathbf{v}^{{MRT}}_2=\bar{p}_2\frac{\mathbf{h}^{H}_{22}}{\|\mathbf{h}^{H}_{22}\|}\\[10pt]
&\displaystyle \mathbf{v}^{{ZF}}_1=\bar{p}_1\frac{\Pi^{\bot}_{\mathbf{h}^{H}_{12}}\mathbf{h}^{H}_{11}}
{\|\Pi^{\bot}_{\mathbf{h}^{H}_{12}}\mathbf{h}^{H}_{11}\|},
&\displaystyle \mathbf{v}^{{ZF}}_2=\bar{p}_2\frac{\Pi^{\bot}_{\mathbf{h}^{H}_{21}}\mathbf{h}^{H}_{22}}
{\|\Pi^{\bot}_{\mathbf{h}^{H}_{21}}\mathbf{h}^{H}_{22}\|}\\
\end{array}
\end{equation}
where $\Pi^{\bot}_{\mathbf{X}}$ represents the orthogonal projection
on to the complement of the column space of $\mathbf{X}$. The
authors show that any point on the Pareto boundary is achievable
with the beamforming strategy
\begin{equation}\label{eqv12}
\begin{array}{l}
\displaystyle \mathbf{v}_1(\lambda_1)=\bar{p}_1\frac{\lambda_1\mathbf{v}_1^{ZF}+(1-\lambda_1)\mathbf{v}_1^{MRT}}
{\|\lambda_1 \mathbf{v}^{ZF}_1+(1-\lambda_1)\mathbf{v}_1^{MRT}\|}\nonumber\\ [10pt]
\displaystyle \mathbf{v}_2(\lambda_2)=\bar{p}_2\frac{\lambda_2\mathbf{v}_2^{ZF}+(1-\lambda_2)\mathbf{v}_2^{MRT}}
{\|\lambda_2 \mathbf{v}^{ZF}_2+(1-\lambda_2)\mathbf{v}_2^{MRT}\|}
\end{array}
\end{equation}
where $0\le \lambda_1, \lambda_2\le 1$. Intuitively, it is clear that $\mathbf{v}_1(\lambda_1),\mathbf{v}_2(\lambda_2)$ should stay in the subspace spanned by the channel vectors $\mathbf{h}^{H}_{11},\mathbf{h}^{H}_{22}$. Since this subspace is spanned by the MRT and ZF beamformers, it is no surprise that $\mathbf{v}_1(\lambda_1),\mathbf{v}_2(\lambda_2)$ can be written as linear combinations of the MRT and ZF beamformers. The novelty of (\ref{eqv12}) lies in the claim that the parameters
$\lambda_1, \lambda_2$ are real numbers and lie in the interval $[0,1]$. Similar to the characterization (\ref{eqFrontier1})-(\ref{eqFrontier2}) for the rate region of a scalar IC, the characterization (\ref{eqv12}) of optimal beamforming strategy can be used to computationally determine the rate region for a 2-user MISO IC.

In \cite{Jorswieck08},
the authors extend their 2-user MISO channel work to a general $K$-user MISO IC. In particular, any point in the achievable rate
region can be achieved using a set of beamformers
$\{\mathbf{v}_k\}_{k=1}^{N}$ that is characterized by $K^2$ complex
numbers $\{\epsilon_{kl}\}_{(k,l)\in\mathcal{K}\times\mathcal{K}}$ as
\[
\begin{array}{l}
\displaystyle \mathbf{v}_k=\sum_{l=1}^{K}\epsilon_{kl}\mathbf{h}^H_{kl},~\forall~k\in\mathcal{K}\nonumber\\ [10pt]
\displaystyle \|\mathbf{v}_k\|^2=\bar{p}_k,~\forall~k\in\mathcal{K} .
\end{array}
\]
However, because of the large number of (complex) parameters involved, this characterization appears less useful computationally in the determination of the rate region. We refer the readers to the web pages of \href{http://user.cs.tu-berlin.de/~jorsey/}{Jorswieck}
and \href{http://www.commsys.isy.liu.se/en/staff/egl}{Larsson}
for more details.
We emphasize again that except for these limited results, the
structure of a general MIMO IC rate region is still unknown when the interference is
treated as noise.


\section{Optimal Resource Allocation in Interference Channel}\label{secOptimal}
As is evident from the discussions in Section~\ref{secInformationTheory},
the most interesting points on the boundaries of the rate
regions can only be achieved by careful resource allocation. In this
section we discuss optimal resource allocation schemes for the
general IC models. Such optimality is closely related to the choice of
a performance metric for the communication
system under consideration.

\subsection{Problem Formulations}\label{subSecFormulation}
A communication system should provide users with QoS guarantees,
fairness through efficient resource utilization.
Mathematically, the resource allocation problem can be formulated
as the problem of optimizing a certain system level utility function subject to
resource budget constraints.

\begin{figure*}[ht]
    \begin{minipage}[t]{0.53\linewidth}
    \centering
    {\includegraphics[width=
1\linewidth]{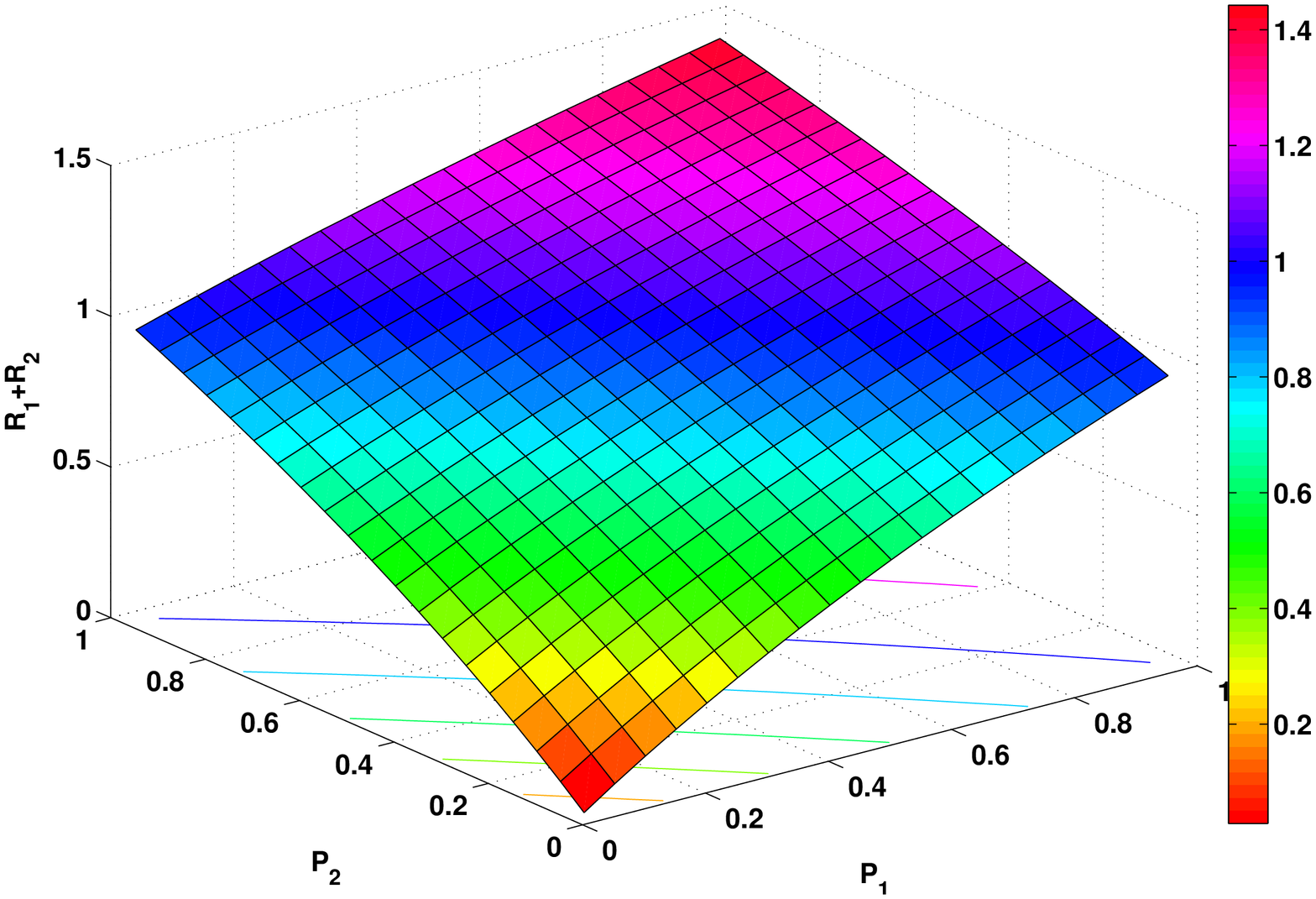}}
\end{minipage}
    \begin{minipage}[t]{0.53\linewidth}
    \centering
    {\includegraphics[width=
1\linewidth]{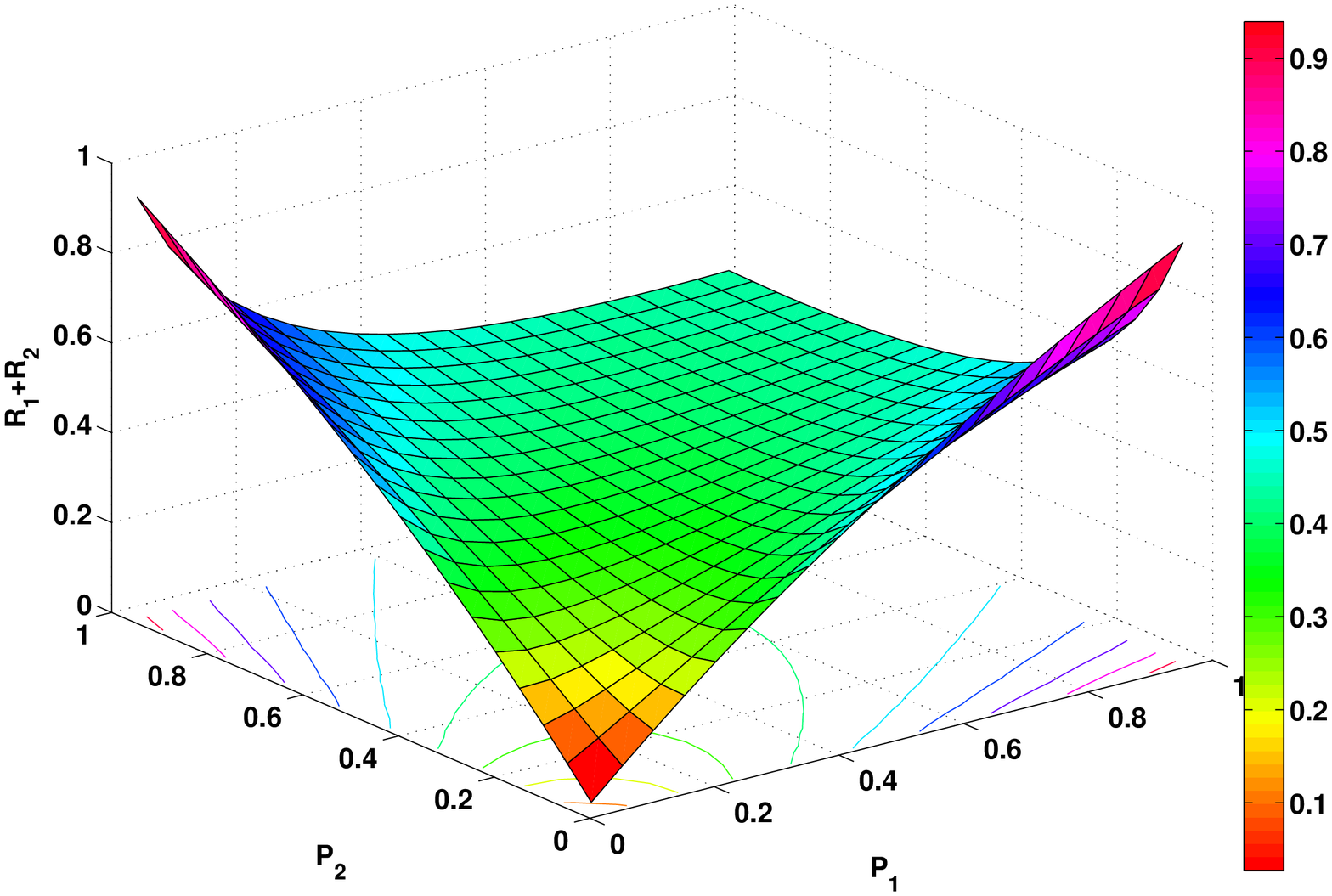}}
\end{minipage}
\caption{{\small The sum rate utility for a 2-user scalar IC with
different interference conditions. $\bar{p}_1=\bar{p}_2=1$,
$|H_{11}|^2=|H_{22}|^2=1$, $|H_{12}|^2=|H_{21}|^2=\alpha$. In the
left panel, $\alpha=0.5$. In the right panel,
$\alpha=5$.}}\label{figSumRateUtilityNonConvex}
    \end{figure*}

A popular family of utility functions is the so called ``$\alpha$-fair"
utility functions, which can be expressed as
\begin{eqnarray}
U\left(\{R_k\}_{k=1}^{K}\right)=\sum_{k=1}^{K}\frac{(R_k)^{1-\alpha}}{1-\alpha}\label{eqAlphaUtility}
\end{eqnarray}
where $R_k$ denotes the transmission rate of user $k$. As pointed
out in \cite{mo2000}, different choices of the parameter $\alpha$
give different priorities to user fairness and overall system
performance. We list four commonly used utility functions
that belong to the family of $\alpha$-fair utility functions:\\
a) {\bf The sum rate utility}: $U_1(\{R_k\}_{k=1}^{K})=\sum_{k=1}^{K}R_k$, obtained by setting $\alpha=0$; \\
b) {\bf The proportional fair utility}: $U_2(\{R_k\}_{k=1}^{K})=\sum_{k=1}^{K}\ln(R_k)$,
obtained by letting $\alpha\to 1$;\\
c) {\bf The harmonic-rate utility}:
$U_3(\{R_k\}_{k=1}^{K})=\left(\sum_{k=1}^{K}R^{-1}_k\right)^{-1}$, obtained by setting $\alpha=2$;\\
d) {\bf The min-rate utility ($\alpha\to \infty$)}: $U_4(\{R_k\}_{k=1}^{K})=\min_{1\le k\le K}R_k$,
obtained by letting $\alpha\to \infty$. \\
In terms of overall system performance, these utility functions can
be ordered as
\begin{eqnarray}
U_1\ge U_2\ge U_3 \ge U_4.
\end{eqnarray}
In terms of user fairness, the order is reversed. We note that
except for the case in which the interference is weak, these utility
functions are {\it nonconcave} in general. For example, in Fig.\
\ref{figSumRateUtilityNonConvex} we plot the sum rate utility for a
2-user scalar IC in cases where the interference is either weak or
strong. {Moreover, in most cases, it is  not possible to represent
these utility functions as concave ones via a nonlinear
transformation. See \cite{Boche11ConvexConcave} for an impossibility
result in scalar interference channel. This is consistent with the
complexity status (NP-hard) of the utility maximization problems
\cite{luo08a,liu11MISO,Razaviyayn11Asilomar} (see discussions in
Section~\ref{subSecComplexity})}.

If we wish to find a resource allocation scheme that maximizes the
system level performance, then we need to determine the conditions
under which the system level problem is easy to solve. Whenever such
conditions are met, efficient system level resource allocation
decision can be carried out by directly solving a convex
optimization problem. Intuitively, when the crosstalk coefficients
are zero or sufficiently small (low interference regime), the
utility functions should be concave. It will be interesting to
analytically determine how small the crosstalk coefficients need to
be in order to preserve concavity.

 From a practical perspective, the conditions for the
concavity of the utility function (in terms of the crosstalk coefficients) are valuable because they
can be used to find high quality approximately optimal resource allocation schemes. In particular,
we can use these conditions to partition the users into small groups within which the interference is
less and resource allocation is easy. Different  groups can be put on orthogonal resource dimensions,
because the groups cause too much interference to each other. Ultimately a good resource allocation
scheme in an interference limited network will likely involve a hybrid scheme whereby some small
groups of users share resources, while different groups are separated from competition.

The lack of concavity (or more generally, the lack of concave
reformulation/transformation) has made it difficult to numerically
maximize these utility functions for resource allocation. To
circumvent the computational difficulties, and to reduce the amount
of channel state information required for practical implementation,
some researchers have proposed to use alternative utility functions
for resource allocation. For example, both the mean squared error
(MSE) and the leakage power cost functions have been proposed as
potential substitutes for the rate-based utility functions listed
above \cite{shi07MMSE} \cite{ululus01} \cite{Sadek07} and
\cite{Ho10}. Recently, a number of studies
\cite{Boche08Calculus,Stanczak07Distributed,Boche11ConvexConcave,Boche10Unify}
have characterized a family of system utility functions that, under
appropriate transformations, admit concave representations. Such
transformations allow the associated utility maximization problems
to be easily solvable. We refer the readers to
\href{http://www.lti.ei.tum.de/index.php?id=boche&L=1}{Holger
Boche}'s web page for details on this topic. Unfortunately these
utility functions are not directly related to individual users'
transmission rates, hence the solutions of the associated
optimization problems tend to give suboptimal system performance (in
terms of the users' achievable rates). We shall not further
elaborate on these resource allocation approaches in this article.
Instead, we will focus on the use of above listed rate-based utility
functions for resource allocation. 

Let us describe several utility maximization problems to be considered in this article.

1) {\bf Utility maximization for the scalar IC model}:
\begin{eqnarray}
&\max_{\{p_k\}_{k}}&U(\{R_k\}^{K}_{k=1})\label{eqUtilityMaximizationScalar}\\
&\textrm{s.t.}&R_k=\log_2\left(1+\frac{|H_{kk}|^2 p_k}{1+ \sum_{l\ne
k}|H_{lk}|^2
p_l}\right),~k=1\cdots,K,\nonumber\\
&&0\le {p_k}\le \bar{p}_k,~k=1,\cdots,K\nonumber.
\end{eqnarray}

2) {\bf Utility maximization for the parallel IC model}:
\begin{eqnarray}
\max_{\{p^n_k\}_{k,n}}&& U(\{R_k\}^{K}_{k=1})\label{eqUtilityMaximizationParallel}\\
\textrm{s.t.}&&R_k=\sum_{n=1}^{N}\log_2\left(1+\frac{|H^n_{kk}|^2
p^n_k}{1+ \sum_{l\ne k}|H^n_{lk}|^2
p^n_l}\right),~k=1,\cdots,K,\nonumber\\
&&\sum_{n=1}^{N}{p^n_k}\le \bar{p}_k,~k=1,\cdots,K,\nonumber\\
&&p^n_k\ge 0,~ (k,n)\in\mathcal{K}\times\mathcal{N}\nonumber.
\end{eqnarray}

3) {\bf Utility maximization for the MISO IC model}:
\begin{eqnarray}
\max_{\{\mathbf{v}_k\}_{k}}&& U(\{R_k\}^{K}_{k=1})\label{eqUtilityMaximizationMISO}\\
\textrm{s.t.}&&
R_k=\log_2\left(1+\frac{|\mathbf{h}_{kk}\mathbf{v}_{k}|^2} {1+
\sum_{l\ne k} |\mathbf{h}_{lk}\mathbf{v}_{l}|^2}\right),~k=1,\cdots,K,\nonumber\\
&&\|\mathbf{v}_k\|^2\le \bar{p}_k,~k=1,\cdots,K\nonumber.
\end{eqnarray}

4) {\bf Utility maximization for the MIMO IC model}:
\begin{eqnarray}
\max_{\{\mathbf{Q}_k\}_{k}}&& U(\{R_k\}^{K}_{k=1})\label{eqUtilityMaximizationMIMO}\\
\textrm{s.t.}&&
R_k=\log_2\det\left(\mathbf{H}_{kk}\mathbf{Q}_{k}\mathbf{H}_{kk}^{H}
\left(\mathbf{I}_{N_r}+ \sum_{l\ne k}
\mathbf{H}_{lk}\mathbf{Q}_{l}\mathbf{H}_{lk}^{H}\right)^{-1}+\mathbf{I}_{N_{r}}\right),~k=1,\cdots,K\nonumber\\
&&\mbox{Tr}(\mathbf{Q}_k)\le \bar{p}_k,~\mathbf{Q}_k\succeq 0, ~k=1,\cdots,K\nonumber.
\end{eqnarray}

5) {\bf Utility maximization for the MIMO IC model (single stream per user)}:
\begin{eqnarray}
\max_{\{(\mathbf{u}_k,\mathbf{v}_k)\}_{k=1}^{K}}&& U(\{R_k\}_{k=1}^{K})\label{eqUtilityMaximizationMIMOSingle}\\
\textrm{s.t.}&&
R_k=\log_2\left(1+\frac{|\mathbf{u}^{H}_k\mathbf{H}_{kk}\mathbf{v}_k|^2}
{\|\mathbf{u}_k\|^2+\sum_{l\ne k}|\mathbf{u}^{H}_k\mathbf{H}_{lk}\mathbf{v}_l|^2}\right), ~k=1,\cdots,K\nonumber,\\
&&\|\mathbf{v}_k\|^2\le \bar{p}_k,~k=1,\cdots,K\nonumber.
\end{eqnarray}

A ``dual" paradigm for the design of the resource allocation
algorithm is to provide QoS guarantees to all the users while
minimizing the total power consumption.  This formulation
traditionally finds its application in voice communication networks where
it is desirable to maintain a minimum communication rate (or SINR level) for each
user in the system. Define $\{\gamma_k\}_{k=1}^{K}$ as the set of SINR targets.
We list several QoS constrained min-power
problem to be considered in this article.

6) {\bf Power minimization for the scalar IC model}:
\begin{eqnarray}
&\min&\sum_{k=1}^{K}p_k \label{eqSumPowerQoSScalar}\\
&\textrm{s.t.}& \mbox{SINR}_k\ge \gamma_k,~k=1,\cdots,K,\nonumber\\
&&\mbox{SINR}_k=\frac{|H_{kk}|^2p_k}{1+\sum_{l\ne k}|H_{lk}|^2 p_l},~k=1,\cdots,K.\nonumber
\end{eqnarray}

7) {\bf Power minimization for the MISO IC model}:
\begin{eqnarray}
\min_{\{\mathbf{v}_k\}_{k=1}^{K}}&& \sum_{k=1}^{K}\|\mathbf{v}_k\|^2\label{eqSumPowerQoSMISO}\\
\textrm{s.t.} &&\mbox{SINR}_k\ge \gamma_k,~k=1,\cdots,K,\nonumber\\
&&\mbox{SINR}_k=\frac{|\mathbf{h}_{kk}\mathbf{v}_{k}|^2} {1+
\sum_{l\ne k}
|\mathbf{h}_{lk}\mathbf{v}_{l}|^2},~k=1,\cdots,K\nonumber.
\end{eqnarray}

8) {\bf Power minimization for the MIMO IC model (single stream per user)}:
\begin{eqnarray}
\min_{\{(\mathbf{u}_k,\mathbf{v}_k)\}_{k=1}^{K}}&& \sum_{k=1}^{K}\|\mathbf{v}_k\|^2\label{eqSumPowerQoSMIMO}\\
\textrm{s.t.}&& \mbox{SINR}_k\ge \gamma_k, ~k=1,\cdots,K\nonumber,\\
&&\mbox{SINR}_k=\frac{|\mathbf{u}^{H}_k\mathbf{H}_{kk}\mathbf{v}_k|^2}
{\|\mathbf{u}_k\|^2+\sum_{l\ne
k}|\mathbf{u}^{H}_k\mathbf{H}_{lk}\mathbf{v}_l|^2},~k=1,\cdots,K\nonumber.
\end{eqnarray}

A hybrid formulation combines the above two approaches. It aims to provide QoS
guarantees while at the same time maximizing a system level utility
function. 
This hybrid formulation is useful in data communication networks where besides the
minimum rate constraints, it is preferable to deliver high system
throughput.  We list two such formulations to be considered later in this article.

9) {\bf Hybrid formulation for the scalar IC model}:
\begin{eqnarray}
&\max&U(\{p_k\}_{k=1}^{K}) \label{eqMixedScalar}\\
&\textrm{s.t.}& \mbox{SINR}_k\ge \gamma_k,~k=1,\cdots,K,\nonumber\\
&&0\le {p_k}\le \bar{p}_k,~k=1,\cdots,K\nonumber.
\end{eqnarray}

10) {\bf Hybrid formulation for the parallel IC model}:
\begin{eqnarray}
\max_{\{p^n_k\}_{k,n}}&& U(\{R_k\}^{K}_{k=1})\label{eqMixedParallel}\\
\textrm{s.t.}&&R_k=\sum_{n=1}^{N}\log_2\left(1+\frac{|H^n_{kk}|^2
p^n_k}{1+ \sum_{l\ne k}|H^n_{lk}|^2
p^n_l}\right),~k=1,\cdots,K,\nonumber\\
&&R_k\ge\zeta_k,~k=1,\cdots,K,\nonumber\\
&&\sum_{n=1}^{N}{p^n_k}\le \bar{p}_k,~k=1,\cdots,K,\nonumber\\
&&p^n_k\ge 0,~ (k,n)\in\mathcal{K}\times\mathcal{N}\nonumber
\end{eqnarray}
where $\{\zeta_k\}_{k=1}^{K}$ is a set of rate targets.

We note that for the latter two formulations, the minimum rate/SINR
requirements
provide fairness to the users, while the optimization objectives are
aimed at efficient utilization of system resource (e.g., spectrum or power).
For both of these two problems, the {\it feasibility} of the set of rate/SINR
targets needs to be carefully examined,
as  the rate/SINR requirements may not be simultaneously satisfiable.

\subsection{Complexity of the Optimal Resource Allocation Problems}\label{subSecComplexity}

{The aforementioned optimal resource allocation problems are
nonconvex. However, the lack of convexity does not necessarily imply
that the problem is difficult to solve. In some cases, it may be
possible for a nonconvex problem to be appropriately transformed
into an equivalent convex one and solved efficiently. A principled
approach to characterize the intrinsic difficulty of an utility
maximization problem is by way of the computational complexity
theory \cite{garey79}.

In the following, we summarize a number of recent studies on the computational complexity status of
these resource allocation problems. These complexity results suggest
that in most cases solving the utility maximization problems to
global optimality is computationally intractable as the number of
users in the system increases. }

Table \ref{tableComplexity} lists the complexity status for resource allocation problems
with specific utility functions for the parallel and MISO IC models.
Note that the scalar IC model is included as a special case.

Table \ref{tableComplexityMIMO}
summarizes the complexity status for the minimum rate utility maximization problem and the sum power
minimization problem with the QoS constraint in MIMO IC model (i.e., problem (\ref{eqUtilityMaximizationMIMOSingle}) with min-rate utility and problem (\ref{eqSumPowerQoSMIMO})).
Note that the results in Table \ref{tableComplexityMIMO} are based on the assumption that all transmitters
and receivers use linear beamformers and that
each mobile receives a single data stream.

{
\begin{table*}[htb]
\caption{Complexity status of utility maximization problems for the parallel and MISO IC models (\cite{luo08a}, \cite{liu11MISO})}
\begin{center}
\small{
\begin{tabular}{|l |c | c | c | c |}
\hline
\backslashbox{ Problem Class }{ Utility Function } & Sum Rate & Proportional Fair & Harmonic Mean & Min-Rate  \\
 \hline
 \hline
Parallel IC, $K$=1,
$N$ arbitrary
&Convex Opt & Convex Opt & Convex Opt &  Convex Opt\\
 \hline
Parallel IC, $K\ge 2$ fixed, $N$ arbitrary &NP-hard & NP-hard& NP-hard & NP-hard \\
 \hline
Parallel IC, $N\ge 2$ fixed, $K$ arbitrary & NP-hard & NP-hard&  NP-hard &   NP-hard\\
 \hline
Parallel IC,  $N=1$, $K$ arbitrary &NP-hard & Convex Opt & Convex Opt &  LP\\
 \hline
 MISO IC,  $N_t\ge 2$, $K$ arbitrary & NP-hard & NP-hard  & NP-hard  &  Poly. Time Solvable\\
 \hline
\end{tabular} } \label{tableComplexity}
\end{center}
\vspace*{-0.3cm}
\end{table*}}

{
\begin{table*}[htb]
\caption{Complexity status of the min-rate utility maximization for
the MIMO IC model (\cite{liu11ICC}), \cite{Razaviyayn11Asilomar}}
\begin{center}
\small{
\begin{tabular}{|l |c | c | c | }
\hline
\backslashbox{ $N_r$ }{ $N_t$ } & $N_t=1$ & $N_t=2$ & $N_t \ge 3$   \\
 \hline
 $N_r=1$& Poly. Time Solvable & Poly. Time Solvable & Poly. Time Solvable\\
 \hline
 $N_r=2$& Poly. Time Solvable& NP-hard &  NP-hard\\
 \hline
 $N_r\ge 3$ & Poly. Time Solvable &  NP-hard &  NP-hard\\
 \hline
\end{tabular} } \label{tableComplexityMIMO}
\end{center}
\vspace*{-0.3cm}
\end{table*}
Recall that the MIMO
IC is a generalization of the Parallel IC (see Section \ref{subsubSecRateRegion}). It follows that
the complexity results in Table \ref{tableComplexity} hold true for the
 MIMO IC model with an arbitrary number of data streams per user.
 We refer the readers to the
\href{http://www.ece.umn.edu/~luozq}{author}'s web page  for
 recent developments in the complexity analysis as well as other resource allocation algorithms.

\subsection{Algorithms for Optimal Resource Allocation}\label{subSecOptimalAlgorithm}

We now describe various utility maximization based algorithms for resource allocation. These algorithms will be grouped and discussed according to their main algorithmic features. Since the min-rate utility function is non-differentiable, it requires a separate treatment that is different from the other utility functions. We begin our discussion with resource allocation algorithms based on the min-rate utility maximization.

\subsubsection{Algorithms for Min-Rate Maximization}\label{subsubMinRate}
Early works on resource allocation aimed to find optimal transmission powers that can
maximize the min-SINR utility. In case of the scalar IC,
this problem can be formulated as
\begin{eqnarray}
&\max_{\{p_k\}_{k\in\mathcal{K}}}& \min_{k\in\mathcal{K}} \mbox{SINR}_k\label{eqMinPowerScalar}\\
&\textrm{s.t.}& p_k\ge 0,~\forall~k\in\mathcal{K},\nonumber\\
&& \mbox{SINR}_k~\mbox{defined in (\ref{eqSINRScalar})}.\nonumber
\end{eqnarray}
In \cite{Foschini93}, \cite{zander92a} and  \cite{zander92b}, the authors studied the
feasibility of this problem and proposed optimal power allocation
strategies for it.  For randomly generated scalar interference channels, they showed that with probability
one, there exists a unique optimum value to the
above problem. This optimal value, denoted as $\gamma^*$, can be expressed as
\begin{eqnarray}
\gamma^*=\frac{1}{\rho(\mathbf{Z})-1}
\end{eqnarray}
where $\rho(\mathbf{Z})$ represents the maximum eigenvalue of the matrix $\mathbf{Z}$;
 $\mathbf{Z}\in\mathcal{R}^{K\times K}$ is a matrix with its $(k,l)^{th}$ element defined
as $[\mathbf{Z}]_{k,l}=\frac{|H_{lk}|^2}{|H_{kk}|^2}$. Distributed power allocation
algorithms for this problem were also developed. For example, \cite{Foschini93} proposed an autonomous
power control (APC) algorithm that
iteratively adjusts the users' power levels as follows
\begin{eqnarray}
\frac{|H_{kk}|^2 p^{(t+1)}_k}{1+\sum_{l\ne k}
|H_{lk}|^2 p^{(t)}_l}=\frac{|H_{kk}|^2 p^{(t)}_k}{1+\sum_{l\ne k}
|H_{lk}|^2 p^{(t)}_l}-\beta(\mbox{SINR}^{(t)}_k-\gamma^*)
\end{eqnarray}
where $\beta$ is a small positive constant and $t$ is the iteration index. We refer the readers to
\cite{Hanly99powercontrol} and the web page of
\href{http://people.eng.unimelb.edu.au/hanly/}{Hanly}
for further discussion of power control techniques for a scalar IC.

For a MISO IC model, the problem of finding optimal transmit beamformers for
the maximization of the min-SINR utility has been considered
by \cite{bengtsson01,Wiesel06}. The corresponding resource allocation problem can be equivalently formulated as
\begin{eqnarray}
&\max_{\{\mathbf{v}_k\}_{k\in\mathcal{K}}}&\gamma\label{eqMAXMINMISO}\\
&\textrm{s.t.}& \mbox{SINR}_k\ge \gamma,~k=1,\cdots,K,\nonumber\\
&&\|\mathbf{v}_k\|^2\le \bar{p}_k,~k=1,\cdots,K,\nonumber\\
&&\mbox{SINR}_k~\mbox{defined in (\ref{eqMISOSINR})}\nonumber.
\end{eqnarray}
This optimization problem (\ref{eqMAXMINMISO}) is nonconvex, but can
be relaxed to a semidefinite program ( or SDP; see \cite{luo06a} for
an introduction to the related concepts and algorithms).
Surprisingly, \cite{bengtsson01} established that the SDP relaxation
for (\ref{eqMAXMINMISO}) is tight; see the subsequent section
``Algorithms for QoS Constrained Power Minimization" for more
discussions. Later, \cite{Wiesel06} further showed that this
nonconvex optimization problem can be solved via a sequence of
second order cone programs (SOCP); see \cite{luo06a} for the
definition of SOCP. The key observation is that for a fixed
$\gamma$, checking the feasibility of (\ref{eqMAXMINMISO}) is an
SOCP, which can be solved efficiently by the standard interior point
methods. Let $\gamma^*$ denote the optimal objective for problem
(\ref{eqMAXMINMISO}), this max-min SINR problem can be solved by a
bisection technique:
\begin{center}
\fbox{\begin{minipage}{4in}
\begin{description}
\item [1)] choose $\epsilon>0$ (termination parameter), $\gamma_l$ and $\gamma_u$ such that $\gamma^*$ lies in
$[\gamma_l,~\gamma_u]$;
\item [2)] let $\gamma_{mid}=(\gamma_l+\gamma_u)/2$;
\item [3)] check the
feasibility of problem (\ref{eqMAXMINMISO}) with $\gamma=\gamma_{mid}$. If feasible,
let $\gamma_l=\gamma_{mid}$, otherwise set $\gamma_u=\gamma_{mid}$.
\item [4)] terminate if $\gamma_u-\gamma_l\le \epsilon$; else go to step 2) and repeat.
\end{description}
\end{minipage}
}
\end{center}
More recently, the max-min fairness resource allocation problem has
been considered by \cite{liu11ICC} for the MIMO IC model.
Unfortunately, the problem becomes NP-hard in this case (see
Table~\ref{tableComplexityMIMO}).


The joint transceiver beamformer design for
the min-SINR maximization problem in a MIMO IC (i.e., problem (\ref{eqUtilityMaximizationMIMOSingle})
with min-rate utility) has
recently been considered in \cite{liu11ICC} . As shown in Section \ref{subSecComplexity}, this problem
is in general NP-hard. Consequently, they
proposed a low-complexity algorithm that converges to a stationary point of this  problem. A key
observation is that when the receive beamformers $\{\mathbf{u}_k\}_{k=1}^{K}$ are fixed,
the considered problem can be written as
\begin{eqnarray}
\max_{\{\mathbf{v}_k\}_{k=1}^{K}}&& \gamma\label{eqMAXMINMIMOFixU}\\
\textrm{s.t.}&&
\frac{|\mathbf{u}^{H}_k\mathbf{H}_{kk}\mathbf{v}_k|^2}
{\|\mathbf{u}_k\|+\sum_{l\ne k}|\mathbf{u}^{H}_k\mathbf{H}_{lk}\mathbf{v}_l|^2}\ge \gamma, ~k=1,\cdots,K\nonumber,\\
&&\|\mathbf{v}_k\|^2\le \bar{p}_k,~k=1,\cdots,K\nonumber
\end{eqnarray}
which has the same form as the MISO min-SINR
problem in (\ref{eqMAXMINMISO}), and thus can be solved using bisection and SOCP.
As a result, the authors propose to alternate between the following two steps
to solve the min-rate maximization problem:
\begin{center}
\fbox{\begin{minipage}{3.5in}
\begin{description}
\item [1)] for fixed $\{\mathbf{u}^{(t)}_k\}_{k=1}^{K}$,
solve (\ref{eqMAXMINMIMOFixU}) via SOCP to obtain $\{\mathbf{v}^{(t+1)}_k\}_{k=1}^{K}$;
\item [2)]
for fixed $\{\mathbf{v}^{(t+1)}_k\}_{k=1}^{K}$, update
$\{\mathbf{u}^{(t+1)}_k\}_{k=1}^{K}$ to the minimum mean square
error (MMSE) receiver:\\
~~~$\mathbf{u}_k^{(t+1)}=\left(\sum_{l=1}^{K}\mathbf{H}_{lk}\mathbf{v}^{(t+1)}_l
(\mathbf{H}_{lk}\mathbf{v}_l^{(t+1)})^{H}+\mathbf{I}\right)^{-1}\mathbf{H}_{kk}\mathbf{v}^{(t+1)}_k$.
\end{description}
\end{minipage}
}
\end{center}
Unlike the MISO min-SINR case,
only {\it local} optimal solutions can be found in the MIMO case. Extending the above algorithm
to the MIMO IC/IBC/IMAC case with multiple data streams per user is not a trivial task.
For a MIMO IC model, the feasibility problem becomes
\begin{eqnarray}
&&R_k=\log_2\det\left(\mathbf{H}_{kk}\mathbf{Q}_{k}\mathbf{H}_{kk}^{H}
\left(\mathbf{I}_{N_r}+ \sum_{l\ne k}
\mathbf{H}_{lk}\mathbf{Q}_{l}\mathbf{H}_{lk}^{H}\right)^{-1}+\mathbf{I}_{N_{r}}\right)\ge \zeta,~k=1,\cdots,K\nonumber\\
&&\mbox{Tr}(\mathbf{Q}_k)\succeq 0,~\mbox{Tr}(\mathbf{Q}_k)\le \bar{p}_k, ~k=1,\cdots,K\nonumber.
\end{eqnarray}
This problem is nonconvex and there is no known convex reformulation for it. Finding
efficient and preferably distributed algorithms for these channel models is
a challenging problem which deserves investigation.

\subsubsection{Algorithms for Weighted Sum-Utility Maximization}

In addition to the min-rate (min-SINR) utility, we can use other utility functions to allocate resources. For instance, let $\{\mu_k\}_{k=1}^{K}$ denote a set of positive weights that represent the relative priorities of the users in the system. Then the weighted sum-rate maximization (WSRM) problem for a parallel IC can be formulated as
\begin{eqnarray}
\max_{\{p^n_k\}_{k,n}}&& \sum_{k=1}^{K}\mu_k R_k\label{eqWSRM}\\
\textrm{s.t.}&&R_k=\sum_{n=1}^{N}\log_2\left(1+\frac{|H^n_{kk}|^2
p^n_k}{1+ \sum_{l\ne k}|H^n_{lk}|^2
p^n_l}\right),\nonumber\\
&&\sum_{n=1}^{N}{p^n_k}\le \bar{p}_k,~p_k\ge 0,~ k=1,\cdots,
K\nonumber.
\end{eqnarray}
This simply corresponds to the problem
(\ref{eqUtilityMaximizationParallel}) with $ \sum_{k=1}^{K}\mu_k
R_k$ as the objective function. WSRM is a central problem for
physical layer resource allocation. Many sum-utility maximization
problems can be reduced to solving a sequence of WSRM problems for
the single channel $N=1$ case, see \cite{luo09duality}.
Unfortunately, the complexity results in Section
\ref{subSecComplexity} indicate that WSRM is in general a hard
problem which can not be solved to global optimality by a polynomial
time algorithm (unless NP=P). As a result, many works are devoted to
finding high quality locally optimal solutions for the WSRM problem.

\medskip
\noindent{\it Algorithms based on Lagrangian dual decomposition}
\smallskip

The linear additive structure of the power budget constraints in the weighted sum-utility maximization problem
(\ref{eqUtilityMaximizationParallel}) can be exploited by Lagrangian dualization. In particular,
\cite{yu06} (see also \cite{luo08a,luo09duality}) considered the Lagrangian
dual relaxation of the utility maximization problem
(\ref{eqUtilityMaximizationParallel}) for the parallel IC model. Let us define the {\it dual function} of
the primal problem (\ref{eqUtilityMaximizationParallel}) as
\begin{eqnarray}
d(\bflambda)=\max_{\{p^n_k\ge
0\}_{k,n}}\left\{U(\{R_k\}^{K}_{k=1})-\sum_{k=1}^{K}\lambda_k
\left(\sum_{n=1}^{N}{p^n_k}-\bar{p}_k\right)\right\}\label{eqDualFunction}
\end{eqnarray}
where $\bflambda=\{\lambda_k\ge 0\}_{k=1}^{K}$ is the set of dual
variables associated with the sum power constraints. Then the {\it
dual problem} of the utility maximization problem can be expressed
as follows
\begin{eqnarray}
\min_{\bflambda}&&~d(\bflambda)\label{eqDualProblem}\\
\textrm{s.t.}~~&& \lambda_k\ge 0,~k=1,\cdots,K.\nonumber
\end{eqnarray}
Denote the optimal objective values of the primal problem (\ref{eqUtilityMaximizationParallel})
and the dual problem (\ref{eqDualProblem}) with $N$ channels as $p^*_N$ and $d^*_N$,
respectively. By the standard duality theory in optimization \cite{boyd04}, we have that the
{\it duality gap} $d^*_N-p^*_N$ satisfies
\begin{eqnarray}
d^*_N-p^*_N\ge 0.
\end{eqnarray}
When the primal problem is convex, strong duality holds and the
inequality becomes equality. When restricted to the FDMA (Frequency
Division Multiple Access) solutions, the Lagrangian dual problem
decomposes across tones and is efficiently solvable
\cite{Hayashi2009} and \cite{luo09duality}. However, when the dual
optimal solutions are not unique, it is difficult to construct a
primal optimal solution for the problem
(\ref{eqUtilityMaximizationParallel}). \cite{luo09duality} proposed
to use an additional randomized step to generate a primal feasible
solution from the dual optimal solution.

When the primal problem is
not restricted to the FDMA solutions, the Lagrangian dual function
is difficult to compute, let alone optimize (see the complexity results in Section~\ref{subSecComplexity}).
\cite{yu06} proposed an iterative spectrum
balancing (ISB) algorithm that alternates between the following two steps
to solve the WSRM problem (\ref{eqWSRM}):

\begin{center}
\fbox{\begin{minipage}{4in}
\begin{description}
\item [1)] given a $\bflambda\ge \mathbf{0}$, use a coordinate ascent strategy to {\it approximately} evaluate
the dual function until convergence;
\item [2)] update $\bflambda$ using the subgradient method or the ellipsoid
method.
\end{description}
\end{minipage}
}
\end{center}

\noindent Due to the inexactness of step 1), this algorithm is not guaranteed to converge to a global optimal solution of the WSRM problem (\ref{eqWSRM}).

A surprising observation in \cite{yu06} is that when $N$ (the number of channels) goes to
infinity, the duality gap vanishes. \cite{luo08a,luo09duality}
rigorously proved this result using Lyapunov theorem in functional
analysis. In particular, Lyapunov's theorem implies that for the continuous formulation of the WSRM problem (infinite number of channels), the rate region is actually convex. With additional steps to estimate of the approximation of Lebesque integrals, \cite{luo09duality} showed that for some constant $L$, an estimate of the duality gap is bounded by
\begin{eqnarray}
0\le d^*_N-p^*_N\le \frac{L}{\sqrt{N}}.
\end{eqnarray}
Clearly the gap vanishes as $N$ goes to infinity. Using this estimate, \cite{luo09duality} further developed a polynomial time approximation scheme to find an optimal FDMA solution for the continuous version of the WSRM problem (\ref{eqWSRM}).

\medskip
\noindent{\it Algorithms based on interference pricing}
\smallskip

In a number of related works \cite{huang06b,yu07,wang08,wu08},
the authors proposed a modified iterative water-filling (M-IWF) algorithm
that iteratively solves $K$ subproblems.
The subproblem related to user $k$ can be expressed as
\begin{eqnarray}
\max_{\{p^n_k\}_{k,n}}&& R_k-\sum_{n=1}^{N}p^n_k T^n_k\label{eqIndividualPricing}\\
\textrm{s.t.}&& \sum_{n=1}^{N}{p^n_k}\le \bar{p}_k, \\
&&p^n_k\ge 0,~n\in \mathcal{N}\nonumber.
\end{eqnarray}
where $T^n_k$ is defined as
\begin{eqnarray}\label{eqTn}
T^n_k=\sum_{l\ne k}\left(\frac{|H^n_{kl}|^2}{\sum_{j\ne
k}|H^n_{jl}|^2p^n_j+1}-\frac{|H^n_{kl}|^2}{\sum_{j=1}^{K}|H^n_{jl}|^2p^n_j+1}
\right).
\end{eqnarray}
This term can be viewed as the {\it interference price} that user
$k$ needs to pay on channel $n$ for the unit of interference it
causes to all other users in the system. In other words, the price
$T^n_k$ corresponds to the marginal decrease in the sum-rate utility
per unit increase in interference power $p^n_k$. If the interference
price is set to zero, then we are led to the standard iterative
water-filling algorithm \cite{yu02a}. The M-IWF algorithm works by
iteratively performing the following steps:
\begin{center}
\fbox{\begin{minipage}{4.5in}
\begin{description}
\item [1)] fix $\left\{T^{n, (t)}_k\right\}_{n,k}$,
each user iteratively computes the optimal solution $\left\{p^{n,(t+1)}_k\right\}_{n,k}$ to the convex subproblem (\ref{eqIndividualPricing})
until convergence;
\item [2)]
update $\left\{T^{n,(t+1)}_k\right\}_{n,k}$ according to (\ref{eqTn}) using $\left\{p^{n,(t+1)}_k\right\}_{n,k}$.
\end{description}
\end{minipage}
}
\end{center}

\noindent
We note that the overall computational complexity of step 1) is $O(T K N \log N)$, where
$T$ is the total number of iterations needed for convergence.
It was conjectured that this
algorithm converges at least to a stationary point of the WSRM problem,
but no formal proof was given.
In \cite{Shi:2009}, the authors successfully established the
convergence (to the stationary point) of this type of pricing
algorithm under the condition that the users act {\it sequentially},
i.e., step 1) of M-IWF, only a {\it single user} solves its
optimization problem (\ref{eqIndividualPricing}). They interpreted
this sequential M-IWF as a successive linear approximation of the
WSRM problem, and showed that the term
$-\sum_{n=1}^{N}\sum_{k=1}^{K}p^n_k T^n_k$ is the first order Taylor
approximation  (up to an additive constant term) of $\sum_{l\ne
k}R_l$, the non-concave part  of the objective function. With this
interpretation, the M-IWF algorithm can be seen as letting each user
sequentially solve a partially linearized version of the WSRM
problem. Since the first order Taylor approximation is a locally
tight approximation of the weighted sum-rate objective function, the
weighted sum-rates computed by the sequential M-IWF algorithm
improve monotonically.  Moreover, since the users update their power
allocations locally, the M-IWF algorithm can be implemented in a
distributed manner as long as the interference prices are exchanged
among the users at each iteration. We shall refer to the sequential
modification of the M-IWF as the multichannel distributed pricing
(MDP) algorithm.

The interference pricing idea has been extended to the MISO IC in
\cite{shi2008pricingmiso,schmidt2008}, and to the
MIMO IC with single stream per user in \cite{shi2009pricingmimo}. The reference
\cite{kim08} considered interference pricing for the general MIMO IC
without the single data stream per user restriction. Similar to the parallel IC situation, the convergence of
the interference pricing algorithm for the MIMO IC has only been analyzed for the sequential user update case.
It will be interesting to see how the pricing technique (and its convergence proof) can be extended
to the MIMO IC/IBC/IMAC models with an {arbitrary} number of streams per user, while allowing simultaneous user updates. A step in this direction was taken by  \cite{venturino10} which
extended the interference pricing technique to the MISO IBC model.
Their algorithm (named Iterative Coordinated BeamForming (ICBF))
calculates proper pricing coefficients that enable each BS to update
their respective beamformers. Convergence was always observed in the
simulation, but no formal proof was given.
A recent survey of various pricing techniques used in wireless
networks can be found in \cite{Schmidt09}. We also refer the readers
to the web pages of \href{http://users.eecs.northwestern.edu/~rberry/}{Berry} and
\href{http://users.eecs.northwestern.edu/~mh/}{Honig} for
other related works on this topic.

\begin{figure*}[htb]
    {\includegraphics[width=
0.7\linewidth]{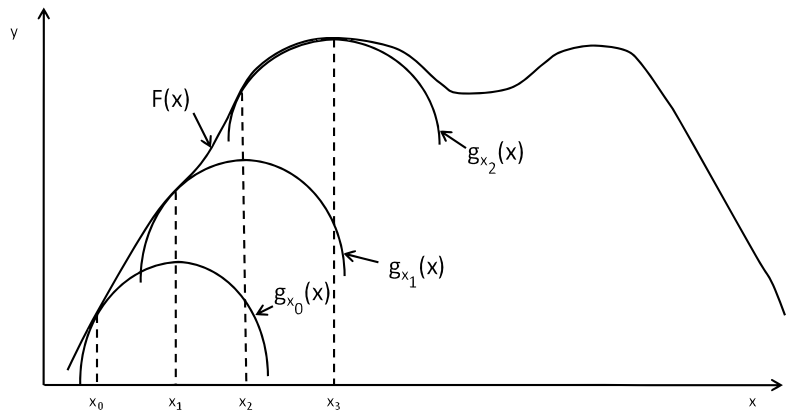} \caption{{\small Graphical
illustration of the family of algorithms using successive convex
approximation. In this example, $x_0$ is the initial point. At
$x_0$, a (strictly) concave function $g_{x_0}(x)$ is used to
approximate the original non-convex function $F(x)$. The optimal
solution of $g_{x_0} (x)$, $x_1$, is found by standard optimization
technique. Then the (strictly) concave function $g_{x_1}(x)$ is
constructed at the point $x_1$. $g_{x_1}(x)$ is then optimized to
obtain the point $x_2$. Continue this process, a stationary solution
of the original function $F(x)$ can be found.}}\label{figSCA} }
    \end{figure*}

\medskip
\noindent{\it Algorithms based on successive convex approximation}
\smallskip

The MDP algorithm belongs to a class of algorithms called successive
convex approximation (SCA). The idea is to
construct and maximize a series of (concave) lower bounds of the
original WSRM problem, so that a high quality solution
can be obtained asymptotically. See Fig. \ref{figSCA} for a graphical
illustration of how this class of algorithms work. In \cite{papand09},
an algorithm called Successive Convex Approximation for Low
complExity (SCALE) is proposed to improve the spectral efficiency of
the DSL network. This algorithm transforms the noncocave sum rate
maximization problem into a series of convex problems by utilizing
the following lower bound
\begin{eqnarray}
&&\alpha\log_2(z)+\beta\le \log_2(1+z)\label{eqLogLowerBound}\\
&&\alpha=\frac{z_0}{1+z_0},~\beta=\log_2(1+z_0)-\frac{z_0}{1+z_0}\log_2
z_0\label{eqApproximationParameter}
\end{eqnarray}
where the inequality (\ref{eqLogLowerBound}) is tight at $z=z_0$.
This lower bound allows the WSRM problem
to be approximated by
\begin{eqnarray}
\max_{\{p^n_k\}_{k,n}}&&
\sum_{k=1}^{K}\mu_k\sum_{n=1}^{N}\alpha^n_k\log_2\left(\frac{|H^n_{kk}|^2
p^n_k}{1+
\sum_{l\ne k}|H^n_{lk}|^2 p^n_l}\right)+\beta^n_k\label{eqSCALEApproximation}\\
\textrm{s.t.}& &\sum_{n=1}^{N}{p^n_k}\le \bar{p}_k,~k=1,\cdots,K\nonumber\\
&&p^n_k\ge 0,~ (k,n)\in\mathcal{K}\times N\nonumber.
\end{eqnarray}
After a log transformation
$\widetilde{p}^n_k=\log_2(p^n_k),~(k,n)\in\mathcal{K}\times N$, this
relaxed problem turns out to be concave in $\widetilde{p}^k_n$. The
SCALE algorithm alternates between the following two steps:
\begin{center}
\fbox{\begin{minipage}{4.5in}
\begin{description}
\item [1)] fix
$\left\{(\alpha^{n,(t)}_k,\beta^{n,{(t)}}_k)\right\}_{k,n}$, solve (\ref{eqSCALEApproximation}) and
obtain $\left\{p^{n,(t+1)}_{k}\right\}_{n,k}$;
\item [2)] update the parameters
$\left\{(\alpha^{n,(t+1)}_k,\beta^{n,(t+1)}_k)\right\}_{k,n}$ according to
(\ref{eqApproximationParameter}) using $\left\{p^{n,(t+1)}_{k}\right\}_{n,k}$.
\end{description}
\end{minipage}
}
\end{center}
Step 1) can be solved either in a centralized fashion using Geometric Programming (GP) technique,
or by solving the
dual problem of (\ref{eqSCALEApproximation}) in a distributed way.
This algorithm is guaranteed to reach a stationary point of the original sum rate
maximization problem. We briefly compare the major differences of the MDP and SCALE algorithms in Table \ref{tableSCALEMDP}.
{
\begin{table}
\caption{Comparison of SCALE and MDP algorithms}

{\small
\begin{tabular}{|c |c | c | c |c| }
\hline
\backslashbox{}{$Properties$} & user update schedule  & approximation methods &
computation per iteration& dual updates  \\
 \hline
 SCALE& simultaneously&concave approximation& iterative & subgradient
 \\
 \hline
 MDP& sequentially& linear approximation& closed form & bisection\\
 \hline
\end{tabular} } \label{tableSCALEMDP}
\end{table}

In \cite{Tsiaflakis08}, a different lower bound is proposed
for the WSRM problem. Specifically, the authors decompose the
objective function as the difference of two concave functions
of $\{p_k^n\}_{n,k}$ (referred to as the ``dc" function)
\begin{eqnarray}
&&\sum_{k=1}^{K}\sum_{n=1}^{N}\log_2\left(1+\frac{|H^n_{kk}|^2
p^n_k}{1+ \sum_{l\ne k}|H^n_{lk}|^2 p^n_l}\right)
\nonumber\\
&=&{\sum_{k=1}^{K}\sum_{n=1}^{N}\log_2\left(1+
\sum_{l\ne k}|H^n_{lk}|^2 p^n_l+|H^n_{kk}|^2 p^n_k\right)}
-\sum_{k=1}^{K}\sum_{n=1}^{N}\log_2\left({1+ \sum_{l\ne
k}|H^n_{lk}|^2 p^n_l}\right).\label{eqDC}
\end{eqnarray}
Similar to the steps of SCA introduced earlier, in each iteration of the
algorithm, the second sum is replaced with its linear lower bound,
and the resulting concave maximization problem is solved. Compared to the MDP algorithm which linearizes
$\sum_{l\neq k}R_l$, this algorithm linearizes all the interference terms in each iteration.
As such, it linearizes more terms than the MDP algorithm per iteration.

%


A related algorithm has been proposed in the recent work \cite{liu11MISO} where the authors
considered the general utility maximization problem in
MISO IC (i.e., problem formulation (\ref{eqUtilityMaximizationMISO})). Besides
providing complexity results, the authors proposed an algorithm that is
able to converge to local optimal solutions for  problem (\ref{eqUtilityMaximizationMISO})
with any {\it smooth} (twice continuously differentiable) utility functions. The basic
idea is to let the users cyclically update their beamformers using projected gradient ascent
algorithm. In particular, at iteration $t$, user $k$ takes a gradient projection step to compute the direction $\mathbf{d}_k^{(t+1)}$ by solving the following problem
\begin{eqnarray}
\max_{{\mathbf{d}_k}}&& \langle\nabla_{\mathbf{v}_k} U(\{\mathbf{v}^{(t)}_l\}^{K}_{l=1}),\mathbf{d}_k\rangle
-\frac{1}{2}\|\mathbf{d}_k\|^2 \label{eqprojection}\\
\textrm{s.t.}&& \|\mathbf{d}_k+\mathbf{v}_k^{(t)}\|^2\le \bar{p}_k,
\ \nonumber
\end{eqnarray}
In contrast to the MDP and SCALE, the subproblem (\ref{eqprojection}) linearizes the entire objective function of (\ref{eqUtilityMaximizationMISO}) at the current point $\{\mathbf{v}_k^{(t)}\}_{k\in\mathcal{K}}$,  and has an additional quadratic regularization term $\frac{1}{2}\|\mathbf{d}_k\|^2$. This subproblem is a convex quadratic minimization problem over a ball. As such, it is easier to solve than the corresponding subproblems of MDP and SCALE which are based on partial linearization of the original WSRM objective function.
We list below the main steps of this cyclic coordinate ascent (CCA) algorithm:
\begin{center}
\fbox{\begin{minipage}{4.5in}
\begin{description}
\item [1)] select a user $k\in\mathcal{K}$ and  compute its gradient projection
 direction $\mathbf{d}^{(t+1)}_k$  by solving (\ref{eqprojection});
 \item [2)] determine stepsize
$\alpha^{(t+1)}_k$ for user $k$ using a line search strategy;
\item [3)] update beamformer:
$\mathbf{v}^{(t+1)}_k=\mathbf{v}^{(t)}_k+\alpha^{(t+1)}_k\mathbf{d}^{(t+1)}_k$, and go to Step 1).
\end{description}
\end{minipage}
}
\end{center}
The CCA algorithm only works in MISO IC case, and it is not clear how to
extend it to the MIMO IC.

\medskip
\noindent{\it Algorithms based on weighted MMSE minimization}
\smallskip

A different weighted sum-rate maximization approach was proposed in
\cite{christensen08} for the MIMO broadcast downlink channel, where
the WSRM problem is transformed to an equivalent weighted sum MSE
minimization (WMMSE) problem with some specially chosen weight
matrices. Since the optimal weight matrices are generally unknown,
the authors of \cite{christensen08} proposed an iterative algorithm
that adaptively chooses the weight matrices and updates the linear
transmit/receive beamformers at each iteration. A nonconvex cost
function was constructed in \cite{christensen08} and shown to
monotonically decrease as the algorithm progresses. But the
convergence of the iterates to a stationary point (or the global
minimum) of the cost function is not known. Later, a similar
algorithm was proposed in \cite{Honig} for the interference channel
where each user only transmits one data stream.

It turns out that this WMMSE based resource allocation approach can
be extended significantly to handle the MIMO-IC and MIMO-IBC/IMAC
models as well as general utility functions. In particular, the
authors of \cite{shi11WMMSE, shi11WMMSE_TSP} established a general
equivalence result between the global (and local) minimizers of the
weighted sum-utility maximization problem (e.g.,
(\ref{eqUtilityMaximizationMIMO}) and (\ref{eqWSRM})) and a suitably
defined weighed MMSE minimization problem. The latter can be
effectively optimized by utilizing the block coordinate descent
technique, resulting in independent, closed form iterative update
across the transmitters and receivers. The resulting algorithm is
named the WMMSE algorithm.

To gain some insight, let us consider the special case of a scalar
IC system where the equivalence of the WSRM problem (\ref{eqWSRM})
and a weighted sum MSE minimization can be seen more directly. Let
$v_j,\;u_k$ denote the complex gains used by
the transmitter $j$ and receiver $k$ respectively. Consider the following weighted sum-MSE minimization problem%
\begin{equation}\label{eq:virtual_IA_1}
\begin{array}{ll}
\displaystyle \min_{\{w_k,u_k,v_k\}_{k=1}^K}\quad &\displaystyle \sum_{k=1}^K \mu_k\left(w_k e_k-\log w_k\right)\\
\textrm{s.t.} \quad &|v_k|^2\leq {\bar p}_k, ~k=1,2,\ldots, K
\end{array}
\end{equation}
where $w_k$ is a positive weight variable, and $e_k$ is the mean
square estimation error
\[
e_k \triangleq |u_k H_{kk}v_k-1|^2+\sum_{j\neq k} |u_k H_{jk}
v_j|^2+|u_k|^2.
\]
To see the equivalence, we can check the first order
optimality condition to find the optimal $w_k$ and $u_k$ 
\begin{equation}\label{eq:f_k}
u_k^{opt} = \frac{\overline{H_{kk} v_k}}{\sum_{j=1}^K
|H_{jk}|^2|v_j|^2+1},\quad  w_k^{opt} = e_k^{-1}, \quad \forall\ k =
1,2,\ldots,K.\nonumber
\end{equation}
Plugging these optimal values in (\ref{eq:virtual_IA_1})  gives the following equivalent optimization problem
\[
\begin{array}{ll}
\displaystyle\max_{\{v_k\}_{k=1}^K} \quad&\displaystyle\sum_{k=1}^K
\mu_k\log\left(1-\frac{|H_{kk}|^2|v_k|^2}{\sum_{j=1}^K |H_{jk}|^2|v_j|^2+1}\right)^{-1}\\
\textrm{s.t.}\quad &|v_k|^2\leq {\bar p}_k,\quad k=1,2,\ldots, K,
\end{array}
\]
which, upon a change of variable $p_k=|v_k|^2$, is equivalent to
\[
\begin{array}{ll}
\displaystyle \max_{\{p_k\}_{k=1}^K} \quad&\displaystyle
\sum_{k=1}^K \mu_k
\log\left(1+\frac{|H_{kk}|^2p_k}{\sum_{j\neq k} |H_{jk}|^2p_j+1}\right)\\
\textrm{s.t.}\quad &p_k\leq {\bar p}_k, \quad k=1,2,\ldots, K.
\end{array}
\]
This establishes the equivalence of the WMMSE problem
(\ref{eq:virtual_IA_1}) and the WSRM problem (\ref{eqWSRM}). More
importantly, the equivalence goes one step further: there is a
one-to-one correspondence between the local minimums of the two
problems (see \cite{shi11WMMSE, shi11WMMSE_TSP}).

The equivalence relation implies that maximizing the weighted sum-rate can be accomplished via iterative weighted MSE minimization. The latter problem is in the space of $(u,v,w)$ and is easier to handle since optimizing each variable while holding the others fixed is convex and easy (e.g., closed form). This property has been exploited in \cite{shi11WMMSE, shi11WMMSE_TSP} to design the WMMSE algorithm. In contrast, the original sum-rate maximization problem (\ref{eqWSRM}) is in the space of $\mathbf{p}$ and is nonconvex, which makes the iterative optimization process difficult.

The general form of the WMMSE algorithm can handle any utility
functions satisfying the following conditions
\begin{eqnarray}
&\mbox{(Separability)}~&U(\{R_k\}_{k=1}^{K})=\sum_{k=1}^{K}u_k(R_k)\label{eqWMMSEUtilitySeparable}\\
&\mbox{(Concavity)}~& -u_k(-\log_2\det(\mathbf{X}))~~\mbox{strictly concave in $\mathbf{X}\succ\mathbf{0}$}~,\forall~k\in\mathcal{K}\label{eqWMMSEUtilityConcave}\\
&\mbox{(Differentiability)}~& u_k(x)~~\mbox{increasing and twice continuously differentiable in $x$}~,\forall~k\in\mathcal{K}\label{eqWMMSEUtilityDiff};
\end{eqnarray}
In addition, it also handles a wide range of channel models, e.g., MIMO and parallel IC/IBC/IMAC.
It is well known that
$R_k=\max_{\mathbf{U}_k}\log_2\det\left(\mathbf{E}_k^{-1}(\mathbf{U}_k,\{\mathbf{V}_k\}_{k=1}^{K})\right)$,
where $\mathbf{E}_k$ is the mean square error (MSE) matrix for user $k$. Define a set of new functions:
$c_k\left(\mathbf{E}_k\right)=
-u_k\left(-\log\det\left(\mathbf{E}_k(\mathbf{U}_k,\{\mathbf{V}_k\}_{k=1}^{K})\right)\right)$, $k=1,\cdots, K$.
Similar to the scalar IC case, the equivalence of the following two optimization problems can be established
\begin{eqnarray}
\min_{\{(\mathbf{U}_k,\mathbf{V})\}_{k=1}^{K}}&&\sum_{k=1}^{K}c_k
\left(\mathbf{E}_k\right)\\
\textrm{s.t.}&&\mbox{Tr}(\mathbf{V}_k\mathbf{V}^H_k)\le \bar{p}_k,~k=1,\cdots,K.\nonumber\\
\min_{\{(\mathbf{U}_k,\mathbf{V}_k,\mathbf{W}_k)\}_{k=1}^{K}}&&
\sum_{k=1}^{K}\mbox{Tr}\left(\mathbf{W}^H_k \mathbf{E}_k\right)+c_k\left(\phi_k(\mathbf{W}_k)\right)-\mbox{Tr}\left(\mathbf{W}^H\phi_k(\mathbf{W}_k)\right)
\label{eqWMMSE}\\
\textrm{s.t.}&&\mbox{Tr}(\mathbf{V}_k\mathbf{V}^H_k)\le
\bar{p}_k,~k=1,\cdots,K.\nonumber
\end{eqnarray}
where $\phi_k(\mathbf{W}_k)$ is the inverse map of $\triangledown
c_k(\mathbf{E}_k)$. The WMMSE algorithm finds a stationary point of
the alternative problem (\ref{eqWMMSE}). In particular, it
alternately updates the three sets of variables
$\{\mathbf{U}_k\}_{k=1}^{K}$, $\{\mathbf{V}_k\}_{k=1}^{K}$ or
$\{\mathbf{W}_k\}_{k=1}^{K}$ for problem (\ref{eqWMMSE}), each time
keeping two sets of variables fixed. The WMMSE algorithm for a MIMO
IC is listed in the following table:

\begin{center}
\fbox{\begin{minipage}{4.5in}
\begin{description}
\item [1)] Initialize $\{\mathbf{V}_k\}_{k\in\mathcal{K}}$ such that
$\mbox{Tr}(\mathbf{V}_k\mathbf{V}^{H}_k)={\bar{p}_k}$;
 \item [2)] {\bf repeat}
\item [3)] ~~~~$\mathbf{W}^{'}_k \leftarrow
\mathbf{W}_k,~\forall~k\in\mathcal{K}$;
\item [4)]~~~~$\mathbf{U}_k \leftarrow
\bigg(\sum_{l\in\mathcal{K}}\mathbf{H}_{lk}\mathbf{V}_l\mathbf{V}^{H}_l\mathbf{H}^H_{lk}+\mathbf{I}\bigg)^{-1}
\mathbf{H}_{kk}\mathbf{V}_k,~\forall~k\in\mathcal{K}$;
\item [5)]~~~~$\mathbf{W}_k\leftarrow\left(\mathbf{I}-\mathbf{U}^H_{kk}\mathbf{H}_{kk}\mathbf{V}_k\right)^{-1},
~\forall~k\in\mathcal{K}$;
\item [6)]~~~~$\mathbf{V}_k\leftarrow \mu_k
\bigg(\sum_{l\in\mathcal{K}}\mu_l
\mathbf{H}^{H}_{kl}\mathbf{U}_l\mathbf{W}_l\mathbf{U}^H_l\mathbf{H}_{kl}+\lambda_k^*\mathbf{I}\bigg)^{-1}
\mathbf{H}^H_{kk}\mathbf{U}_k\mathbf{W}_k,~\forall~k\in\mathcal{K}$;
\item [7)] {\bf until}~$\left|\sum_{l\in\mathcal{K}}
\log\det\left(\mathbf{W}_l\right)-\sum_{l\in\mathcal{K}}\log\det\left(\mathbf{W}^{'}_l\right)\right|\le
\epsilon$.
\end{description}
\end{minipage}
}
\end{center}
We note that in Step 6), $\lambda_k^*\ge 0$ is the Lagrangian
multiplier for the constraint
$\mbox{Tr}(\mathbf{V}_k\mathbf{V}^H_k)\le \bar{P}_k$. This
multiplier can be found easily by bi-section method. Also, notice
that all updates are in closed form (except for $\lambda_k^*$) and
can be performed simultaneously across users.


To compare the performance and the efficiency of various resource
allocation methods, we consider a simple simulation experiment
involving a parallel-IC and a MIMO-IC. We first specialize the WMMSE
algorithm to the parallel-IC scenario and compare it with SCALE and
MDP algorithms described earlier.  To specialize the WMMSE algorithm
for a parallel IC, let us restrict the transmit/receive matrices for
each user to be diagonal. That is, the beamforming directions are
fixed to be unit vectors and we only optimize power loading factors
on the parallel channels. Let $\mathbf{v}_k\in\mathcal{C}^{N\times
1}$ denote the user $k$'s transmit filter vector, with ${v}^n_k$
corresponding to the complex scaling coefficient to be used for the
data stream on channel $n$. Similarly, the receive filter vector and
the weight vector are denoted by $\mathbf{u}_k,
\mathbf{w}_k\in\mathcal{C}^{N\times 1}$  respectively. Then the
WMMSE algorithm for the parallel IC channel can be described as

\begin{center}
\fbox{\begin{minipage}{4.5in}
\begin{description}
\item [1)] Initialize $\{\mathbf{v}_k\}_{k\in\mathcal{K}}$ such that
$\sum_{n\in\mathcal{N}}{v}^n_k={\bar{p}_k}$;
 \item [2)] {\bf repeat}
\item [3)] ~~~~$({w}^n_k)^{'} \leftarrow
{w}^n_k,~\forall~(n,k)\in\mathcal{N}\times\mathcal{K}$;
\item [4)]~~~~${u}^n_k \leftarrow
\bigg(\sum_{l\in\mathcal{K}}|{H}^n_{lk}|^2|{v}^n_l|^2+1\bigg)^{-1}
{H}^n_{kk}{v}^n_k,~\forall~(n,k)\in\mathcal{N}\times\mathcal{K}$;
\item [5)]~~~~${w}^n_k\leftarrow\bigg(1-\overline{u}^n_{k}{H}^n_{kk}{v}^n_k\bigg)^{-1},
~\forall~(n,k)\in\mathcal{N}\times\mathcal{K}$;
\item [6)]~~~~${v}^n_k\leftarrow \mu_k
\frac{\overline{H}^n_{kk}{u}^n_k{w}^n_k}{\sum_{l\in\mathcal{K}}\mu_l
|{H}^n_{kl}|^2|{u}^n_l|^2{w}^n_l+\lambda_k^*}
,~\forall~(n,k)\in\mathcal{N}\times\mathcal{K}$;
\item [7)] {\bf
until}~$\left|\sum_{l\in\mathcal{K}}\sum_{n\in\mathcal{N}}
\log({w}^n_l)-\sum_{l\in\mathcal{K}}\sum_{n\in\mathcal{N}}\log\left(({{w}^{n}}_l)^{'}\right)\right|\le
\epsilon$.
\end{description}
\end{minipage}
}
\end{center}


In the simulation, we set the weights $\{\mu_k\}_{k\in\mathcal{K}}$
all equal to $1$, and set the maximum power
$\bar{p}_{k}=10^{\mbox{SNR}/10}$ for all the users. We set the
stopping criteria as $\epsilon=0.01$ for all algorithms. The channel
coefficients are generated from the complex Gaussian distribution
$\mathcal{CN}(0,1)$. For MIMO IC, all the transmitters and receivers
are assumed to have the same number of antennas.

We first investigate the performance of SCALE, MDP and the parallel
version of the WMMSE algorithm for a parallel IC. Fig.\
\ref{figCompareRate} illustrates the sum rate performance of
different algorithms when $K=[10,~20]$ and $N=32$. We see that these
algorithms all have similar performance across all the SNR values.
Fig.\ \ref{figCompareTime} shows the averaged CPU time comparison of
these three algorithms under the same termination criteria and the
same accuracy for the search of Lagrangian variables. We observe
that the WMMSE requires much less computational time compared to the
other two algorithms when the number of users becomes large. Note
that the first step in the SCALE algorithm is implemented using the
subgradient and the fixed point iterations suggested in
\cite[Section IV-A]{papand09}. The stepsizes for the subgradient
method as well as the number of the fixed point iterations need to
be tuned appropriately to ensure fast convergence.

    \begin{figure*}[ht]
    \begin{minipage}[t]{0.45\linewidth}
    \centering
     {\includegraphics[width=
1\linewidth]{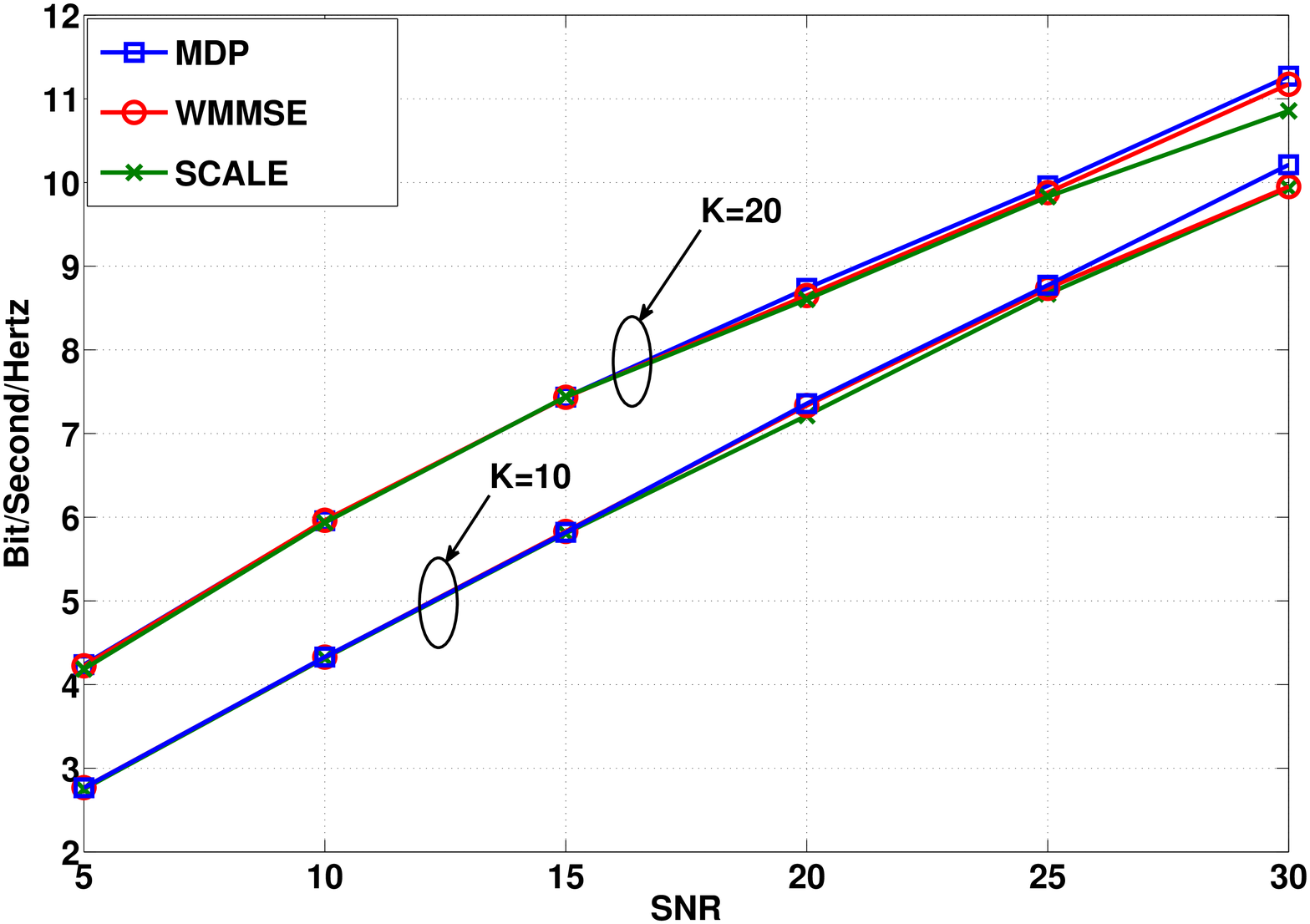} \caption{\small{Comparison of
the averaged sum rate performance versus SNR of different algorithms
in parallel IC. $K=[10,~20]$, $N=32$. Each curve in the figure is
averaged over 100 random channel
realizations}.}\label{figCompareRate} }
\end{minipage}\hfill
    \begin{minipage}[t]{0.45\linewidth}
    \centering
    {\includegraphics[width=
1\linewidth]{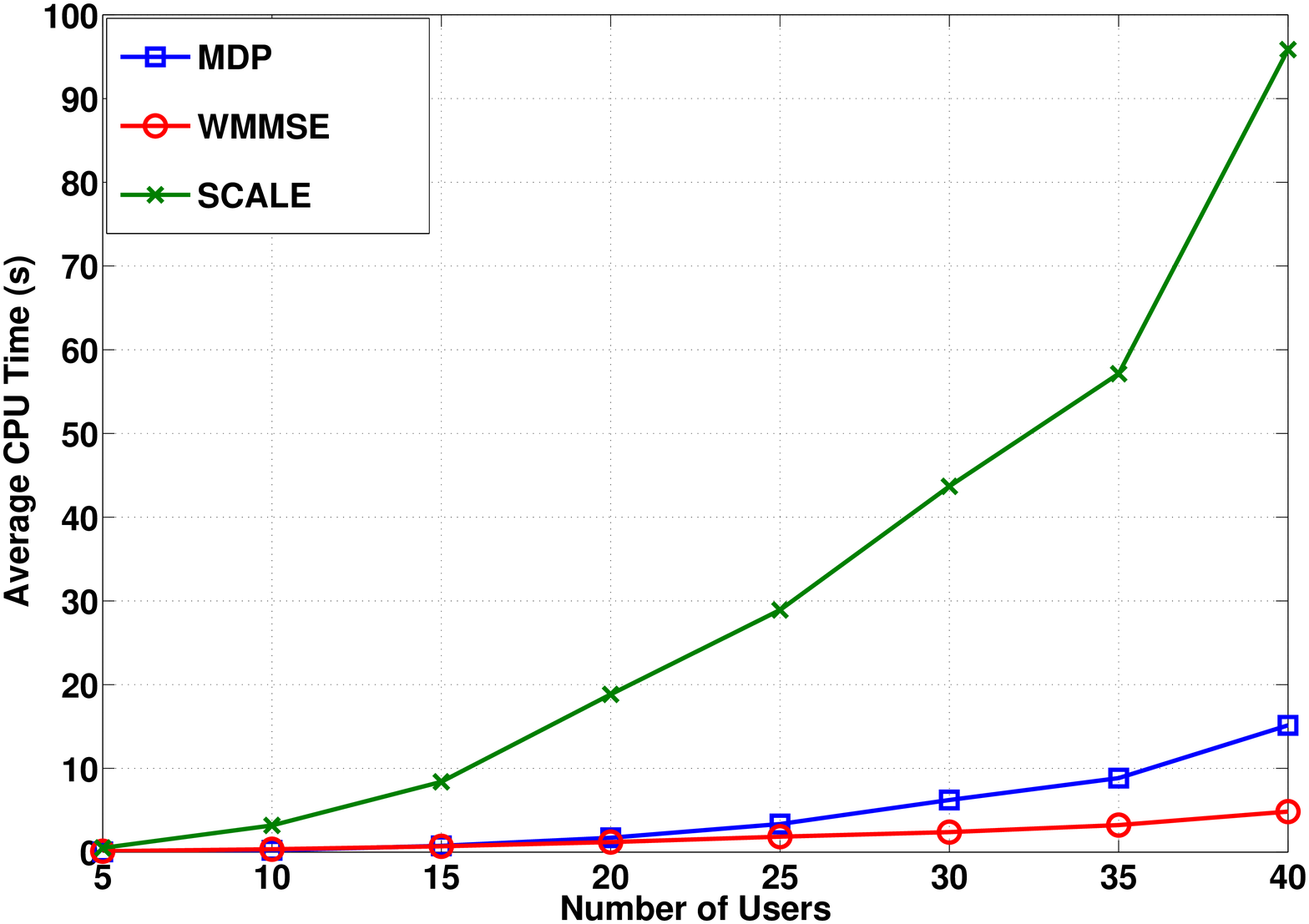} \caption{{\small Comparison of
the averaged CPU time versus the number of users of different
algorithms in parallel IC. $N=32$, $SNR=10$. Each curve in the
figure is averaged over 100 random channel
realizations.}}\label{figCompareTime} }
\end{minipage}
    \end{figure*}

Next we examine the performance of the WMMSE and the MIMO
distributed pricing (MIMO-DP) algorithm developed in \cite{kim08} in
the context of a MIMO-IC. Fig. \ref{figCompareRateMIMO} illustrates
the sum rate performance of the two algorithms when $K=[3,~10]$ and
$N_r=N_t=3$. Fig. \ref{figCompareTimeMIMO} shows the averaged CPU
time comparison of the two algorithms. We again observe that the
WMMSE requires much less computational time compared with the
MIMO-DP algorithm when the number of users becomes large.

       \begin{figure*}[ht]
    \begin{minipage}[t]{0.48\linewidth}
    \centering
     {\includegraphics[width=
1\linewidth]{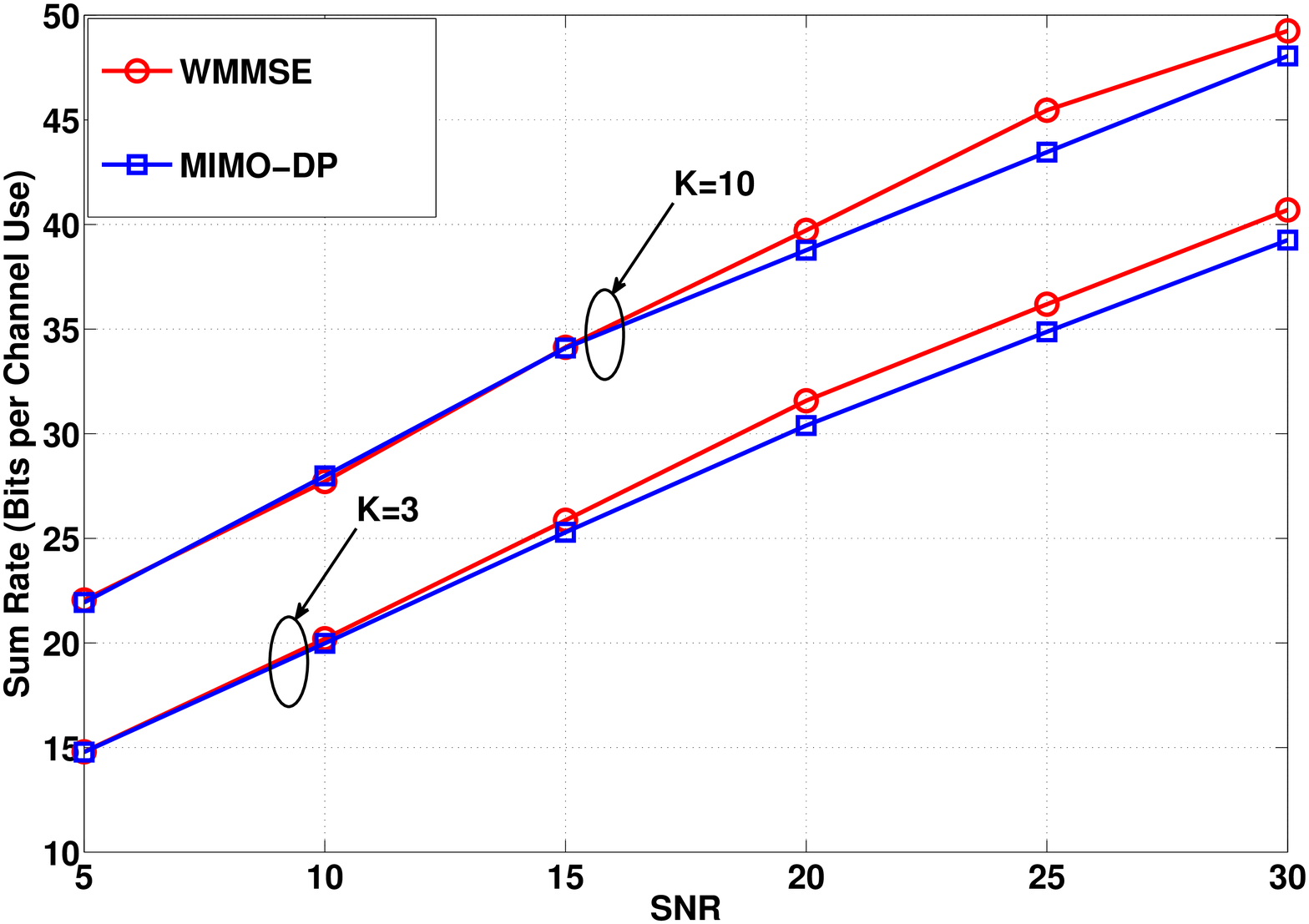} \caption{\small {Comparison of
the averaged sum rate performance versus SNR of different algorithms
in MIMO IC. $K=[3,~10]$, $N_r=N_t=3$. Each curve in the figure is
averaged over $100$ random channel
realizations}}\label{figCompareRateMIMO} }
\end{minipage}\hfill
    \begin{minipage}[t]{0.48\linewidth}
    \centering
    {\includegraphics[width=
1\linewidth]{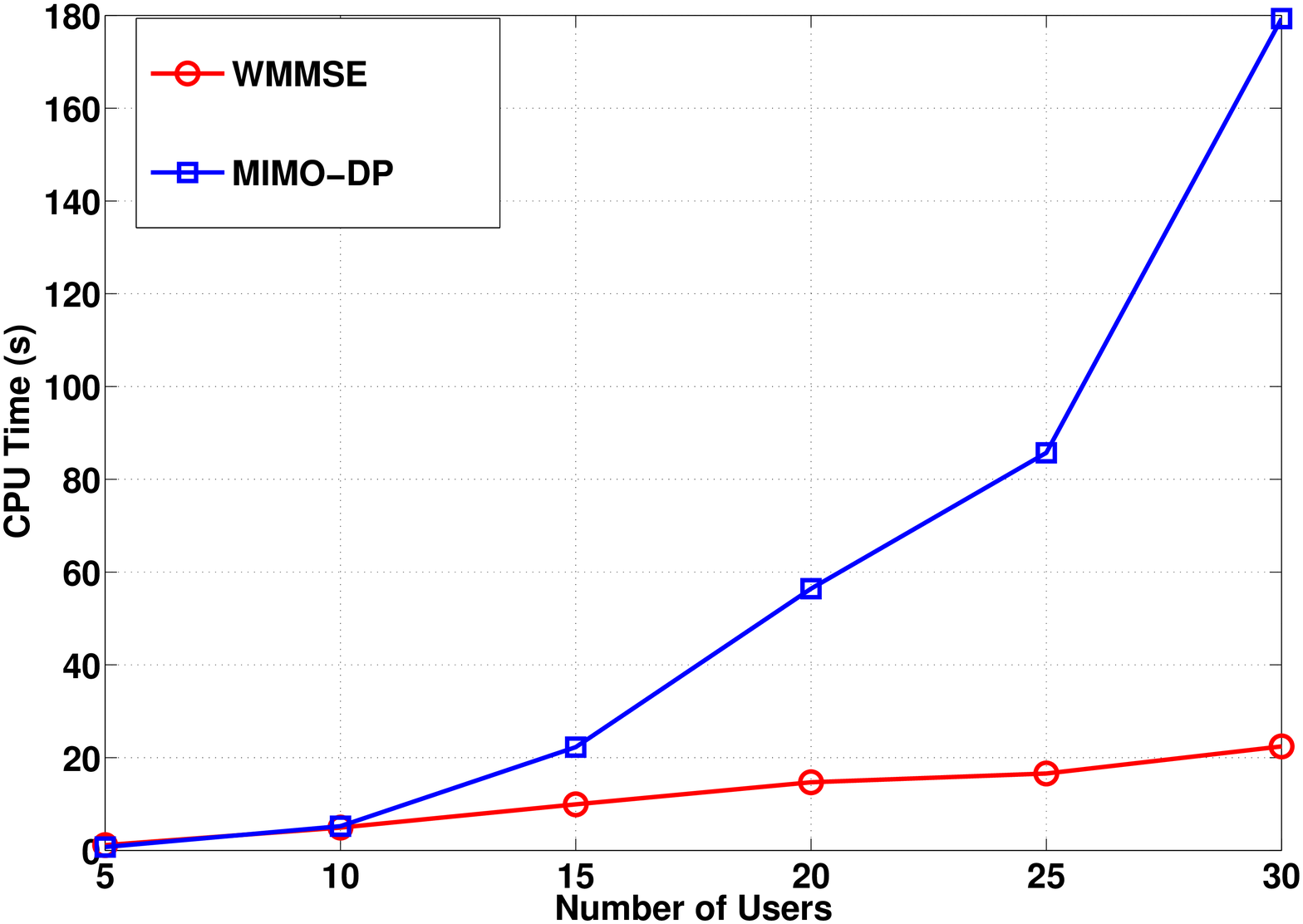} \caption{\small {Comparison of
the averaged CPU time versus the number of users of different
algorithms in MIMO IC. $N_r=N_t=3$, $\mbox{SNR}=10$. Each curve in
the figure is averaged over $100$ random channel
realizations}}\label{figCompareTimeMIMO} }
\end{minipage}
    \end{figure*}

Different from many algorithms discussed earlier (e.g., CCA, MDP), the WMMSE algorithm
allows all transmitters/receivers to update their beamformers simultaneously.
This feature leads to simple implementation and fast convergence. It will be
interesting to see how this algorithm can be further extended to include other
utility functions such as the min-rate utility, and to other formulations
like QoS constrained power minimization. Also, further research is needed to uncover the full algorithmic potential of WMMSE algorithm for a wide range of applications including joint base station assignment, power control, and beamforming.

\medskip
\noindent{\it Algorithms for cross layer resource allocation}
\smallskip

We briefly mention a few cross-layer resource allocation algorithms
which require solving a weighted sum-utility  problem at each step.
These algorithms jointly optimize physical layer as well as the
media access (MAC) layer resources to improve the overall system
performance.

Recently, \cite{yu11} considered the joint MAC layer scheduling and
physical layer beamforming and power control in a multicell
OFDMA-MIMO network. The algorithm assigns the users to the BSs
according to their individual priority and channel status. The
beamformers are updated using a MSE duality results developed in
\cite{dahrouj10} and  \cite{song07} for multicell network. The
transmit powers of the BSs are updated by using the Newton's method.
In \cite{venturino09,venturino07} and \cite{kim09}, the authors
considered the WSRM problem in a multicell downlink OFDMA wireless
network. They proposed to let the BSs alternate between the
following two tasks to achieve high system throughput: 1) optimally
schedule the users on each channel; 2) jointly optimize their
downlink transmit power using a physical layer resource allocation
algorithm such as M-IWF or SCALE. Reference
\cite{Razaviyayn11grouping} proposed to adaptively group users into
non-interfering groups, and optimize the transceiver structure and
the group membership jointly. Such grouping strategy results in fair
resource allocation, as cell-edge users with weak channels are
protected from the strong users. A generalized version of the WMMSE
algorithm has been developed to perform such joint optimization. In
all these works the resulting resource allocation schemes achieved a
weighted sum-utility that is significantly higher than what is
possible with only performing physical layer beamforming/power
allocation.

Joint admission control and downlink beamforming is another example
of cross layer resource allocation. For a single cell MISO network,
this problem has been considered in \cite{Matskani08,Matskani09}. 
A related problem is the joint BS selection and power
control/beamforming problem. This problem has been addressed in the
traditional CDMA based network (see, e.g., \cite{hanly95, yates95b,
Rashid98jiont}), in OFDMA networks (e.g., \cite{hong11_infocom,
gao11, Perlaza09}) and  in a more general MIMO-HetNet in which all
BSs operate on the same frequency bands \cite{hong12_icassp,
Sanjabi12}. An interesting research direction to pursue is to
effectively incorporate these higher layer protocols to boost the
system performance for a MIMO and parallel IBC/IMAC network.

\subsubsection{Algorithms for QoS Constrained Power Minimization}\label{subsubQoS}

For the scalar IC model, the QoS
constrained min-power problem as formulated in (\ref{eqSumPowerQoSScalar}) has been considered in \cite{hanly95,hanly95thesis}
and \cite{yates95b}. They
 derived conditions for the existence of a feasible power
allocation given a set of SINR targets. Define a $K\times K$ matrix $\mathbf{A}$ as follows
\begin{eqnarray}
[\mathbf{A}]_{k,l}= \left\{ \begin{array}{ll}
\frac{\gamma_k|H_{lk}|^2}{|H_{kk}|^2} &\textrm{if } l\ne k\\
0  &\textrm{otherwise } \\
\end{array} \right..
\end{eqnarray}
If $\rho(\mathbf{A})<1$, an optimal power allocation
$\mathbf{p}=\left[p_1,\cdots,p_K\right]^{T}$ can be found as follows
\begin{eqnarray}
\mathbf{p}=\left(\mathbf{I}_{K}-\mathbf{A}\right)^{-1}\mathbf{b}
\end{eqnarray}
where $\mathbf{b}=\left[\frac{\gamma_1}{|H_{11}|^2},\cdots,
\frac{\gamma_K}{|H_{KK}|^2}\right]^{T}$. {The convexity of the
feasible SINR region for this problem has been established in
\cite{Boche04Convexity, Stanczak07Convexity}.}

{Alternatively, \cite{yates95} has provided a framework that allows
the users to compute the optimal solution of problem
(\ref{eqSumPowerQoSScalar}) {\it distributedly} by the following
fixed point iteration
\begin{eqnarray}
p^{(t+1)}_k=\gamma_k\left(\frac{{1+\sum_{l\ne k}|H_{lk}|^2
p^{(t)}_l}}{|H_{kk}|^2}\right), k=1,\cdots,K.
\end{eqnarray}
This algorithm is shown to have linear rate of convergence, that is,
\begin{eqnarray}
\lim{\sup}_{t\to\infty}\frac{\|\mathbf{p}^{(t+1)}-\mathbf{p}^*\|}{\|\mathbf{p}^{(t)}-\mathbf{p}^*\|}=c<1
\end{eqnarray}
where $\mathbf{p}^*$ is the optimal solution of the min-power
problem, and $c<1$ is some positive constant. Recently
\cite{Boche08Superlinear} has proposed a different algorithm based
on a Newton-type update that exhibits even faster (super-linear)
rate of convergence. This algorithm can be applied to more general
scenarios when the receivers are equipped with multiple antennas.}

When the set of SINR targets cannot be supported by the system (that
is, the problem (\ref{eqSumPowerQoSScalar}) is infeasible), the call
admission control mechanism should be invoked. A couple of recent
works \cite{Mitliagkas11, liu12ICASSP} have considered the problem
of joint admission and power control arises in the QoS constrained
power minimization problem. See \cite{Ahmed05} for a survey on the
general topic of call admission control.

For a MISO IC, \cite{Bengtsson99optimaldownlink,bengtsson01} considered the min-power
transmit beamforming problem under QoS constraints (problem (\ref{eqSumPowerQoSMISO})).
Define $\mathbf{V}_k=\mathbf{v}_k\mathbf{v}^H_k$ and
$\mathbf{G}_{lk}=\mathbf{h}^H_{lk}\mathbf{h}_{lk}$,
this problem can be equivalently formulated as
\begin{eqnarray}
\min_{\{\mathbf{V}_k\}_{k=1}^{K}}&&\sum_{k=1}^{K}\mbox{Tr}(\mathbf{V}_k)\label{eqMinPowerMISO}
\\
\textrm{s.t.}~ & &\mbox{Tr}(\mathbf{G}_{kk}\mathbf{V}_k)-
\gamma_k\sum_{l\ne k}\mbox{Tr}(\mathbf{G}_{lk}\mathbf{V}_l)=\gamma_k,~k=1,\cdots,K\nonumber\\
& &\mbox{rank}(\mathbf{V}_k)=1, \mathbf{V}_k\succeq 0,~k=1,\cdots,K\nonumber.
\end{eqnarray}
 Relaxing the rank constraint,
this problem is a convex semidefinite program and can be
solved efficiently. Interestingly, the authors showed that a  rank-one solutions must
exist for the relaxed problem, revealing a certain hidden convexity in this problem.
The following procedure can be used to construct a
rank-1 solution from an optimal solution $\{\mathbf{V}^*_k\}_{k=1}^{K}$ of the relaxed problem.

\begin{center}
\fbox{\begin{minipage}{4in}
\begin{description}
\item [1)]  Take $\mathbf{e}_k\in \mbox{span}(\mathbf{V}^*_k)$;
\item [2)] Define ${\bfeta}= [\gamma_1,\cdots,\gamma_K]^{T}$;
\item [3)] Define a matrix $\mathbf{F}$ with its elements
as
\begin{eqnarray}
[\mathbf{F}]_{kl}= \left\{ \begin{array}{ll}
-\gamma_k\mathbf{e}^H_l \mathbf{G}_{kk}\mathbf{e}_l &\textrm{if } l\ne k,\\
\mathbf{e}^H_k \mathbf{G}_{kk}\mathbf{e}_k  &\textrm{otherwise } \\
\end{array} \right.;
\end{eqnarray}
\item [4)] Find $\mathbf{p}=\mathbf{F}^{-1}{\bfeta}$;
\item [5)] Obtain $\mathbf{v}^*_k=\sqrt{[\mathbf{p}]_k}\mathbf{e}_k$.
\end{description}
\end{minipage}
}
\end{center}
The approach works for the IBC model as well. It can also be
extended to include other resource allocation options, such as
admission control and base station assignment;  see \cite{stridh06},
\cite{Bengtsson01BSAssignment} and \cite{Rashid98jiont} for details.

\subsubsection{Algorithms for Hybrid Formulations}\label{subsubMixed}

For a scalar IC, \cite{Chiang07Geometric} proposed to use a
technique called {\it geometric programming} (GP) to find an
approximate solution to the WSRM problem with QoS constraint. They
showed that after approximating the rate function
$\log(1+\mbox{SINR}_k)$ by $\log(\mbox{SINR}_k)$, the WSRM problem
becomes a GP and can be solved efficiently. {Moreover, with this
approximation, the resource allocation problem falls into the family
of problems considered in \cite{Boche10Unify,Boche11ConvexConcave},
which can be transformed into equivalent convex optimization
problems. For this family of problems, fast and distributed
algorithms based on certain Newton-type iteration have been proposed
in \cite{Wiczanowski08}.}

However, this approximation is not so useful in practice because 1)
it is accurate only in high SINR region; 2) it always leads to a
solution for which all
 links are active. The latter feature is undesirable because
having all links active can be highly suboptimal when interference is strong.
In fact, the main difficulty with WSRM is precisely how to identify
 which links should be shut off, an important option that is excluded by the GP approximation approach.

Recognizing such problems, the same authors further proposed in
\cite{Chiang07Geometric} a successive convex approximation (GP-SCA)
method that aims at finding a stationary solution to the original
WSRM problem. In particular, let $e_{lk}(\mathbf{p})=|H_{lk}|^2
p_l$,
 $f_k(\mathbf{p})={1+\sum_{l=1}^{K}e_{lk}(\mathbf{p})}
$ and $g_k(\mathbf{p})={1+\sum_{l\ne k}|H_{lk}|^2 p_l}$, for $k=1,\cdots,K$. Utilizing the
arithmetic-geometric mean inequality, the users' rate functions
can be lower-approximate as
\begin{eqnarray}
\log_2(1+\mbox{SINR}_k)&=&\log_2\left(\frac{f_k(\mathbf{p})}{g_k(\mathbf{p})}\right)\label{eqFractional}\\
&\ge& \log_2\left(\frac{\prod_{l=1}^{K}\left(\frac{e_{lk}(\mathbf{p})}{\alpha_l}\right)^{\alpha_l}
\times \left(\frac{1}{\beta}\right)^{\beta}}{g_k(\mathbf{p})}\right),\label{eqGPApproximate}\\
\mbox{where},~~\alpha_l&=&\frac{e_{lk}(\widehat{\mathbf{p}})}{f_k(\widehat{\mathbf{p}})},~
\beta=\frac{1}{f_k(\widehat{\mathbf{p}})}.
\end{eqnarray}
This lower bound is again concave (upon performing a log-transformation),
and it is tight when $\mathbf{p}=\widehat{\mathbf{p}}$. The QoS constrained WSRM
problem with the approximated objective (\ref{eqGPApproximate})
can be again solved by a GP. A similar alternating procedure as the one we
have introduced for the SCALE algorithm can be used to compute a stationary
solution to the WSRM problem with QoS constraint.

\medskip
\noindent{\it Algorithms based on global optimization}
\smallskip

There are a number of attempts to find globally optimal solution for the WSRM problem.
However, these algorithms are all based on implicit enumeration (not surprising in light of the complexity results in Section~\ref{subSecComplexity}). As a result, they can only solve small scale problems and are unlikely to be suitable for implementation in practical applications. However, this does not mean that global optimization algorithms for WSRM are useless. For one thing, they can be a valuable tool to benchmark various low-complexity suboptimal approaches for resource allocation (e.g., those described earlier in this section).

For a scalar IC,
\cite{qian09mapel} proposed to use an existing algorithm (\cite{frenk06} and \cite{nguyen03})
for nonconvex fractional programming to find the {\it global} optimal solution
for the hybrid problem (\ref{eqMixedScalar}).  Specifically, introducing a set of auxiliary variable $\{z_k\}_{k=1}^{K}$,
the scalar WSRM problem with SINR constraint can be formulated into the following equivalent form
\begin{eqnarray}
&\max_{\mathbf{z},\mathbf{p}}&\prod_{k=1}^{K}(z_k)^{w_k}\label{eqMAPELReformulation}\\
&\textrm{s.t.}&0\le {z}_k\le \frac{f_k(\mathbf{p})}{g_k(\mathbf{p})},\nonumber\\
&&0\le p_k\le\bar{p}_k,~\frac{f_k(\mathbf{p})}{g_k(\mathbf{p})}\ge {\gamma_k},\forall~k\in\mathcal{K}\nonumber.
\end{eqnarray}
This reformulated problem has a concave objective (upon a log transformation), and a nonconvex
feasible set $\mathcal{G}$. The global optimization algorithm of \cite{frenk06} and \cite{nguyen03}
solves the reformulated problem via some convex optimization problems over a sequence {\it shrinking} convex sets $S_1\supset S_2\cdots\supset \mathcal{G}$. The worst case complexity of this algorithm is exponential.

Several other global optimization methods have been proposed
to solve the utility maximization problems for more general IC models. For example, \cite{qian09}, \cite{xu08} and \cite{tan09infocom,Tan11} considered the parallel IC model, and \cite{Jorswieck10}
treated the two user-MISO IC model. In
particular, the algorithm proposed in \cite{xu08} utilized the dc structure (\ref{eqDC}) of
the weighted sum rate function, and applied a branch-and-bound (BB) algorithm
to find global optimal solution to the WSRM problem.
Due to their exponentially increasing complexity, these algorithms are only suitable for benchmarking
resource allocation algorithm for networks with relatively small number of links. For example, the work of \cite{qian09,qian09mapel} compared their
global algorithms for a small parallel IC with $N=4$, $K=2$, and a scalar IC with $K$ up to $10$.

An important open problem is how to develop efficient algorithms (suitable for large networks) that can find (provably) tight upper bounds for the system performance.

\subsubsection{Algorithms for robust resource allocation}
All of the aforementioned resource management schemes require
perfect channel state information (CSI) at the transmitter side.
However, in practice the CSI obtained at the transmitter is
susceptible to various sources of uncertainties such as quantization
error, channel estimation error or channel aging. These
uncertainties may significantly degrade the performance of resource
allocation schemes that are designed using perfect CSI. As a result,
{\it robust} designs are needed for practical resource management.

Several recent contributions considered {robust} linear transmitter
design in a MISO channel with a single transmitter and multiple
receivers. Let $\mathbf{h}_k\in\mathcal{C}^{N_t}$ and
$\hat{\mathbf{h}}_k\in\mathcal{C}^{N_t}$ denote the {\it true}
channel  and the {\it estimated} channel between the transmitter and
the $k$th receiver, respectively. Let $\mathcal{U}_k$ denote the
uncertainty set of channel $\mathbf{h}_k$, which is the set of
possible values that $\mathbf{h}_k$ may take after obtaining the
estimated channel $\hat{\mathbf{h}}_k$. Consider the following
specific form of uncertainty set
\begin{equation}
\mathcal{U}_k(\mathbf{\Delta}_k)=\left\{\mathbf{h}_k\mid \mathbf{h}_k=\hat{\mathbf{h}}_k+\mathbf{\Delta}_k,
\|\mathbf{\Delta}_k\|\le \delta_k\right\}\label{eqUncertaintySet}
\end{equation}
where $\mathbf{\Delta}_k$ is the vector of estimation error and
$\delta_k$ is the uncertainty bound. One of the most popular
formulation of the robust design is the following QoS constrained
min power problem
\begin{eqnarray}
&\displaystyle\min_{\{\mathbf{v}_k\}_{k=1}^{K}}&\quad\sum_{k=1}^{K}\|\mathbf{v}_k\|^2\label{eqRobustQoS}\\
&\textrm{s.t.} & \quad
\frac{|\mathbf{h}_k\mathbf{v}_k|^2}{\sum_{l\ne
k}|\mathbf{h}_k\mathbf{v}_l|^2+1}\ge
\gamma_k,~\forall~\mathbf{h}_k\in\mathcal{U}_k(\mathbf{\Delta}_k),
~k=1,\cdots,K.\nonumber
\end{eqnarray}
This formulation aims at minimizing the total transmission power
while ensuring that the SINR constraints are satisfied {\it under
all possible channel uncertainties}. Define
\begin{eqnarray}
&\mathbf{V}&=\left[\mathbf{v}_1,\cdots,\mathbf{v}_k\right]\nonumber\\
&\underline{\mathbf{h}}_k&=\left[\textrm{Re}\{\mathbf{h}_k\}\quad
\textrm{Im}\{\mathbf{h}_k\}\right]\nonumber\\
&\underline{\mathbf{V}}&=\left[\begin{array}{ll} \displaystyle
\textrm{Re}\{\mathbf{V}\}&
\textrm{Im}\{\mathbf{V}\}\nonumber\\
\displaystyle -\textrm{Im}\{\mathbf{V}\}& \textrm{Re}\{\mathbf{V}\}
\end{array}\right]\nonumber\\
&\underline{\mathbf{v}}_k&=\left[\begin{array}{l} \displaystyle
\textrm{Re}\{\mathbf{v}_k\}\nonumber\\
\displaystyle -\textrm{Im}\{\mathbf{v}_k\}
\end{array}\right]\nonumber.
\end{eqnarray}
In \cite{shenouda07} problem (\ref{eqRobustQoS}) has been
reformulated into the following semi-infinite SOCP
\begin{eqnarray}
&\displaystyle\min_{\underline{\mathbf{V}},t}&\quad t\label{eqSOCPRelax}\\
&\textrm{s.t.}&\quad
\left\|\textrm{vec}\left(\left[\underline{\mathbf{v}}_1,\cdots,
\underline{\mathbf{v}}_K\right]\right)\right\|\le t\nonumber\\
&&\quad
\left\|\left[\underline{\mathbf{h}}_k\underline{\mathbf{V}},~1\right]\right\|
\le\sqrt{1+\gamma_k^{-1}}\underline{\mathbf{h}}_k\underline{\mathbf{v}}_k,
~\forall~\underline{\mathbf{h}_k}\in\mathcal{U}_k(\mathbf{\Delta}_k),~k=1,\cdots,K.\label{eqSOCConstraint}
\end{eqnarray}

This reformulation is a \emph{convex restriction} to the original
problem (\ref{eqRobustQoS}) in that the the complex magnitude
$|\mathbf{h}_k\mathbf{v}_k|$ in the constraint is replaced by the
lower bound equal to its real part
$\textrm{Re}(\mathbf{h}_k\mathbf{v}_k)=\underline{\mathbf{h}}_k\underline{\mathbf{v}}_k$.
However, due to the presence of $\underline{\mathbf{h}}_k$ on both
sides of the SOC constraint (\ref{eqSOCConstraint}), this
reformulated problem is still difficult to solve. A conservative
design is then developed by assuming independent uncertainties for
$\underline{\mathbf{h}}_k$ on the left and right hand sides of each
constraint in (\ref{eqSOCConstraint}). With such an assumption,
problem (\ref{eqSOCConstraint}) can be transformed to the following
SDP problem and solved efficiently using standard interior point
method.
\begin{eqnarray}
&\displaystyle\min_{\underline{\mathbf{V}},\bfeta,\bfkappa, t}&\quad t
\label{eqRobustSOCPSDPRelaxation}\\
&\textrm{s.t.}&\quad\left\|\textrm{vec}\left(\left[\underline{\mathbf{v}}_1,\cdots,
\underline{\mathbf{v}}_K\right]\right)\right\|\le t\nonumber\\
&&\quad\left[\begin{array}{lll} \displaystyle
\kappa_k-\eta_k&\mathbf{0}&
\left[\underline{\hat{\mathbf{h}}}_k\underline{\mathbf{V}},~1\right]\nonumber\\
\displaystyle \mathbf{0}& \eta_k\mathbf{I}_{2N_t}&
\delta_k[\underline{\mathbf{V}},\mathbf{0}]\nonumber\\
\left[\underline{\hat{\mathbf{h}}}_k\underline{\mathbf{V}},~1\right]^{T}&
\delta_k[\underline{\mathbf{V}},\mathbf{0}]^{T}&
\kappa_k\mathbf{I}_{2K+1}
\end{array}\right]\succeq 0,~k=1,\cdots,K\nonumber\\
&&\quad\left[\begin{array}{ll} \displaystyle
\sqrt{1+\gamma_k^{-1}}\underline{\hat{\mathbf{h}}}_k\underline{\mathbf{v}}_k-\kappa_k&
\delta_k \sqrt{1+\gamma_k^{-1}} \underline{\mathbf{v}}^T_k\nonumber\\
\displaystyle \delta_k \sqrt{1+\gamma_k^{-1}}
\underline{\mathbf{v}}_k&\left(\sqrt{1+\gamma_k^{-1}}\underline{\hat{\mathbf{h}}}_k\underline{\mathbf{v}}_k
-\kappa_k\right)\mathbf{I}_{2N_t}
\end{array}\right]\succeq 0,~k=1,\cdots,K\nonumber.
\end{eqnarray}
Instead of solving (\ref{eqSOCPRelax}) by the SDP relaxation
(\ref{eqRobustSOCPSDPRelaxation}), \cite{Vucic2009} proposed to
solve (\ref{eqSOCPRelax}) by {\it 1)} directly applying the
ellipsoid method from convex optimization and {\it 2)} approximating
(\ref{eqSOCPRelax}) by a robust MSE constrained min power problem.
Let $u_k\in\mathcal{C}$ denote the scalar receive filter used at
receiver $k$. Let $\mathbf{e}_k$ denote the unit vector with its
$k$th element being $1$. Define the MSE of the $k$th user as
\begin{eqnarray}
\textrm{MSE}_k\left(\{\mathbf{v}_k,u_k\}_{k=1}^{K}\right)=|u_k|^2
\left(\left(\mathbf{h}_k\mathbf{V}-\frac{1}{u_k}\mathbf{e}_k\right)
\left(\mathbf{h}_k\mathbf{V}-\frac{1}{u_k}\mathbf{e}_k\right)^{H}+1\right).
\end{eqnarray}
The robust MSE constrained min power problem is given as
\begin{eqnarray}
&\displaystyle\min_{\{\mathbf{v}_k, u_k\}_{k=1}^{K}}&\quad
\sum_{k=1}^{K}\|\mathbf{v}_k\|^2\label{eqRobustMSEConstrainedMinPower}\\
&\textrm{s.t.}&\quad\textrm{MSE}_k\left(\{\mathbf{v}_k,u_k\}_{k=1}^{K}\right)\le
\frac{1}{1+\gamma_k},~\forall~{\mathbf{h}_k}\in\mathcal{U}_k(\mathbf{\Delta}_k),~k=1,\cdots,K.
\end{eqnarray}
This problem is convex and can be equivalently formulated as an SDP
problem and efficiently solved by interior point methods. It is
shown in \cite{Vucic2009} that both the ellipsoid method approach
and the robust MSE constrained reformulation approach achieve better
performance than the SDP relaxation
(\ref{eqRobustSOCPSDPRelaxation}) in terms of various system level
performance measures.

As noted earlier, the original min power SINR constrained problem
(\ref{eqRobustQoS}) is not equivalent to the formulation
(\ref{eqSOCPRelax}), as the latter replaces the nonlinear term
$|\mathbf{h}_k\mathbf{v}_k|$ by a linear lower bound
$\textrm{Re}(\mathbf{h}_k\mathbf{v}_k)=\underline{\mathbf{h}}_k\underline{\mathbf{v}}_k$.
Implicit in this reformulation is the additional requirement that
$\textrm{Re}(\mathbf{h}_k\mathbf{v}_k)$ is positive for all the
channels $\mathbf{h}_k\in \mathcal{U}_k(\delta_k)$. Recently, the
authors of \cite{song11} showed that the direct SDP relaxation of
the original problem (\ref{eqRobustQoS}) is actually tight as long
as the size of the uncertainty set is sufficiently small. This
implies that robust resource allocation for MISO channels can be
solved to global optimality in polynomial time, provided the channel
uncertainty is small. More precisely, define
$\mathbf{V}_k=\mathbf{v}_k\mathbf{v}^H_k$, and
$\mathbf{X}_k=\frac{1}{\gamma_k}\mathbf{V}_k-\sum_{l\ne
k}\mathbf{V}_l$, the problem (\ref{eqRobustQoS}) can be equivalently
reformulated as
\begin{eqnarray}
&\displaystyle\min_{\{\mathbf{V}_k, \kappa_k\}_{k=1}^{K}}&\quad\sum_{k=1}^{K}\textrm{Tr}(\mathbf{V}_k)\label{eqRobustQoSReformulation}\\
&\ \ \textrm{ s.t.} & \quad \kappa_k\ge 0,~k=1,\cdots,K\nonumber\\
&&\quad\left[\begin{array}{ll} \displaystyle
\mathbf{X}_k+\kappa_k\mathbf{I}&\mathbf{X}_k\hat{\mathbf{h}}^H_k\nonumber\\
\displaystyle \hat{\mathbf{h}}_k\mathbf{X}^H_k&
\hat{\mathbf{h}}_k\mathbf{X}^H_k\hat{\mathbf{h}_k}^H-1-\kappa_k\delta_k^2
\end{array}\right]\succeq 0,~k=1,\cdots,K\nonumber.\\
&& \mathbf{V}_k\succeq 0,\quad\textrm{rank}(\mathbf{V}_k)=1,
~k=1,\cdots,K.\nonumber
\end{eqnarray}
When the rank constraints are dropped, this problem becomes the
following SDP and can be efficiently solved.
\begin{equation}
\begin{array}{lll}\displaystyle
&\displaystyle\min_{\{\mathbf{V}_k, \kappa_k\}_{k=1}^{K}}&\quad\sum_{k=1}^{K}\textrm{Tr}(\mathbf{V}_k)\nonumber\\
\displaystyle
&\ \ \ \textrm{s.t.} & \quad \kappa_k\ge 0,~k=1,\cdots,K\nonumber\\
\displaystyle &&\quad\left[\begin{array}{ll} \displaystyle
\mathbf{X}_k+\kappa_k\mathbf{I}&\mathbf{X}_k\hat{\mathbf{h}}^H_k\nonumber\\
\displaystyle \hat{\mathbf{h}}_k\mathbf{X}^H_k&
\hat{\mathbf{h}}_k\mathbf{X}^H_k\hat{\mathbf{h}_k}^H-1-\kappa_k\delta_k^2
\end{array}\right]\succeq 0,~k=1,\cdots,K\nonumber.\\
\displaystyle && \mathbf{V}_k\succeq 0, ~k=1,\cdots,K.\nonumber
\end{array}
\end{equation}
Let $\bfdelta=[\delta_1,\cdots,\delta_K]$. Let $P_{\bfdelta}$ denote
the above SDP problem when the bounds on the uncertainty set is
$\bfdelta$. Let $P^*({\bfdelta})$ denote the optimal value of the
the problem $P_{\bfdelta}$. Suppose that for some
choice of uncertainty bounds
$\bar{\bfdelta}=\left[\bar{\delta}_1,\cdots,\bar{\delta}_K\right]^T>0$,
the problem $P_{\bar{\bfdelta}}$ is strictly feasible. Define the
set
\begin{equation}
\Omega(\bar{\bfdelta})=\left\{\bfdelta\ \big|\ \delta_k\le\bar{\delta}_k~\textrm{and}~\delta_k<
\sqrt{\frac{\gamma_k}{P^*\left(\bar{\bfdelta}\right)}},~k=1,\cdots,K\right\}.
\end{equation}
Then, according to \cite{song11},  for any vector of uncertainty bounds
$\bfdelta\in\Omega(\bar{\bfdelta})$, the problem $P_{\bfdelta}$ is
feasible. Moreover, its optimal solution
$\{\mathbf{V}^*_k\}_{k=1}^{K}$ satisfies
$\textrm{rank}(\mathbf{V}^*_k)=1,~k=1\cdots,K$, and it must be the
optimal solution of the original problem (\ref{eqRobustQoS}).

Alternative system level objectives and constraints can be
considered to result in different formulations of the robust
resource allocation problem. For example, reference
\cite{shenouda08} considered robust design for both the averaged sum
MSE minimization problem and the worst case sum MSE minimization
problem. Reference \cite{Tajer11} considered the worst case weighted
sum rate maximization problem and min-rate maximization problem. The
authors of \cite{zheng08} considered the robust beamformer design in
a cognitive radio network in which there are additional requirements
that the transmitter's interference to the primary users should be
kept under a prescribed level. However, most of the above cited
works focus on robust design in a single cell network with a single
transmitter. The extensions to the general MIMO IC/IBC/IMAC will be
interesting.


\begin{flushleft}
\begin{sidewaystable}
\caption{Comparisons of resource allocation algorithms}
\scriptsize{
\begin{tabular}{|c |c | c | c |c|c |c | c | c |c|c| }
\hline
{\bf Algorithm} & {\bf Optimality} & {\bf Complexity} &{\bf Convergence}&
{\bf Coordination} & {\bf Message}& {\bf Channel} &{\bf Update}&{\bf Problem}  \\
 &  & {\bf Per Iteration} & {\bf Status}&
{\bf Level} & {\bf Exchange}& {\bf Model} &{\bf Schedule} &{\bf Formulation}  \\
\hline
\hline
APC & Global& $O(K)$&Yes& Distributed& $O(K)$& Scalar IC &Sequential& Min SINR\\
(\cite{Foschini93})& & && & &  &&  \\
\hline
BB & Global&Lower Bounded By &Yes& Centralized& $N/A$& Parallel IC &N/A&Sum Rate \\
(\cite{xu08})& &$O\left((3+B/2)K^3+3K^2/2+C\right)$ && & &  && \\
\hline
Bisection-SOCP& Global& $O\left(K^{3.5}N_t^{3.5}\log(\frac{1}{\epsilon})\right)$&Yes& Centralized& N/A & MISO IC&N/A&Min SINR \\
(\cite{liu11MISO})& & && & &  && \\
\hline
CCA & Local& $O(N_t K^2)$&Yes& Distributed& $O(K^2)$& MISO IC &Sequential&Smooth \\
(\cite{liu11MISO})& & && & &  &&Utility \\
\hline
ICBF & Unknown& $O(T K^2 N^3_t )$&Unknown& Distributed& $O(K^2)$& MISO IC &Sequential&Sum Rate\\
(\cite{venturino10})& & && & &MISO IBC/IMAC  && \\
\hline
ISB & Unknown& $O(BNK)$&Unknown& Distributed& $O(K^2 N)$&  Parallel IC   &Sequential&Sum Rate\\
(\cite{yu06})& & && & & && \\
\hline
GP & Unknown& $O(K^3)$ &Yes &Centralized& N/A & Parallel IC&N/A&Mixed\\
(\cite{Chiang07Geometric})&  & (Scalar IC Case) && &  & Scalar IC &&Sum Rate \\
\hline
NFP & Global& Upper Bounded By &Yes&Centralized& N/A& Parallel IC &N/A&Mixed\\
(\cite{nguyen03})& &$K^t$ && & & Scalar IC && Sum Rate\\
\hline
MDP & Local& $O(KN\log N+K^2 N)$&Yes& Distributed& $O(K^2 N)$& MISO IC  &Sequential&Sum Rate\\
(\cite{Shi:2009})& & (Parallel IC Case)&& & & Parallel IC && \\
\hline
MIMO-DP & Unknown& $O\left(K^3(N_t N^2_r+N^2_t N_r)+K^2N^3_r\right)$&Unknown& Distributed& $O(K^2 N^2_r )$& MIMO IC &Sequential&Sum Rate\\
(\cite{kim08})& & && & &  && \\
\hline
M-IWF & Unknown& $O(TKN\log N+K^2 N)$&Unknown& Distributed& $O(K^2 N)$& Parallel IC&Simultaneous&Sum Rate\\
(\cite{yu07})& & && & &  && \\
\hline
SCALE-Dual & Local& $O(TKN+N K^2)$&Yes& Distributed& $O(K^2 N)$& Parallel IC&Simultaneous&Sum Rate \&\\
(\cite{papand09})& & && & &  && Min Power\\
\hline
SCALE-GP& Local& $O\left(K^4 N^4\log(KN/\epsilon)\right)$&Yes& Centralized& N/A & Parallel IC&N/A&Sum Rate \&\\
(\cite{papand09})& & && & &  && Min Power\\
\hline
WMMSE-MIMO& Local& $O\left(K^2(N_t N^2_r+N^2_t N_r+N^3_t)+KN^3_r\right)$&Yes& Distributed& $O(K^2 N^2_r)$& MIMO IC &Simultaneous& Utility Satisfy \\
(\cite{shi11WMMSE_TSP})& &(MIMO IC Case) && & & MIMO IBC/IMAC && (\ref{eqWMMSEUtilitySeparable})-(\ref{eqWMMSEUtilityDiff}) \\
\hline
WMMSE-Parallel& Local& $O\left(K^2 N^3\right)$&Yes& Distributed& $O(K^2 N^2)$& Parallel IC &Simultaneous& Utility Satisfy \\
(\cite{shi11WMMSE_TSP})& &(MIMO IC Case) && & & Parallel IBC/IMAC && (\ref{eqWMMSEUtilitySeparable})-(\ref{eqWMMSEUtilityDiff}) \\
\hline
\hline
\end{tabular} } \label{tableAlgorithmComparison}
\end{sidewaystable}
\end{flushleft}

To close this section, we summarize the properties of most of the
algorithms discussed in this section in Table
\ref{tableAlgorithmComparison}. These algorithms usually admit
certain forms of decentralized implementation, in which the
computational loads are distributed to different entities in the
network. We emphasize that the per-iteration computational
complexity and the amount of message exchanges are important
characteristics for practical implementation of these distributed
algorithms. Efficient computation ensures real time implementation,
while fewer number of message exchanges per iteration implies less
signaling overhead. In Table \ref{tableAlgorithmComparison}, these
characteristics are listed for each of the algorithms. We note that
the computational complexity and the required message exchanges are
calculated on a per iteration basis, where in one iteration each
user $k\in\mathcal{K}$ completes one update. Also note that in Table
\ref{tableAlgorithmComparison}, the variable $T$ in ICBF, SCALE-Dual
and M-IWF represents the number of {\it inner} iterations needed;
the variable $\epsilon$ in Bisection-SOCP and SCALE-GP represents
the required precision for their respective inner solutions; the
variable $t$ in MAPEL represents the iteration index; the variable
$B$ in BB and ISB represents the maximum number of transmitted bits
allowed for each subchannel;  the variable $C$ in the BB algorithm
represents its computation overhead.

\section{Distributed Resource Allocation in Interference
Channel}\label{secDistributed}

Most of the algorithms introduced in the previous section are either centralized or require certain
level of user coordination. Such coordination may be costly
in infrastructure based networks, and is often infeasible for fully
distributed networks. In this section we discuss fully distributed resource allocation algorithms that
require no user coordination.

\subsection{Game Theoretical Formulations}\label{subGameFormulation}


If users cannot exchange information explicitly, it is no longer possible
to allocate resources using the maximizer of a system wide utility function. Instead, we need to
rely on alternative solution concepts for distributed resource allocation. One such concept
that is particularly useful in our context is the renowned notion of Nash equilibrium (NE) for a noncooperative game; see \cite{basar99} and \cite{osborne94}, and \href{http://academicearth.org/courses/game-theory}{the Yale Open Course} online. In a noncooperative game, there are a number of players, each seeking to maximize its own utility function by choosing a strategy from an individual strategy set. However, the utility of one player depends on not only the strategy of its own, but also those of others in the system. As a result, when players have conflicting utility functions, there is usually no joint player strategy that will simultaneously maximize the utilities of all players. For such a noncooperative game, a NE solution is defined as a tuple of joint player strategies in which
no single player can benefit by changing its own strategy unilaterally.

Mathematically, a $K$-person noncooperative game in the strategic
form is a three tuple $(\mathcal{K},\chi,\mathbf{U})$, in which
$\mathcal{K}=\{1,\cdots,K\}$ is the set of players of the game;
$\chi=\prod_{k=1}^{K}\chi_k$ is the joint strategic space of all the
players, with $\chi_k$ being player $k$'s individual strategy space;
$\mathbf{U}=[{U}_1,\cdots,{U}_K]$, where
$U_k(\mathbf{x}_k,\mathbf{x}_{-k}): \chi\mapsto \mathcal{R}$ is user
$k$'s utility function. In the above definition we have used
$\mathbf{x}_k\in\chi_k$ to denote player $k$'s strategy,
$\mathbf{x}_{-k}=\{\mathbf{x}_l\}_{l\ne k}$ to denote the strategies
of all remaining users. It is clear that player $k$'s strategy
depends on its own strategy $\mathbf{x}_k\in\chi_k$ as well as those
of others $\mathbf{x}_{-k}\in\chi_{-k}$. A NE of the game
$\mathcal{G}$ is defined as the set of joint strategies of all the
players $\mathbf{x}^*\in\chi$ such that the following inequality is
satisfied simultaneously for all players $k\in\mathcal{K}$
\begin{eqnarray}
U_k(\mathbf{x}^*_k,\mathbf{x}^*_{-k})\ge
U_k(\mathbf{x}_k,\mathbf{x}^*_{-k}),~\forall~\mathbf{x}_k\in\chi_k.
\label{eqNE}
\end{eqnarray}
Clearly at a NE, the system is stable as none of the players has any intention to switch to
a different strategy. We define a {\it best
response function} for each player in the game, as its best strategy
when all other players have their strategies fixed
\begin{eqnarray}
BR_k(\mathbf{x}_{-k})=\arg\max_{\mathbf{x}_k\in\chi_k}U_k(\mathbf{x}_k,\mathbf{x}_{-k}).
\end{eqnarray}

Using this definition, a NE of the game $\mathcal{G}$ can be
alternatively defined as
\begin{eqnarray}
\mathbf{x}^*\in BR_k(\mathbf{x}^*_{-k}),~\forall~k=1,\cdots,K.
\label{eqNEBR}
\end{eqnarray}

Fig. \ref{figNE} is an illustration of the NE point of a game with
2-player and affine best response functions. This figure also
shows how a sequence of best response may enable the players to
approach the NE.

\begin{figure*}[htb]
    {\includegraphics[width=
1\linewidth]{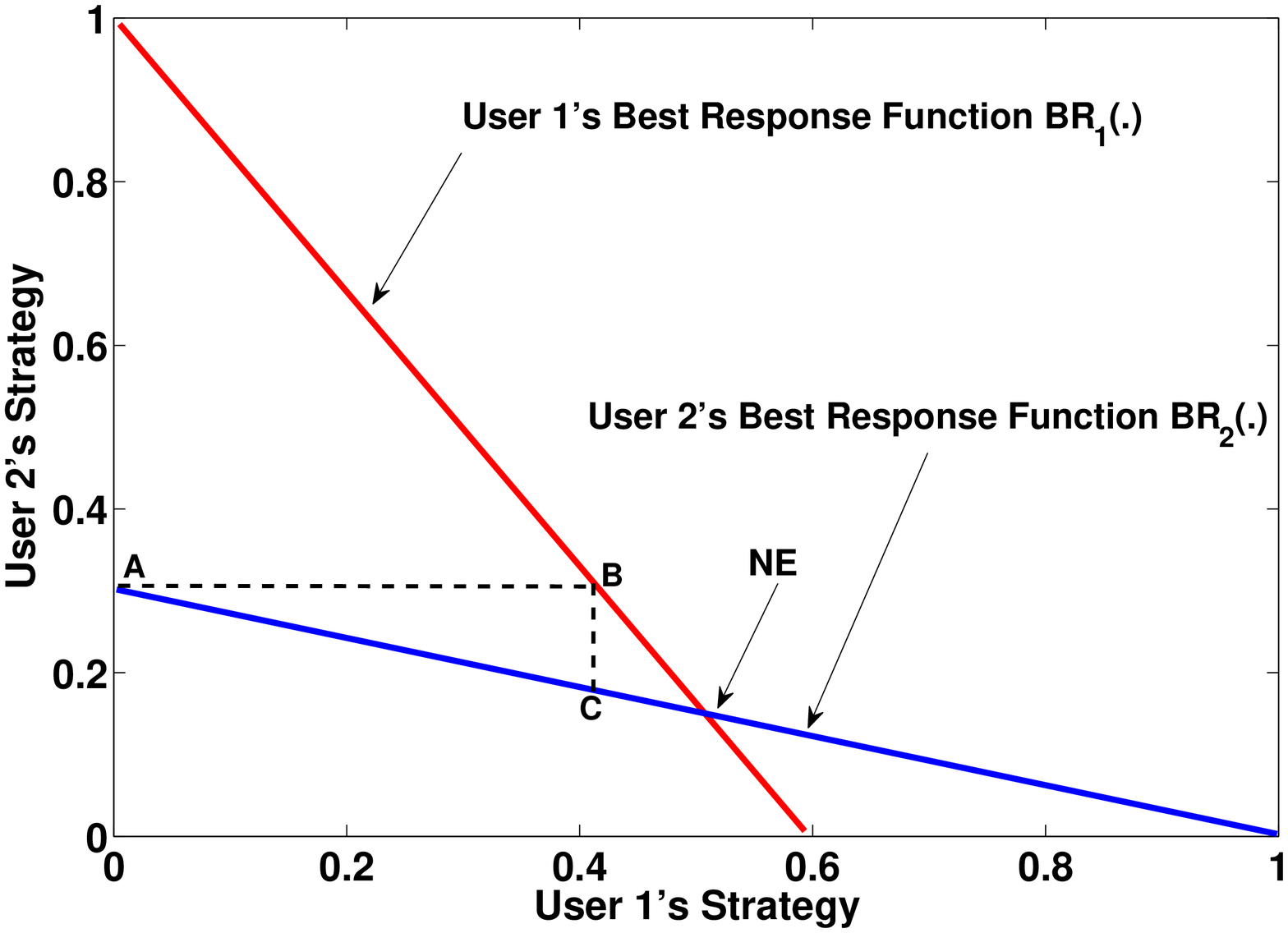} \caption{Illustration of the Nash
Equilibrium (NE) of a 2-user scalar game. This figure also shows the
process of a sequence of best responses that reach the NE. The
function $BR_k(\cdot)$ represents user $k$'s best response function.
Suppose both users initially choose $0$. User $2$ acts first and
chooses point $A$ which is its best response, User $1$
acts next and chooses its best response point $B$.
User $2$ acts again and chooses point $C$. Continuing iteratively in this fashion,
the NE will be reached in the limit.
}\label{figNE} }
    \end{figure*}

Let us illustrate the
notion of NE in our 2-user scalar IC model (\ref{eqTwoUser}).
Suppose these two users are the players of a game, and their
strategy spaces are $\chi_k=\{p_k\mid 0\le p_k\le\bar{p}_k\}$, $k=1,2$. Assume
that the users' utility functions are their maximum transmission
rates defined in (\ref{eqRateTwoUser}). Thus, for this example,
user 1's best response function admits a particular simple
expression
\begin{eqnarray}
BR_1(p_2)&=&\arg\max_{0\le p_1\le
\bar{p}_1}\log_2\left(1+\frac{|H_{11}|^2
p_1}{1+|H_{21}|^2 p_2}\right)\nonumber\\
&=&\bar{p}_1.
\end{eqnarray}
This says that {\it regardless} of user $2$'s transmission strategy,
user $1$ will transmit with full power. The same can be
said about user $2$. Consequently, the only NE of this game is the
transmit power tuple $(\bar{p}_1, \bar{p}_2)$. Obviously, assuming
that each user is indeed selfish and they intend to maximize their own
utility, the NE point $(\bar{p}_1, \bar{p}_2)$ can be
implemented {\it without} any explicit coordination between the
users. Now let us assess the efficiency of such power allocation
scheme in terms of system sum rate. In Fig. \ref{figNEInefficient1} and Fig. \ref{figNEInefficient2},
we plot the rate region boundary and the NE points for different interference levels.
We see that when interference is low, the NE corresponds exactly to the maximum sum rate point.
However, when interference is strong, the NE scheme is inferior
to the time sharing scheme in which the users transmit with full power
in an orthogonal and interference free fashion
(e.g., TDMA or FDMA). Nonetheless, it should be pointed out
that the NE point can be reached without user coordination,
while the time sharing scheme requires the users to synchronize
their transmissions.

We refer the readers to \href{http://ieeexplore.ieee.org/xpl/tocresult.jsp?isnumber=5230827}
{the September 2009 issue of IEEE Signal Processing Magazine} for the applications of game theory
to wireless communication and signal processing.

%
%

       \begin{figure*}[ht]
    \begin{minipage}[t]{0.48\linewidth}
    \centering
     {\includegraphics[width=
1\linewidth]{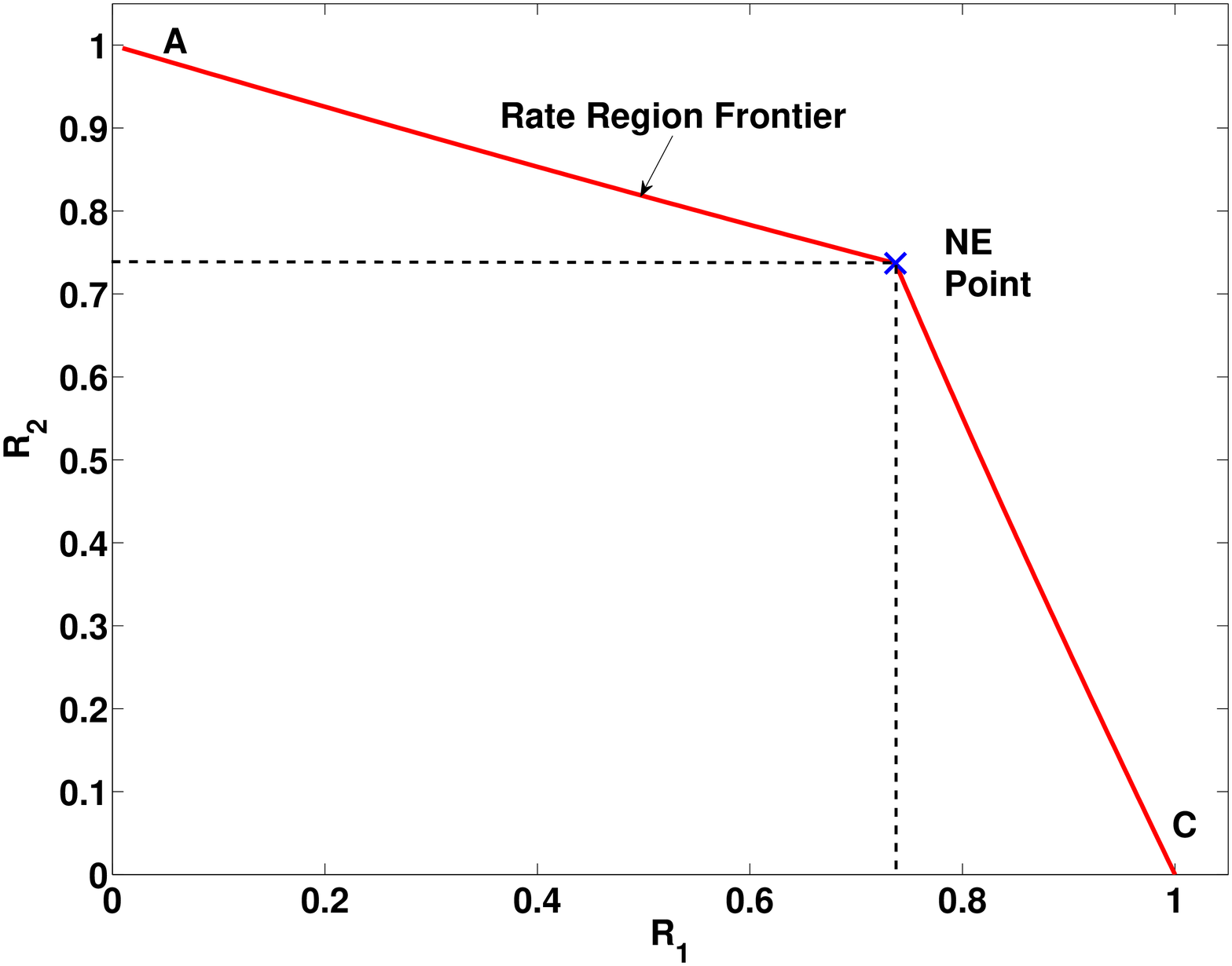} \caption{\small {An
illustration of the efficiency of the NE for 2-user IC when the
interference is weak. $\bar{p}_1=\bar{p}_2=1$,
$|H_{11}|^2=|H_{22}|^2=1,$ $|H_{12}|^2=|H_{21}|^2=0.5$,
$\sigma^2_1=\sigma^2_2=1$. At point $A$ and $C$, a single user
transmits using full power; at the NE point, both users transmit
using full power.}}\label{figNEInefficient1} }
\end{minipage}\hfill
    \begin{minipage}[t]{0.48\linewidth}
    \centering
    {\includegraphics[width=
1\linewidth]{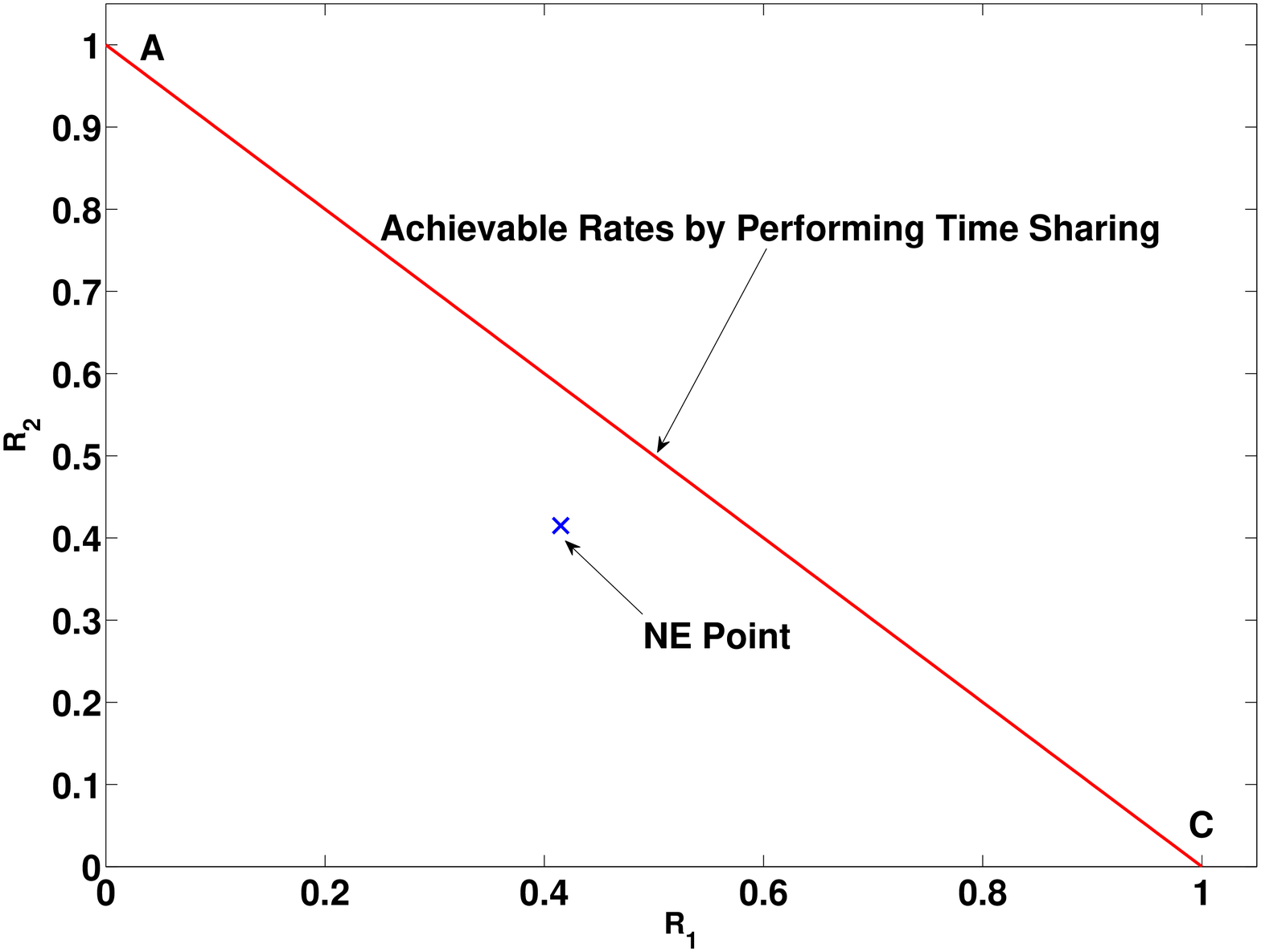} \caption{\small {An
illustration of the inefficiency of the NE for 2-user IC when the
interference is strong. $\bar{p}_1=\bar{p}_2=1$,
$|H_{11}|^2=|H_{22}|^2=1,$ $|H_{12}|^2=|H_{21}|^2=2$,
$\sigma^2_1=\sigma^2_2=1$. At point $A$ and $C$, a single user
transmits using full power; at the NE point, both users transmit
using full power.}}\label{figNEInefficient2} }
\end{minipage}
    \end{figure*}


\subsection{Distributed Resource Allocation for Interference
Channels}\label{subDistributedResourceAllocation}

Early works on distributed physical layer resource allocation in
wireless networks largely deal with the scalar IC models. 
\cite{saraydar01}, \cite{saraydar02} and
\cite{goodman00powercontrol} are among the first to cast the general
scalar power control problem in a game theoretic framework. They
proposed to quantify the tradeoff between the users' QoS
requirements and energy consumption by a utility function in the
form:
\begin{eqnarray}
U_k(p_k,\mathbf{p}_{-k})=\frac{R_k f(\mbox{SINR}_k)}{p_k}
\end{eqnarray}
where $R_k$ is user $k$'s fixed transmission rate, and
$f(\mbox{SINR}_k)$ is a function of user $k$'s SINR that
characterizes its bit error rate (BER). They showed that any NE
point is inefficient in the Pareto sense, i.e., it is possible to
increase the utility of some of the terminals without hurting any
other terminal. To improve the efficiency of the power control game,
they proposed to charge the users with a price that is proportional
to their transmit powers. Specifically, each users' utility function
now becomes $U_k(p_k, \mathbf{p}_{-k})=R_k
f\left(\mbox{SINR}_k\right)/ p_k- \alpha_k p_k$, where $\alpha_k$ is
a positive scalar that can be appropriately chosen by the system
operator. They showed that this modified game always admits a NE,
and proposed an algorithm that allows the users to reach one of the
NEs by adapting their transmit powers in the best response fashion.
\cite{Meshkati06} and \cite{Meshkati07} extended the above works to
the multi-carrier data network. They defined the utility function
for each user as
\begin{eqnarray}
U_k(\mathbf{p}_k,\mathbf{p}_{-k})=\frac{\sum_{n=1}^{N}R^n_k f_k(\mbox{SINR}^n_k)}{\sum_{n=1}^{N}p^n_k}.
\end{eqnarray}
where the function $f_k(\cdot)$ represents the BER of user $k$, and it
incorporates the underlying structure of different linear receivers.
However, such multi-carrier power control game is more complicated
than the scalar power control game introduced early, and in
certain network configurations it is possible that no NE exists (see
\cite{Meshkati06}).

An alternative approach in distributed power control is to directly optimize
the individual users' transmission rates. Consider a $K$-user $N$-channel parallel IC model.
Assume that each user $k\in\mathcal{K}$ is interested in maximizing its transmission rate,
and again assume that its total transmission power budget is $\bar{p}_k$.
Then the users' utility functions as well as their feasible spaces can be expressed as
\begin{eqnarray}
U_k(\mathbf{p}_k,\mathbf{p}_{-k})&=&\sum_{n=1}^{N}\log_2\left(1+\frac{|H^n_{kk}|^2 p^n_k}
{1+\sum_{l\ne k}|H^n_{lk}|^2 p^n_l}\right),~k=1,\cdots,K,\label{eqIWFUtility}\\
\chi_k&=&\left\{\mathbf{p}_k\ \big|\ \sum_{n=1}^{N}p^n_k\le \bar{p}_k, p^n_k\ge 0,~n=1,\cdots,N\right\},~k=1,\cdots,K\label{eqIWFeasibleSpace}.
\end{eqnarray}
Fixing $\mathbf{p}_{-k}$, user $k$'s best response solution $\mathbf{p}_k^*$
is the classical {\it water-filling} (WF) solution
\begin{eqnarray}
p^{n,*}_k=\left[\frac{1}{\lambda_k}-\frac{1+\sum_{l\ne k}|H^n_{lk}|^2 p^n_l}{|H^n_{kk}|^2}\right]^{+},~n=1,\cdots,N\label{eqWFSolution}
\end{eqnarray}
where $\lambda_k\ge 0$ is the dual variable ensuring the sum power
constraint, and the operator $[x]^+=\max\{x,0\}$. Figure
\ref{figIWF} illustrates the WF solution for user $k$. We note that
in order to compute the WF solution, user $k$ needs to know the
terms $\{1+\sum_{l\ne k}|H^n_{lk}|^2 p^n_l\}_{n=1}^{N}$, which is
simply the set of noise-plus-interference (NPI) levels on all its
channels. They can be measured {\it locally} at its receiver.
    \begin{figure*}[htb]
    \centering
    {\includegraphics[width=
0.8\linewidth]{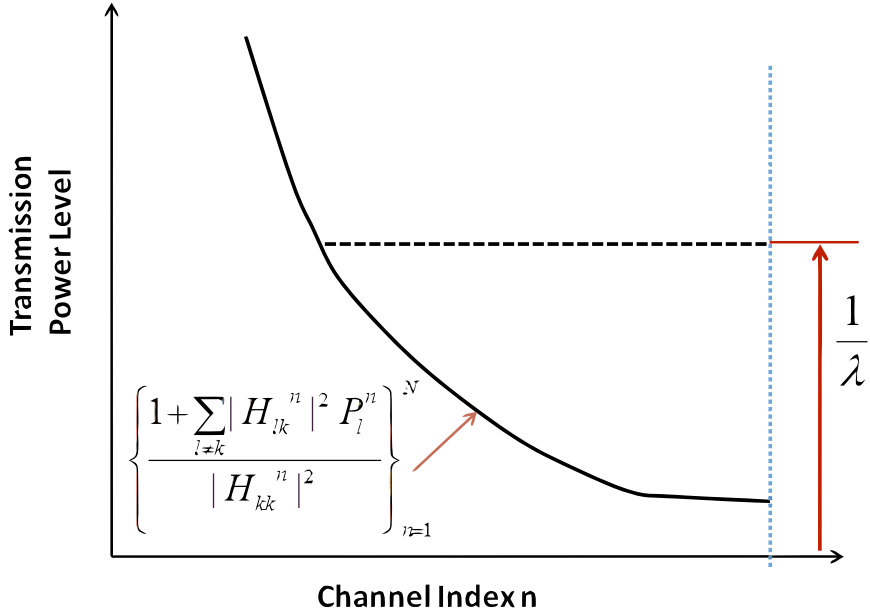} \caption{Illustration of the
Water-Filling computation for user $k$. }\label{figIWF} }
\end{figure*}
We refer to this game as a WF game.

\cite{yu02a} is the first to formulate the distributed power control
problem as a WF game. The authors proposed an iterative
water-filling algorithm (IWFA) in which the following two steps are
performed iteratively:
\begin{center}
\fbox{\begin{minipage}{4.5in}
\begin{description}
\item [1)] each user
$k\in\mathcal{K}$ measures its NPI power
$\{1+\sum_{l\ne k}|H^{n}_{lk}|^2 p^{n,(t)}_l\}_{n=1}^{N}$;
\item [2)] each
user $k\in\mathcal{K}$ computes its power allocation
$\mathbf{p}^{(t+1)}_k$ according to (\ref{eqWFSolution}).
\end{description}
\end{minipage}
}
\end{center}

Variations of the IWFA algorithm allow the users to update using different
schedules. The following three update schemes have been proposed: a) {\it simultaneous update},
in which all the users update their power in each iteration, see \cite{yu02a} and \cite{scutari08b};
b) {\it sequential update}, in which a single user updates in each iteration,
see \cite{luo06b} and \cite{scutari08b}; c) {\it asynchronous
update}, in which a random fraction of users update in each iteration, and they
are allowed to use {\it outdated} information in their computation, see
\cite{scutari08c}. Regardless of the specific update schedule used, the IWFA
algorithm is a distributed algorithm because only local NPI measurements are
needed for the users to perform their independent power update.

The properties of the WF game as well as the convergence conditions
of the IWFA have been extensively studied. The original work has
only provided sufficient conditions for the convergence of the IWFA
in a 2-user network. Subsequent works such as \cite{luo06b},
\cite{shum07}, \cite{Cendrillon07}, \cite{scutari08a} and
\cite{scutari08b} generalized this result to networks with arbitrary
number of users. \cite{luo06b} characterized the NE of the
water-filling game as the solution to the following affine
variational inequality (AVI)
\begin{eqnarray}\label{avi}
(\mathbf{p}^{'}-\mathbf{p})(1+\mathbf{M}\mathbf{p})\ge 0, ~\forall~\mathbf{p}^{'}\in\chi,
\end{eqnarray}
where $\mathbf{M}$ is a block partitioned matrix with its
$(ij)^{th}$ block defined as
$\mathbf{M}^{ij}=\mbox{diag}\left(\frac{|H^1_{ji}|^2}{|H^1_{ii}|^2},\cdots,
\frac{|H^N_{ji}|^2}{|H^N_{ii}|^2}\right)$. Using the AVI
characterization (\ref{avi}), they also showed that the sequential
version of the IWFA corresponds to the classical projection
algorithm whose convergence to the {unique} NE of the water-filling
game is guaranteed if the following contraction condition is
satisfied
\begin{eqnarray}
\rho\left(\left(\mathbf{I}-\bfUpsilon_{low}\right)^{-1}\bfUpsilon_{upp}\right)<1\label{eqConditionSequential}
\end{eqnarray}
where $\bfUpsilon_{low}$ and $\bfUpsilon_{upp}$ is the strictly lower and strictly upper triangular part
of a $K\times K$ matrix $\bfUpsilon$ given by
\begin{eqnarray}
[\bfUpsilon]_{q,r}\triangleq \left\{ \begin{array}{ll}
\max_n\left\{\frac{|H^n_{rq}|^2}{|H^n_{qq}|^2}\right\}&\textrm{if } r\ne q\\
0 &\textrm{otherwise. } \\
\end{array} \right. \label{eqDefUpsilon}
\end{eqnarray}
Also using the AVI characterization  (\ref{avi}), the authors of
\cite{scutari08a} and \cite{scutari08b} further proved that the
condition
\begin{eqnarray}
\rho\left(\bfUpsilon\right)<1\label{eqConditionSimultaneous}
\end{eqnarray}
is sufficient for the convergence of the IWFA as well as the uniqueness of the NE. We refer the readers to
\cite{scutari08d} for a detailed comparison of various conditions for the convergence of
the IWFA. It is worth noticing that all the sufficient conditions for the
convergence of IWFA require that the interference among the users are weak. For example,
a sufficient condition for (\ref{eqConditionSimultaneous}) is
\begin{eqnarray}
\sum_{l\ne k} |H^n_{kl}|^2 s_k
<{|H^n_{ll}|^2}s_l,~\forall~k\in\mathcal{K},~n\in\mathcal{N}
\end{eqnarray}
where $s_k>0, k=1,\cdots,K$ is a set of constant scalars.
Intuitively, this condition says that at the receiver of each user
$k\in\mathcal{K}$, the power of the useful signal should be larger
than the power of total interference. When the interference is
strong, IWFA diverges. \cite{leshem09} provided an example in which
{\it all} forms of  IWFA diverge, regardless of their update
schedules. We remark that extending the IWFA so that it converges in
less stringent conditions that do not require the interference to be
weak is still an open problem. Without any algorithmic
modifications, the standard IWFA is only known to converge
(\cite{luo06b}) when the crosstalk coefficients are symmetric
$$\frac{|H^n_{rq}|^2}{|H^n_{qq}|^2}=\frac{|H^n_{qr}|^2}{|H^n_{rr}|^2},\quad \forall\ r\neq q,\ \forall\ n,
$$
regardless the interference levels.

The IWFA  has been recently generalized  to
MIMO IC model. In the MIMO WF game, the strategy of each user $k\in\mathcal{K}$
is its transmission covariance matrix $\mathbf{Q}_k$. The rate utility function and strategy set
for user $k$ can be expressed as
\begin{eqnarray}
U_k(\mathbf{Q}_k,\mathbf{Q}_{-k})&=&\log_2\det\left(\mathbf{H}_{kk}\mathbf{Q}_{k}\mathbf{H}_{kk}^{H}
\left(\mathbf{I}_{N_r}+ \sum_{l\ne k}
\mathbf{H}_{lk}\mathbf{Q}_{l}\mathbf{H}_{lk}^{H}\right)^{-1}+\mathbf{I}_{N_{r}}\right)\label{eqMIMOWFUtility}\\
\chi_k&=&\left\{\mathbf{Q}_k: \mbox{Tr}(\mathbf{Q}_k)\le \bar{p}_k, \mathbf{Q}_k\succeq 0\right\}\label{eqMIMOWFFeasible}.
\end{eqnarray}
In this case, each user $k$'s best response
$BR_{k}(\mathbf{Q}_{-k})$ is again a water-filling solution, see
\cite{telatar99}. \cite{Arslan07} suggested that in each iteration,
the users' covariance can be updated as
\begin{eqnarray}
\mathbf{Q}^{(t+1)}_k=\alpha_{t}\mathbf{Q}^{(t)}_k+(1-\alpha_t)BR_{k}(\mathbf{Q}^{t}_{-k})\label{eqAveragedIWFA}
\end{eqnarray}
where $\{\alpha_t\}_{t=1}^{\infty}$ is a set of constants that
satisfy $\alpha_t\ge 0$, $\lim_{t\to\infty}\alpha_t=0$ and
$\lim_{t\to\infty}\sum_{t=1}^{T}\alpha_t<\infty$. They claimed that
their algorithm converges when the interference is weak, but no
specific conditions are given. This work has been generalized by
\cite{scutari08d} and \cite{scutari09a}, in which rigorous
conditions for the convergence of the MIMO IWFA  have been derived.
In particular, consider a MIMO network in which $N_t=N_r$ and the
direct link channel matrices $\{\mathbf{H}_{kk}\}_{k=1}^{K}$ are all
nonsingular. Define a $K\times K$ matrix $\mathbf{S}$ as
\begin{eqnarray}
[\mathbf{S}]_{q,r}\triangleq \left\{ \begin{array}{ll}
\rho\left(\mathbf{H}^H_{rq}\mathbf{H}^{-H}_{qq}\mathbf{H}^{-1}_{qq}\mathbf{H}_{rq}\right)&\textrm{if } r\ne q\\
0 &\textrm{otherwise. } \\
\end{array} \right. 
\end{eqnarray}
Then the condition $\rho\left(\mathbf{S}\right)<1$ is sufficient for the convergence of
the sequential/simultaneous/asynchronous MIMO IWFA. This condition is again a weak interference
condition, and future work is needed to extend the MIMO IWFA to
work in networks without this restriction. We refer the readers to web pages of
\href{http://www.comm.utoronto.ca/~weiyu/}{Yu},
\href{http://www.ee.ust.hk/~palomar/Home.html}{Palomar}
and \href{http://ise.illinois.edu/research/faculty/pang.html}{Pang}
for other works related to the WF games and IWFA.

The above parallel and MIMO WF games have been extended in several
directions. A series of recent works considered the robustness issue
in a WF games. For instance, \cite{gohary09} considered the WF game
in the presence of a jammer. Let us denote user $0$ as the jammer
and denote its transmission power as
$\mathbf{p}_0=\left[p^1_0,\cdots,p^N_0\right]^{T}$. The rate utility
function of a normal user $k$ ($k\ne 0$) becomes
\begin{eqnarray}
U_k(\mathbf{p}_k,\mathbf{p}_{-k},\mathbf{p}_0)=\sum_{n=1}^{N}\log_2\left(1+\frac{|H^n_{kk}|^2 p^n_k}
{1+\sum_{l\ne k}|H^n_{lk}|^2 p^n_l+|H^n_{0k}|^2 p^n_0}\right).
\end{eqnarray}
Suppose the jammer's objective is to minimize the utility of the whole system. This can be reflected by
its utility function and the strategy set
\begin{eqnarray}
U_0(\mathbf{p}_0,\mathbf{p})&=&-\sum_{k=1}^{K}U_k(\mathbf{p}_k,\mathbf{p}_{-k},\mathbf{p}_0)\label{eqJammerUtility}\\
\chi_0&=&\left\{\mathbf{p}_0: \sum_{n=1}^{N}p^n_0\le \bar{p}_0, p^n_0\ge 0, \forall~n\in\mathcal{N}\right\}.
\end{eqnarray}
\cite{gohary09} proposed a generalized IWFA (GIWFA) algorithm in
which the normal users and the jammer all selfishly maximize their
respective utility functions. Notice that the selfish maximization
problems are all convex. The convergence condition for the GIWFA is
\begin{eqnarray}
\rho\left(\left(\mathbf{I}-\bfUpsilon_{low}\right)^{-1}\bfUpsilon_{upp}\right)\le \frac{1}{1+a}-c<1\label{eqConditionGIWFA}
\end{eqnarray}
where the matrix $\bfUpsilon$ is defined in (\ref{eqDefUpsilon}),
and $a>0$ and $c>0$ are constants related to the system parameters.
Clearly this condition is more restrictive than those of the original IWFA, for example the condition
(\ref{eqConditionSequential}). This
is partly because the presence of the jammer introduces {\it uncertainty} to
the NPI that each normal user experiences.

Uncertainty of the NPI is also caused by events such as sudden
changes in the number of users in the system or errors of
interference measurement at the receivers. \cite{setoodeh09} seek a
formulation that takes into consideration the worst case NPI errors.
Let ${I}^n_k$ denote the power of NPI that user $k$ should have
experienced on channel $n$ if no measurement errors occur. Let
$\tilde{{I}}^{n}_k={I}^n_k-\Delta{I}^{n}_k$ be the measured NPI
value, with $\Delta{I}^{n}_k$ representing the NPI uncertainty. Let
$\Delta\mathbf{I}_k= \left[\Delta{I}^{n}_1,
\cdots,\Delta{I}^{n}_K\right]^{T}$ and suppose it is bounded, i.e.,
$\|\Delta\mathbf{I}_k\|\le \epsilon_k$ for some $\epsilon_k>0$. In
this robust WF game, the objectives of the users are to maximize
their {\it worst case} transmission rate. In other words, user $k$'s
utility function can be expressed as
\begin{eqnarray}
U_k(\mathbf{p}_k,\mathbf{p}_{-k})=\min_{\|\Delta\mathbf{I}_k\|\le \epsilon_k}\sum_{n=1}^{N}
\log_2\left(1+\frac{|H^n_{kk}|^2 p^n_k}{\tilde{{I}}^{n}_k+\Delta{I}^{n}_k}\right).
\end{eqnarray}
This formulation trades performance in favor of robustness, thus the
equilibrium solution obtained is generally less efficient than that
of the original IWFA. In \cite{hong11a}, an averaged version of IWFA
was proposed which converges when the error of the NPI
$\Delta\mathbf{I}_k$ satisfies certain conditions. In
\cite{gohary09b}, the authors provided a probabilistically robust
IWFA to deal with the quantization errors of the NPI at the receiver
of each user. In this algorithm, users allocate their powers to
maximize their total rate for a large fraction of the error
realization.

Another thread of works  such as \cite{xie10}, \cite{wu08},
\cite{hong11b}, \cite{wang11robust} and \cite{pang09} generalized
the original WF game and the IWFA to interfering cognitive radio
networks (CRN). In a CRN, the secondary users (SUs) are allowed to
use the spectrum that is assigned to the primary users (PUs) as long
as the SUs do not create excessive interference to the primary
network. Suppose the secondary network is a $K$-user $N$-channel
parallel IC. Let  $\mathcal{Q}=\{1,\cdots,Q\}$ denote the set of PUs
in the network. Let $|G^n_{kq}|^2$ denote the channel gain from SU
$k$ to PU $q$ on channel $n$. The following aggregated interference
constraints are imposed on the secondary network (these constraints
are also referred to as the interference temperature-constraints,
see \cite{fcc03b} and \cite{zhang10})
\begin{eqnarray}
\sum_{k=1}^{K}|G^n_{kq}|^2 p_k\le
\bar{I}^{n}_{q},~\forall~(q,n)\in\mathcal{Q}\times\mathcal{N}\label{eqInterferenceConstraints}
\end{eqnarray}
where $\bar{I}^{n}_q$ represents the maximum aggregated interference
allowed at the receiver of PU $q$ on channel $n$. The original WF
algorithm needs to be properly modified to strictly enforce these
interference constraints in the equilibrium. \cite{xie10} formulated
the power allocation problem in this CRN as a competitive market
model. In this model, each channel has a fictitious price per unit
power, and the users must purchase the transmission power on each
channel to maximize their data rates. \cite{scutari10} and
\cite{scutari09a} systematically studied the WF game with
interference constraints. For each primary user $q$, they introduced
a set of interference prices
$\bfnu_q=\left[\nu_q^1,\cdots,\nu_q^N\right]^{T}$. Each SU is
charged for their contribution of total interference at
PUs' receiver. 
Specifically, a SU $k$'s utility function and
feasible set is defined as
\begin{eqnarray}
&&U_k(\mathbf{p}_k,\mathbf{p}_{-k},\bfnu)=R_k(\mathbf{p}_k,\mathbf{p}_{-k})-
\sum_{q=1}^{Q}\sum_{n=1}^{N}\nu^n_q|G^n_{kq}|^2 p^n_k,\\
&&\chi_k=\left\{\mathbf{p}_k:\sum_{n=1}^{N}p^n_k\le \bar{p}_k, p^{n}_{k}\ge 0,~n=1,\cdots,N\right\}
\end{eqnarray}
where $R_k(\mathbf{p}_k,\mathbf{p}_{-k})$ is user $k$'s transmission
rate. The NE of this interference-constrained WF game is the tuple
$(\bfnu^*,\mathbf{p}^*)$ that satisfies the following conditions
\begin{eqnarray}
&&\mathbf{p}_k^*=\max_{\mathbf{p}_k\in\chi_k}U_k(\mathbf{p}_k,\mathbf{p}^*_{-k},\bfnu^*),\nonumber\\
&&\sum_{k=1}^K |G^n_{kq}|^2 p^{n,*}_k \le \bar{I}_q^n,~(q,n)
\in\mathcal{Q}\times\mathcal{N}\nonumber\\
&&\nu^{n,*}_q(\bar{I}_q^n-\sum_{k=1}^K |G^n_{kq}|^2 p^{n,*}_k)=0, \
\nu^{n,*}_q\ge 0, \ ~(q,n) \in\mathcal{Q}\times\mathcal{N}.
\nonumber
\end{eqnarray}
\cite{scutari10} and \cite{scutari09a} derived the conditions for
the existence and uniqueness of the NE for this game. Introduce a
$N\times N$ matrix $\widehat{\bfUpsilon}$ {\small
\begin{eqnarray}
[\widehat{\bfUpsilon}]_{l,k}\triangleq \left\{ \begin{array}{ll}
-\max_{n\in\mathcal{N}}\left\{\frac{|H^n_{lk}|^2}{|H^n_{kk}|^2}\times \widehat{\mbox{innr}}^n_{lk}\right\} &\textrm{if } k\ne l\\
1  &\textrm{otherwise } \\
\end{array} \right.
\end{eqnarray}}
where $ \widehat{\mbox{innr}}^n_{lk}\triangleq
1+\sum_{m\in\mathcal{K}}|H^n_{km}|^2 \bar{p}_k\nonumber$. Then the
condition $\widehat{\bfUpsilon}\succeq 0$ guarantees the uniqueness
of the NE. A set of distributed algorithms that alternately update
the users' power allocation and the interference prices were
proposed to reach the NE of this game. The MIMO generalization has
been considered in \cite{Scutari10MIMOCRN}, whereby both the SUs and
PUs are equipped with multiple antennas. Another extension
\cite{wang11robust} considered the possibility that the SU-PU
channels may be uncertain, and formulated a robust WF game that
ensures the SU-PU interference constraints are met even in the worst
case channel conditions.

All the above mentioned WF games can be categorized as {\it rate
adaptive} (RA) games, in which the users selfishly maximize their
own data rates. One drawback of the RA formulation is that
individual users have no QoS guarantees. Alternatively, a {\it fixed
margin} (FM) formulation allows each user to minimize its
transmission power while maintaining its QoS constraint. The FM
formulation is more difficult to analyze due to the {\it coupling}
of the users' strategy spaces resulted from the QoS constraint. For
the parallel IC model, the utility function and the strategy set for
user $k$ in a FM formulation can be expressed as
\begin{eqnarray}
U_k(\mathbf{p}_k)&=&-\sum_{n=1}^{N}p^n_k\\
\chi_k(\mathbf{p}_{-k})&=&\left\{\mathbf{p}_k: R_k(\mathbf{p}_k,\mathbf{p}_{-k})\ge \zeta_k, p^n_k\ge 0,
n\in\mathcal{N}\right\}
\end{eqnarray}
where $\zeta_k$ is the rate target for user $k$. The solution to the individual
users' utility maximization problems, assuming the feasibility of the rate targets, is again a water-filling solution
\begin{eqnarray}
p^{n,*}_k=\left[{\lambda_k}-\frac{1+\sum_{l\ne k}|H^n_{lk}|^2 p^n_l}{|H^n_{kk}|^2}\right]^{+},
~n=1,\cdots,N\label{eqWFSolutionFM}
\end{eqnarray}
where $\lambda_k$ is the water-level that is associated with user $k$'s rate constraint.

A NE of this FM game (which is usually referred to
as the {\it generalized NE} due to the coupling of the users' strategy spaces) is defined as a power
vector $\mathbf{p}^*$ that satisfies
\begin{eqnarray}
\mathbf{p}^*_k\in\arg\max_{\mathbf{p}_k\in\chi_k(\mathbf{p}^*_{-k})}
U_k\left(\mathbf{p}_k,\mathbf{p}^*_{-k}\right),~k=1,\cdots,K.
\end{eqnarray}
Similar to the min-power QoS constrained formulation discussed in the previous section,
the first thing we need to characterize for this FM game is the feasibility of a given
set of rate targets. The following condition is among many of those that
have been derived in \cite{pang08} which guarantee the existence of a bounded
power allocation achieving the given set of rate targets
\begin{eqnarray}
\sum_{l\ne k}\frac{|H^n_{kl}|^2}{|H^n_{kk}|^2}<\frac{1}{\exp\{\zeta_k\}-1},~(n,k)\in\mathcal{N}\times\mathcal{K}.
\end{eqnarray}
This condition is again a weak interference condition. The following condition is sufficient
for the uniqueness of the (generalized) NE of the FM game
\begin{eqnarray}
\sum_{l\ne k}\max_{n\in\mathcal{N}}\left\{\frac{|H^n_{kl}|^2}{|H^n_{kk}|^2}\right\}
<\frac{\beta}{\exp\{\zeta_l\}-1},~k\in\mathcal{K}
\end{eqnarray}
where $\beta<1$ is related to the set of given rate targets. This
condition is also sufficient for the convergence of a FM-IWFA in
which the users sequentially or simultaneously update their power
using the WF solution. Algorithmic extension of this work to the CRN
with interference constraints of the form
(\ref{eqInterferenceConstraints}) has been considered recently in
\cite{wu09}. It remains to see how the FM games and their
theoretical properties (e.g., uniqueness of the NE, convergence of
the FM-IWFA) can be extended to MIMO CRNs with interference
constraints.

We remark that the NE points of the various RA based WF games
introduced in this section is generally {\it inefficient}, in the
sense that the sum rate of the users is often smaller compared with
that of the socially optimal solutions\footnote{However, note that
in a MAC channel, which is a special case of the IC, the NEs are
indeed efficient. See \cite{yu04}. In this case, the sequential
version of the IWFA converge to a joint strategy that maximizes the
system sum rate.}. In Fig. \ref{figIWFSCALE}, we illustrate such
inefficiency of the NE in a parallel IC with $K=2$, $N=32$ and
randomly generated channel coefficients. We plot the NE point of the
WF game as well as the rate region boundary achieved by the SCALE
algorithm. In order to improve the efficiency of the NE, user
coordination must be incorporated into the original WF game. The
pricing algorithms such as MDP,  M-IWF or WMMSE introduced in the
previous section are examples of such extensions. In those
algorithms, system efficiency is improved due to explicit message
exchange and cooperation among the users. Careful analysis is needed
to identify the tradeoffs between the improvement of the system sum
rate and the signalling overhead. Evidently, when the total number
of users in the system is large, a complete cooperation of all users
is too costly. An interesting problem is to decide how to partition
the users into collaborative groups in a way that strike an optimal
tradeoff between system performance and coordination overhead.

\begin{figure*}[htb]
    {\includegraphics[width=
1\linewidth]{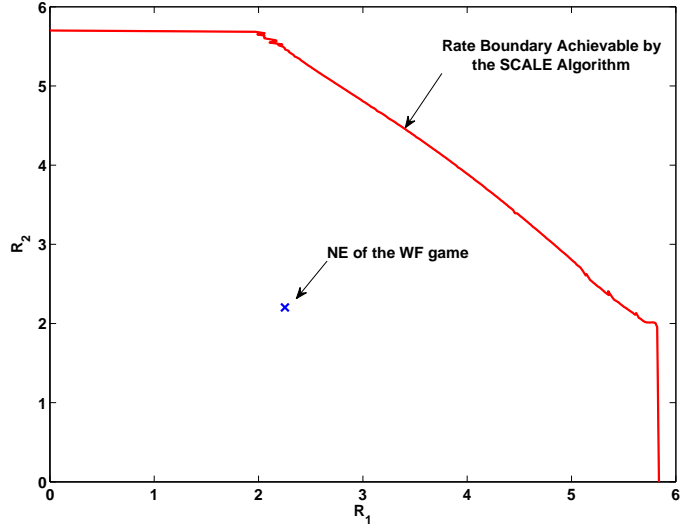} \caption{Illustration of the inefficiency
of the NE point for a WF game with $K=2$, $N=32$ and $\bar{p}_1=\bar{p}_2=1$. }\label{figIWFSCALE}
}
\end{figure*}

%

\section{Resource Allocation via Interference Alignment}

Theoretically, the optimal resource allocation for a MIMO interference channel is related to the characterization of the capacity region of an interference channel, i.e., determining the set of {\color{black} rate tuples} that can be achieved by the users simultaneously.
In spite of intensive research on this subject over the past three decades, 
the capacity region of interference channels is still unknown (even for small number of users).
The lack of progress to characterize the capacity region of the MIMO interference channel has motivated researchers to derive various approximations of the capacity region. For example, the maximum total degrees of freedom (DoF) corresponds to the first order approximation of sum-rate capacity {\color{black}in the high SNR regime}.
Specifically, in a $K$-user interference channel, we define the degrees of freedom region as the following \cite{Jafar1}:
\begin{equation}\label{DoF}
\begin{array}{ll}\mathcal{D} = \bigg\{&\!\!\displaystyle (d_1,d_2,\ldots,d_K) \in \mathbb{R}_+^K\ \; \bigg|\ \; \forall  (w_1,w_2,\ldots,w_K) \in \mathbb{R}_+^K, \; \\
&\displaystyle \sum_{k=1}^K w_k d_k \leq  \limsup_{{\rm SNR} \rightarrow \infty} \left[\sup_{\mathbf{R} \in \mathcal{C}} \frac{1}{\log{\rm SNR}} \sum_{k=1}^K w_kR_k\right]\ \ \bigg\}, 
\end{array}
\end{equation}
where $\mathcal{C}$ is the capacity region and $R_k$ is the rate of user~$k$.
The total DoF in the system can be defined as the following:
\begin{equation}
\eta = \max_{(d_1,d_2, \ldots,d_K) \in \mathcal{D}} d_1 + d_2 + \ldots + d_K. \nonumber
\end{equation}
Roughly speaking, the total DoF is the number of independent data streams that can be communicated
in the channel without interference. 

For various channel models, the DoF region or the total DoF have been characterized recently. In particular,
for a point-to-point MIMO channel with $M$ antennas at the transmitter and $N$~antennas at the receiver, the total DoF is $\eta = \min \{M,N\}$. Different approaches such as SVD precoder or V-BLAST can be used to achieve this DoF bound. For a 2-user MIMO fading interference channel with user $k$  equipped with $M_k$ transmit antennas and $N_k$ receive antennas ($k=1,2$), \cite{JafarFakhK2} proves that the maximum total DoF is 
\begin{equation}
\eta = \min \left\{M_1 + M_2, N_1 + N_2 , \max\{M_1,N_2\}, \max\{M_2,N_1\}\right\}. \nonumber
\end{equation}
Therefore, for the case of $M_1 = M_2 = N_1 = N_2$, the total DoF in
the system is the same as the single user case. In other words, we
do not gain more DoF by increasing the number of users from one to
two. Interestingly, if {\color{black} generic} channel extensions
(drawn from a continuous probability distribution) are allowed
either across time or frequency, \cite{Jafar1} showed that the total
DoF is $\eta = {KM}/{2}$ for a $K$-user MIMO interference channel,
where $M$ is the number of transmit/receive antennas per user.
{\color{black}This  surprising result implies that each user can
effectively utilize half of the total system resources in an
interference-free manner by {aligning the interference at all
receivers}}\footnote{\color{black}The idea of interference alignment
was introduced in \cite{waterloo,JafarXfirst, Birk} and the
terminology ``interference alignment" was first used in
\cite{IATerminology}.}. Moreover, this total DoF can be achieved by
using a carefully designed linear beamforming strategy.

Mathematically, a linear beamforming strategy for a $K$-user MIMO IC can be described by the transmit beamforming matrices $\{\mathbf{V}_k\}_{k\in{\cal K}}$ and the receive beamforming matrices $\{\mathbf{U}_k\}_{k\in{\cal K}}$. The receiver~$k$ estimates the transmitted data vector $\mathbf{s}_k$ as follows 
\begin{equation}
 \mathbf{x}_k = \mathbf{V}_k \; \mathbf{s}_k, \;\;\;\;\;\;\;
\hat{\mathbf{s}}_k = \mathbf{U}_k^H \mathbf{y}_k\nonumber
\end{equation}
where the power of the data vector $\mathbf{s}_k \in \mathcal{R}^{d_k \times 1}$ is normalized such that $E[ \mathbf{s}_k \mathbf{s}_k^H ] = \mathbf{I}$, and $\hat{\mathbf{s}}_k$ is the estimate of $\mathbf{s}_k$ at the $k$-th receiver. The matrices $\mathbf{V}_k \in \mathcal{C}^{M_k \times d_k}$ and $\mathbf{U}_k \in \mathcal{C}^{N_k \times d_k}$ are the beamforming matrices at the $k$-th transmitter and receiver respectively, where $M_k$ ($N_k$) is the number of antennas at transmitter $k$ (respectively receiver $k$).  
Without channel extension, the linear interference alignment conditions can be described by the following zero-forcing conditions \cite{Jafarbound,IANPhard}
\begin{eqnarray}
&&\mathbf{U}_k^H\mathbf{H}_{jk}\mathbf{V}_j=0,\;\;\;\;\  k=1,\cdots,K, \  \forall\ j\neq k, \label{IA1}\\
&&\mbox{rank}\left(\mathbf{U}_k^H
\mathbf{H}_{kk}\mathbf{V}_k\right)=d_k,\quad \ k=1,\cdots,K.
\label{IA2}
\end{eqnarray}
The first equation guarantees that all the interfering signals
at receiver $k$ lie in the subspace orthogonal to $\mathbf{U}_k$, while
the second one assures that the signal subspace $\mathbf{H}_{kk}
\mathbf{V}_k$ has dimension $d_k$ and is linearly independent of
the interference subspace. Clearly, as the number of users $K$ increases, the number of constraints on the beamformers $\{\bU_k,\bV_k\}$ increases quadratically in $K$, while the number of design variables in $\{\bU_k,\bV_k\}$ only increases linearly. This suggests the above interference alignment can not have a solution unless $K$ or $d_k$ is small.

If the interference alignment conditions (\ref{IA1}) and (\ref{IA2}) hold for some linear beamforming matrices
$\{\mathbf{V}_k,\mathbf{U}_k\}_{k\in\mathcal{K}}$, then transmitter $k$ can use $\bV_k$ to send $d_k$ independent data streams to receiver $k$ (per channel use) without any interference. Thus, $d_k$ represents the DoF achieved by the $k$-th transmitter/receiver pair in the information theoretic sense of (\ref{DoF}). In other words, the vector $(d_1,d_2,...,d_K)$ in (\ref{IA1}) and (\ref{IA2}) represents the tuple of DoF achieved by linear interference alignment. Intuitively, the larger the values of $d_1,d_2$,...,$d_K$, the more difficult it is to satisfy the interference alignment conditions (\ref{IA1}) and (\ref{IA2}).

In principle, we can allocate resources by maximizing the total achievable DoF. In particular, for a specific channel realization $\{\mathbf{H}_{kj}\}_{k,j\in\mathcal{K}}$, we need to find the beamforming matrices $\{\bV_k,\bU_k\}$ to maximize the total DoF while satisfying (\ref{IA1})--(\ref{IA2}).
\[
\begin{array}{ll}\displaystyle
\max_{\{\mathbf{U}_k,\mathbf{V}_k\}_{k=1}^K} \;\;\ &\displaystyle\sum_{k=1}^K{d_k}\nonumber\\
\mbox{subject to}
&\displaystyle\mathbf{U}_k^H\mathbf{H}_{kj}\mathbf{V}_j =
0\;\;\;\;\;\;k=1,..,K,
 \;\;j\neq k \nonumber\\
&\mbox{rank}\left(\mathbf{U}_k^H\mathbf{H}_{kk}\mathbf{V}_k\right)=d_k\;\;\;\;\;\;k=1,..,K
\end{array}
\]
Unfortunately, according to \cite{IANPhard}, this problem is NP-hard. So we are led to find suboptimal solution for this problem. However, no efficient algorithms have been developed to approximately solve this problem at this point.

Instead of maximizing the total DoF, we can focus on a seemingly simpler problem: for a given channel realization $\{\mathbf{H}_{kj}\}_{k,j\in\mathcal{K}}$ and a fixed DoF tuple $\mathbf{d}=(d_1,...,d_K)$, check if there exist linear beamformers $\{\bV_k,\bU_k\}_{k\in\mathcal{K}}$ satisfying the alignment conditions (\ref{IA1})--(\ref{IA2}). Notice that the conditions (\ref{IA1})--(\ref{IA2}) are quadratic polynomial equations, which are difficult to solve in general. However, if we fix either $\{\bU_k\}_{k\in\mathcal{K}}$
or $\{\bV_k\}_{k\in\mathcal{K}}$, the quadratic equations become linear and can be solved via the linear least squares. This suggests the following alternating directions method for solving (\ref{IA1})--(\ref{IA2}) (for a fixed $\mathbf{d}$):
 \begin{center}
\fbox{\begin{minipage}{4.5in}
\begin{description}
\item [1)]
Fix the transmit beamformers $\{\bV_k\}_{k\in\mathcal{K}}$. Each receiver $k$ solves the following optimization problem
{\color{black}\begin{equation}\label{subproblem}
\begin{array}{ll}
\min &\displaystyle \mbox{Tr} \left(\mathbf{U}_k^H \mathbf{Q}_k \mathbf{U}_k\right)\nonumber\\
 \mbox{s.t.} &\displaystyle \mathbf{U}_k^H \mathbf{U}_k = \mathbf{I}_{d_k} \nonumber
\end{array}
\end{equation}}
where $\mathbf{Q}_k = \sum_{j \neq k} \frac{p_j}{d_j} \mathbf{H}_{kj}\mathbf{V}_j \mathbf{V}_j^H \mathbf{H}_{kj}^H$, with $I_k = \mbox{Tr} \left(\mathbf{U}_k^H \mathbf{Q}_k \mathbf{U}_k\right)$ being the total received interference power, and ${\bar p}_j$ being the power budget of $j$-th transmitter.
\item [2)] Fix $\{\bU_k\}_{k\in\mathcal{K}}$ and update the transmit beamformers $\{\bV_k\}_{k\in\mathcal{K}}$ in a symmetric fashion as in step 1) (by exchanging the roles of transmitter and receiver, and replacing the channel matrices $\{\mathbf{H}_{k,j}\}_{k,j\in\mathcal{K}}$ by $\{\mathbf{H}^H_{kj}\}_{kj\in\mathcal{K}}$)
\item [3)] Repeat steps 1) and 2) until convergence.
\end{description}
\end{minipage}
}
\end{center}
Notice that the optimal solution $\mathbf{U}_k^*$ for
(\ref{subproblem}) is given by the eigen-vectors of  $\mathbf{Q}_k$
corresponding to the $d_k$-smallest eigen-values. The above
algorithm is proposed first in \cite{Jafar2} and later in
\cite{Heath09}, albeit from a different perspective. Obviously, this
algorithm cannot converge if the DoF vector $\mathbf{d}$ is not
achievable. However, even if $\mathbf{d}$ is achievable, there has
been no formal analysis that shows this alternating direction
algorithm indeed will converge.

The lack of formal convergence proof may not be surprising. In fact, according to \cite{IANPhard}, even checking the feasibility of
(\ref{IA1})--(\ref{IA2}) is NP-hard when each transmitter/receiver is equipped with at least 3
antennas. Hence, for a given channel realization, assigning DoFs to the users in a manner that ensures feasibility is not easy. However, when the number of antennas at each transmitter/receiver is at most 2, the problem of checking feasibility is polynomial time solvable (\cite{IANPhard}).

Now let us turn our attention to the generic solvability of the
interference alignment problem (\ref{IA1})-(\ref{IA2}). In other
words, we focus on the existence  of a beamforming solution to the
quadratic polynomial equations (\ref{IA1})-(\ref{IA2}) when the
channel matrices are randomly generated. To this end, it is natural
to count the number of scalar equations and the number of scalar
variables in the conditions (\ref{IA1})-(\ref{IA2}). It is tempting
to conjecture that there is an interference alignment solution if
and only if the number of constraints is no larger than the number
of variables (see \cite{Jafarbound}). Recently, \cite{RLL} and
\cite{Berkeley-IA} have settled this conjecture completely in one
direction, and partially in the other direction. They derive a
general condition, described below, that must be satisfied by any
DoF tuple $(d_1,d_2,...,d_K)$ achievable through linear interference
alignment.

Let us denote the polynomial equations in (\ref{IA2}) by the index set
$$\mathcal{J} \triangleq \{(k,j)\mid 1\le k\neq j\le K\}.$$
The following result (\cite{RLL} and \cite{Berkeley-IA}) provides an upper bound on the total achievable DoF when no channel extension is allowed.
Consider a $K$-user flat fading MIMO interference channel where the channel matrices $\{\mathbf{H}_{ij}\}_{i,j=1}^K$ are generic ({\color{black}e.g., drawn from a continuous probability distribution}). Assume no channel extension is allowed. Then any tuple of degrees of freedom $(d_1,d_2,...,d_K)$ that is achievable through linear interference alignment (\ref{IA1}) and (\ref{IA2}) must satisfy the following inequalities
\begin{eqnarray}
 &&\min \{M_k, N_k\} \geq d_k  , \quad \forall\; k, \label{I1} \\
 &&\max \{M_k,N_j\} \geq d_k+d_j ,  \quad \forall \;k,j, k\neq j, \label{I2}\\
 &&\sum_{k: (k,j)\in \mathcal{I}} (M_k - d_k)d_k + \sum_{j: (k,j)\in \mathcal{I}} (N_j - d_j)d_j \ge \sum_{(k,j)\in\mathcal{I}} d_k d_j,\quad \forall \; \mathcal{I}  \subseteq \mathcal{J}.\label{IAbound}
\end{eqnarray}

Roughly, the left hand side of (\ref{IAbound}) is equal to the number of independent scalar variables in (\ref{IA1})-(\ref{IA2}) and the right hand side of (\ref{IAbound}) corresponds to the number of constraints in (\ref{IA1}). Thus, the necessity of
condition~(\ref{IAbound}) for the existence of a feasible alignment scheme can be understood by counting the dimensions. However, a formal proof of this condition requires the use of field extension theory (\cite{RLL}). We remark that condition~(\ref{IAbound}) can be used to bound the total DoF achievable in a MIMO interference channel. In particular, the following upper bounds follow directly from condition~(\ref{IAbound}).
\begin{enumerate}
\item [(a)] In the case of $d_k=d$ for all $k$, interference alignment is impossible unless
\[
d\le\frac{1}{K(K+1)}\sum_{k=1}^K(M_k+N_k).
\]
\item [(b)] In the case of $M_k+N_k=M+N$, interference alignment requires
\[
\left(\sum_{k=1}^Kd_k\right)^2+\sum_{k=1}^Kd_k^2\le (M+N)\sum_{k=1}^Kd_k
\]
which further implies
\[
\sum_{k=1}^Kd_k < (M+N).
\]
\end{enumerate}


The principal assumption enabling the surprising result of \cite{Jafar1} is that the channel extensions are exponentially long in $K^2$ and are {\color{black} generic} (e.g., drawn from a continuous probability distribution). If no channel extensions are allowed, part (b) above shows that the total achievable DoF in a MIMO interference channel is bounded by a constant $M+N-1$, regardless of how many users are present in the system. While this bound is an improvement over the single user case which has a maximum DoF of $\min\{M,N\}$, it is significantly weaker than the maximum achievable total DoF of $K/2$ for a diagonal frequency selective (or time varying) interference channel with independent channel extensions. The latter grows linearly with the number of users in the system \cite{Jafar1}.

If channel extensions are restricted to have a polynomial length or are not {\color{black} generic}, the total DoF for a MIMO interference channel is still largely unknown even for the Single-Input-Single-Output (SISO) interference channel. This is an interesting open problem. For the 3-user special case, reference \cite{BT} provided a characterization of the total achievable DoF as a function of the diversity.

Conversely, if  all users have the same DoF $d$ and the number of antennas $M_k$, $N_k$ are divisible by $d$ for each $k$, then condition (\ref{IAbound}) for each subsystem of (\ref{IA1})-(\ref{IA2}) is also sufficient for the feasibility of interference alignment for {\color{black} generic choice of channel coefficients (e.g., drawn from a continuous probability distribution)}. If in addition, $M_k=M$ and $N_k=N$ for all $k$ and $M, N$ are divisible by $d$, then these results imply that interference alignment is achievable if and only if $(M+N)\ge d(K+1)$. Moreover, the reference \cite{Berkeley-IA} considered the symmetric case with $M_k = N_k = M$, $d_k = d$ for all $k$, and
proved that the feasibility of interference alignment in this case is equivalent to $2M \ge d(K + 1)$, regardless of the divisibility of $M$ by $d$. When $K$ is odd and $2M =d(K + 1)$, then $d$ divides $M$, so this result and Theorem 2 are in agreement. However, the case when $K$ is even is not covered by Theorem 2.

To summarize, the initial work \cite{Jafar1} is exciting and suggests that it may be possible to allocate resources in a MIMO IC based on DoF.  However, the complexity and design of the interference alignment schemes have presented several challenges to the practicality of this approach for resource allocation.
\begin{itemize}
\item For a given channel realization, to determine whether a given set of DoF tuple is achievable is NP-hard (i.e., exponential effort is likely to be required for large number of users).
\item Without channel extensions, the average DoF per user is shown to be at most $2M/(K+1)$, which is significantly smaller than $M/2$ when there are a large number of independent channel extensions (see \cite{Jafar1}). Here $M$ is the number of antennas at each transmitter and receive.
Notice that the average per user DoF of $2M/(K+1)$ approximately doubles that of the orthogonal approaches (e.g. TDMA or FDMA).
\item It requires too many channel extensions to reap the DoF benefit promised by \cite{Jafar1}.
\item It requires full CSI, which can be difficult for large networks.
\item It often requires selecting a set of feasible DoFs for the users a priori, which is difficult.
\end{itemize}
At this point, interference alignment appears most useful for a small system (e.g., 3-4 links) where a closed form interference alignment solution exists \cite{Jafar1}, and when using no or a small number of channel extensions. For a large network, direct maximization of the weighted sum-rate (or weighted sum utility maximization) seems to offer more potential for resource allocation and interference mitigation. For one thing, it requires the same amount of CSI, and yet can offer more sum-rate performance  across all SNR regime than that of interference alignment. Moreover, it does not require selecting a DoF for each user in advance. As for future work,
we suggest further investigation of the benefits of interference alignment for a small system with a few channel extensions.

\vfill
\bibliographystyle{model2-names}
\bibliography{ref}

\end{document}